\documentclass[12pt]{iopart}

\usepackage{iopams}  
\usepackage{graphicx}

  \expandafter\let\csname equation*\endcsname\relax
  \expandafter\let\csname endequation*\endcsname\relax

\usepackage{amsmath,amsfonts,amssymb}
\usepackage[margin=1in]{geometry}
\usepackage{array}
\usepackage{caption}
\usepackage{tikz}
\usepackage{listings}
\usetikzlibrary{snakes}
\usetikzlibrary{plotmarks}
\usepackage{bm}   
\usepackage{cite} 

\begin{document}

\title[$Q$-colourings of the triangular lattice]
{$\bm{Q}$-colourings of the triangular lattice: \\ 
Exact exponents and conformal field theory}

\author{Eric Vernier$^{1,2,3}$, Jesper Lykke Jacobsen$^{1,2,3}$,
and Jes\'us Salas$^{4,5}$}
\address{${}^1$LPTENS, \'Ecole Normale Sup\'erieure -- 
PSL Research University, 24 rue Lhomond, F-75231 Paris Cedex 05, France}
\address{${}^2$Institut de Physique Th\'eorique, CEA Saclay, 91191 
Gif Sur Yvette, France}
\address{${}^3$Sorbonne Universit\'es, UPMC Universit\'e Paris 6, 
CNRS UMR 8549, F-75005 Paris, France} 
\address{${}^4$ Grupo de Modelizaci\'on, Simulaci\'on Num\'erica y 
Matem\'atica Industrial, Universidad Carlos III de Madrid, Avda.\/ 
de la Universidad, 30, 28911 Legan\'es, Spain} 
\address{${}^5$ Grupo de Teor\'{\i}as de Campos y F\'{\i}sica 
Estad\'{\i}stica, Instituto Gregorio Mill\'an, Universidad Carlos III de 
Madrid, Unidad Asociada al Instituto de Estructura de la Materia, CSIC, 
Madrid, Spain}  

\eads{\mailto{evernier@sissa.it}, \mailto{jesper.jacobsen@ens.fr}, 
      \mailto{jsalas@math.uc3m.es}}

\begin{abstract}
We revisit the problem of $Q$-colourings of the triangular lattice using a 
mapping onto an integrable spin-one model, which can be solved exactly using 
Bethe Ansatz techniques. In particular we focus on the low-energy excitations
above the eigenlevel $g_2$, which was shown by Baxter to dominate the
transfer matrix spectrum in the Fortuin-Kasteleyn (chromatic polynomial)
representation for $Q_0 \le Q \le 4$, where $Q_0 = 3.819\,671\cdots$.
We argue that $g_2$ and its scaling levels define a conformally
invariant theory, the so-called regime~IV, which provides the actual 
description
of the (analytically continued) colouring problem within a much wider range,
namely $Q \in (2,4]$. The corresponding conformal field theory is identified
and the exact critical exponents are derived. We discuss their implications
for the phase diagram of the antiferromagnetic triangular-lattice Potts model
at non-zero temperature. Finally, we relate our results to
recent observations in the field of spin-one anyonic chains.
\end{abstract}

%
%
\section{Introduction}
\setcounter{footnote}{1}

The four-colour problem is probably the best known problem of graph
theory, and has fascinated generations of mathematicians and laymen
since it was first stated in a letter from Augustus de Morgan to Sir
William Rowan Hamilton dated 23 October 1852. The suspense was
resolved in late July 1976 when Appel and Haken \cite{AH77} announced
the proof of the four-colour theorem: Every planar graph admits a
vertex 4-colouring.

A quantitative version of the problem was proposed by Birkhoff
\cite{Birkhoff12} in 1912: Given a simple unoriented 
graph $G$ and a set of $Q$ colours, how many proper vertex colourings 
does $G$ admit?
This defines the chromatic polynomial $\chi_G(Q)$. The four-colour
theorem can then be stated: If $G$ is planar, then $\chi_G(4) > 0$.

Although initially defined for $Q \in \mathbb{N}$, it is easily proved
\cite{Birkhoff12,Whitney32a,Whitney32b} that
$\chi_G(Q)$ is in fact a polynomial in $Q$, and can as such be
evaluated for any $Q \in \mathbb{C}$. 
In particular, if $G=(V,E)$ is a simple unoriented graph with vertex set
$V$ and edge set $E$, its chromatic polynomial can be written as
\begin{equation}
\chi_G(Q) \;=\; \sum\limits_{A\subseteq E} (-1)^{|E|} \, Q^{k(A)} \,,
\label{def.chiG} 
\end{equation}
where the sum is over all spanning subgraphs $(V,A)$ of $G$, and $k(A)$ is
the number of connected components of $(V,A)$.
This suggests an algebraic or even
analytic approach to the colouring problem. 
There exist many studies of the location in
$\mathbb{C}$ of the roots of $\chi_G(Q)$, henceforth called chromatic roots.
These studies concern either specific graphs, or all planar graphs, or
other infinite families of graphs.

The chromatic polynomial is a specialisation of the more general 
Tutte polynomial \cite{Tutte54}. Moreover, in statistical mechanics, 
the chromatic polynomial arises as a special case of the $Q$-state Potts-model 
partition function \cite{Potts52}. 
(See section~\ref{sec:potts}.) 

Many studies have aimed at elucidating the phase diagram of 
the chromatic polynomial,  
the nature of the different phases, and the transitions between them. 
Although a few results are available for non-planar graphs
\cite{JacobsenSalas13}, most of the effort has concentrated on two dimensions 
($G$ planar or embedded in a cylinder or a torus), a case to which we 
specialise henceforth.

Two lines of investigation have proven particularly powerful. In the first, 
the phase diagram is inferred from the study 
of chromatic roots for $Q \in \mathbb{C}$
(see e.g.\/ the series of papers 
\cite{SalasSokal01,JacobsenSalas01,JacobsenSalasSokal03,JacobsenSalas06,%
JacobsenSalas07})
for recursive graph families, corresponding to (finite or infinite) pieces
of the square and triangular lattices with various boundary conditions. 
Similar but quantitatively more precise results have
recently been obtained via a topologically weighted graph polynomial
\cite{JacobsenScullard12,JacobsenScullard13,Jacobsen14,Jacobsen15} on more 
general families of graphs.

In the second series of studies, the colouring problem has been exactly 
solved in the thermodynamic limit along the chromatic line 
$Q\in{\mathbb R}$ for the triangular lattice \cite{Baxter86,Baxter87}.
In this context, solvability of this model means that the exact free energy
density (per site) 
\begin{equation}
f_G(Q) \;=\; \frac{1}{|V|} \, \log \chi_G(Q) 
\label{eq.freeChiG}
\end{equation}
can be computed in closed form. On this solvable line, $\chi_G(Q)$ can be 
transformed into an equivalent vertex model with a Yang--Baxter integrable 
$R$-matrix, leading to an infinite family of commuting transfer matrices. 
Therefore, the triangular-lattice chromatic polynomial is integrable 
for any $Q\in\mathbb{R}$.

\begin{figure}
\begin{center}
\includegraphics[width=200pt]{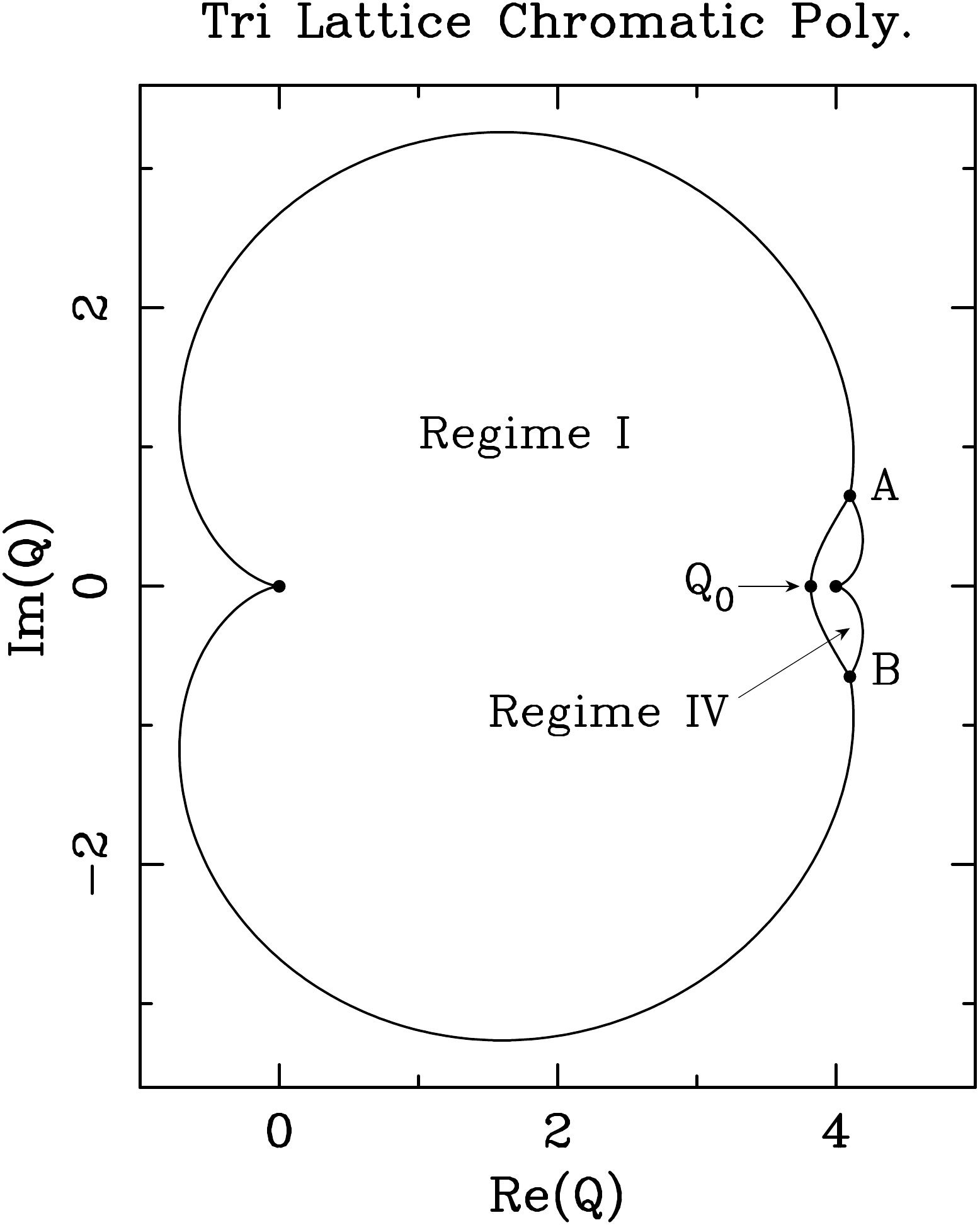}
\end{center}
\caption{
Phase diagram of the triangular-lattice chromatic polynomial in
the complex $Q$-plane \cite{Baxter87}. 
We show the three phases of the model; in each of them, one
of the three functions $g_i(Q)$ found by Baxter \cite{Baxter86,Baxter87}
dominates, and the curves correspond to the loci where two of them coincide
in absolute value: $|g_i| = |g_j| > |g_k|$. At the bifurcation points $A$ and 
$B$ ($Q \approx 3.859\, 627 \pm 0.203\, 154\,i$), 
all three functions are equimodular.
The outer phase (where $g_1$ dominates) is non-critical. The other two 
phases are critical: the ground state of regime~I (resp.\/ regime~IV) 
is given by $g_3$ (resp.\/ $g_2$). But one can extend analytically the 
ground state and its excitations of either of these two regimes 
inside the other,
so that in each regime, we have the superposition of two distinct 
critical spectra. The colouring problem is actually described by the 
(analytically continued) ground state and excitations of regime~IV.  
Furthermore, in regime~I at $Q=B_k$ \eqref{eq.Bk}, 
both the ground state $g_3$ and its scaling levels will disappear from the 
partition function (\ref{def.chiG}).
}
\label{fig:trichrompolypd}
\end{figure}

The chromatic polynomial on the triangular lattice
has only been studied directly through the computation of the bulk free 
energy \cite{Baxter86,Baxter87}, or indirectly via a
series of transformations to an O($n$) loop model on the hexagonal lattice 
\cite{Nienhuis82}. Baxter found \cite{Baxter86} that in the 
complex $Q$-plane the free energy density \eqref{eq.freeChiG} is described 
by three distinct analytic functions $g_i(Q)$ ($i=1,2,3$).  
In addition, he computed the phase diagram of the chromatic polynomial
in the complex $Q$-plane (see figure~\ref{fig:trichrompolypd}), by finding  
on which region each function $g_i$ was dominant, 
and locating the loci where
two of these functions are dominant and coincide in absolute 
value \cite{Baxter87}.

In this general picture, it is clear that the chromatic polynomial is critical 
only in the interval $Q\in [0,4]$. Outside this interval, the model is 
non-critical, meaning that correlations of local observables decay 
exponentially fast with the distance. However, the determination of the 
critical exponents (which are expected to vary continuously, as in other 
integrable systems with a free deformation parameter---here $Q$), 
and the identification of the corresponding conformal 
field theory (CFT) are two important ingredients that are so far still missing. 

A major difference with other well-known integrable models is that
the critical interval, $Q \in [0,4]$, allows for not just one, but 
{\em two} analytically 
unrelated expressions $g_2$ and $g_3$ \cite{Baxter86} for the free energy 
(\ref{eq.freeChiG}). 
Each one defines a different ``phase'' or regime. We shall call regime~I
the critical theory described by the ground state $g_3$ and the set of 
conformally invariant scaling levels
above it. Similarly regime~IV will denote the critical theory corresponding
to the ground state $g_2$. 

It is important to understand that this definition of regimes~I and~IV
does not rely on any considerations
whether the corresponding ground state eigenvalue---that is, $g_3$ 
for regime~I, and $g_2$ for regime~IV---is dominant in
absolute terms. Baxter has indeed argued \cite{Baxter87} that
$g_3$ is dominant (i.e., the largest in absolute value among all
transfer matrix eigenvalues) on the interval 
$0 \le Q \le Q_0$, with $Q_0 = 3.819\,671\cdots$, whereas $g_2$
is dominant on $Q_0 \le Q \le 4$---but this is {\em not} the point here.
We shall instead show that regime~I makes sense for all $0 \le Q < 4$, 
whereas regime~IV makes sense (at least) for $2 < Q \le 4$.
The scaling levels of either regime can in principle be followed (by analytic
continuation) in the Bethe Ansatz (BA) computations, or by direct 
diagonalisation of the transfer matrix (TM),
even into a range where the regime's ground state is sub-dominant with respect
to that of the other regime.
So one of our principal findings is that regimes~I and~IV are two superimposed 
critical 
theories that coexist, so to speak, as a pair of interchangable Russian dolls.

This complication is presumably concomitant with the fact that the standard 
programme in statistical mechanics to find phase transitions and determine 
their critical properties in a given system, has not yet been carried out 
for this model. One of the main results of the present work is to identify the 
critical exponents and CFT of both regimes.
Interestingly, the more exotic regime~IV is {\em not} present in the 
closely related hexagonal-lattice O($n$) loop model, and it turns out to be 
related with a CFT describing a spin-one anyonic chain \cite{Ard}. 

Using different techniques, including both BA and   
TM computations, we moreover conclude that regime~IV is 
the physical meaningful continuation of the critical colouring problem to 
$Q \in \mathbb{R}$, whereas regime~I can best be described (from this
point of view) as some universal and unphysical ``junk'' that has 
no relation to $Q$-colourings of the triangular lattice. 
In particular, we shall see that regime~I---namely its ground state 
$g_3$ and the
corresponding excitations---will disappear from the partition function 
(\ref{def.chiG}) 
at values of $Q$ that are relevant to the colouring problem; 
i.e., the Beraha numbers $B_k$
\begin{equation} 
 B_k \;=\; 4 \, \cos^2 \left( \frac{\pi}{k} \right) \,, 
 \qquad k \;=\; 2,3,4,\ldots \,.
 \label{eq.Bk}
\end{equation}
In some sense, that will be made precise below, regime~IV at the Beraha 
numbers therefore provides the maximal physical extension of the colouring 
problem
with local observables. This is why a large part of this work is dedicated 
to unravelling the critical properties of regime~IV.

Before going on, we outline the contents of the paper. 
In section~\ref{sec:pd}, we will introduce the more general 
$Q$-state Potts model, which, in addition to the number of states or 
colours $Q$, has a temperature-like parameter $v$. In this section, 
we will describe the main features of the phase diagram in the real 
$(Q,v)$-plane
of the $Q$-state Potts model on a 2D regular lattice, using the 
best-known square lattice case as a guide. We will make emphasis on its
antiferromagnetic (AF) and unphysical regimes. This section will provide
the necessary background to discuss in detail the lesser known phase diagram 
of the triangular-lattice Potts model. 

After this introductory section, we will start to discuss our 
results. Section~\ref{sec:mappings} constitutes the first step in our 
programme. It provides a series of mappings that connect the 
triangular-lattice chromatic polynomial $\chi_G(Q)$ with a spin-one vertex 
model on the square lattice. The former model is defined on a strip graph
with cylindrical boundary conditions. A key point in all these mappings is 
the introduction of a \emph{twist} that ensures that the original 
boundary conditions are treated correctly. This ingredient was
not essential to obtain the bulk free energies \cite{Baxter86,Baxter87} 
(and hence was not discussed in those papers);
but it has proven to be important in order to obtain critical exponents.

Our Bethe-Ansatz work on the above-mentioned spin-one vertex model is
presented in section~\ref{sec:BA}. Results for regime~I can be taken over, 
without much ado, from
existing work on an integrable 7-vertex model \cite{BatchelorBlote} on the 
honeycomb lattice, upon imposing the correct twist that ensures its equivalence
with the triangular-lattice Potts model. 
The main endeavour however concerns regime~IV, where two Fermi 
seas are present. In section~\ref{sec:criticalcontent}, we obtain 
(using analytic techniques and some numerical checks) the conformal weights 
of various operators, culminating in the precise identification of the 
corresponding coset CFT for regime~IV. The corresponding rational 
theories take the form of RSOS models, which we identify with recently studied 
spin-one anyonic chains \cite{Ard} in section~\ref{sec:anyonic}. Finally, 
section~\ref{sec:conclusion} contains our conclusions.

%
%
\section{Basic setup}
\label{sec:pd}
\setcounter{footnote}{1}

In this section, we will first describe the $Q$-state Potts model in
the spin and Fortuin--Kasteleyn (FK) representations.
In the AF case, the zero-temperature limit of the
partition function specialises to the already 
defined chromatic polynomial \eqref{def.chiG}. 
We will then discuss the physics of the best-known example: the square-lattice 
Potts model. Many features of the corresponding phase diagram are generic
to all two-dimensional lattices, including the existence of an unusual
phase called the Berker--Kadanoff phase. A more detailed discussion
of this phase is contained in a separate subsection. 
Finally, we will review what is known about the triangular-lattice Potts 
model.  

%
%

\subsection{The $Q$-state Potts model} 
\label{sec:potts}

In statistical mechanics, the chromatic polynomial arises as a special case 
of the Potts model \cite{Potts52}. Let $G=(V,E)$ be an unoriented graph 
with vertex set $V$ and edge set $E$, and attach to each vertex $i \in V$ 
a spin variable $\sigma_i \in \{1,2,\ldots,Q\}$, where $Q \in \mathbb{N}$. 
The partition function of 
the Potts model (in the \emph{spin representation}) then reads
\begin{equation}
 Z^{\rm Potts}_G(Q,K) \;=\; \sum_{\{\sigma_i\}} \, 
    \prod_{\langle i,j\rangle} \,
    \mathrm{e}^{K \, \delta_{\sigma_i, \sigma_j}} \,,
 \label{eq:ZPotts}
\end{equation}
where $K = J/T$, the interaction energy between adjacent spins is 
$-J \delta_{\sigma_i,\sigma_j}$, and $T$ denotes the temperature. It is easy 
to see that in the AF ($J < 0$) case, the colouring
problem arises in the $T \to 0$ limit, namely
\begin{equation}
 \chi_G(Q) \;=\; \lim_{K \to -\infty} Z^{\rm Potts}_G(Q,K) \,.
\end{equation}

Fortuin and Kasteleyn \cite{FK1972} proved that the partition function 
$Z^{\rm Potts}_G$ \eqref{eq:ZPotts} could be rewritten as:
\begin{equation}
 Z^{\rm Potts}_G(Q,K) \;=\; \sum_{A \subseteq E} v^{|A|} \, Q^{k(A)} 
                      \;\equiv\; Z^{\rm FK}_G(Q,v) \,,
\label{def.Z_FK}
\end{equation}
where $|A|$ is the number of edges in the subset $A \subseteq E$, 
$k(A)$ denotes the number of connected components in the spanning subgraph 
$(V,A)$, and the temperature-like variable $v$ is given by
\begin{equation}
v \;=\; e^{J} - 1 \,.
\label{def:v}
\end{equation}
Therefore, the Potts-model partition function is a polynomial in the 
variables $Q,v$, and the polynomial character of its chromatic specialisation, 
$\chi_G(Q) = Z^{\rm FK}_G(Q,-1)$, is now obvious [cf.~\eqref{def.chiG}].
This FK representation of the Potts model is useful to make sense of 
non-integer values of $Q$ and/or imaginary values of the coupling constant $J$ 
(i.e., $v<-1$). In this representation the AF regime 
corresponds to $v\in[-1,0)$, while the ferromagnetic one is given by 
$v\in (0,\infty)$.  

As for the chromatic polynomial, many studies have aimed at elucidating 
the phase diagram of the $Q$-state Potts model, mostly in two dimensions. 
In particular, the zeros of the partition function have been investigated (see 
\cite{ChangSalasShrock02,ChangJacobsenSalasShrock04} and references therein)
for recursive strip graphs of the square and triangular lattices with various 
boundary conditions.
In addition, the use of the topologically weighted graph polynomial defined in 
\cite{JacobsenScullard12,JacobsenScullard13,Jacobsen14,Jacobsen15}, has 
provided more precise results on more general families of graphs. 

\medskip

\noindent
{\bf Remarks.} 
1. The Potts-model partition function \eqref{eq:ZPotts} can be defined 
on any unoriented graph $G$, not necessarily simple. 

2. In graph theory the two-variable polynomial $Z^{\rm Potts}_G(Q,v)$ 
[cf.~\eqref{def.Z_FK}] is known---after an innocuous change of variables---as 
the Tutte polynomial \cite{Tutte54}. 

3. The authors of Refs.~\cite{ChangSalasShrock02,ChangJacobsenSalasShrock04}  
considered the partition-function zeros for integer values of $Q\ge 2$ 
(resp.\/ real values of $v\ge -1$) in the complex $v$ (resp.\/ $Q$) plane. 

%
%
\subsection{The square lattice}
\label{sec:pd_sq}

The phase diagram of the Potts model is best 
understood for the square-lattice model. In particular, Baxter 
\cite{Baxter73,Baxter82}
was able to solve exactly this Potts model in the thermodynamic limit on
specific curves in the $(Q,v)$-plane. On these 
solvable curves the model becomes integrable in the sense discussed in the
Introduction. In particular, the square-lattice Potts model is integrable 
along the selfdual curves \cite{Baxter73}
\begin{equation}
 v \;=\; \pm \sqrt{Q}
 \label{FM_square}
\end{equation}
and along two mutually dual AF curves \cite{Baxter82}
\begin{equation}
 v_\pm \;=\; 2 \pm \sqrt{4-Q} \,.
 \label{AF_square}
\end{equation}

Integrability studies lead first to the exact expression of the bulk free 
energy, from whose singularities some features about the phase transitions 
can be inferred. It is found that when moving across the self-dual curves 
\eqref{FM_square} in the $v$-direction,
the system undergoes a first-order (resp.\/ second-order) phase transition
when $Q > 4$ (resp.\/  when $0 \le Q \le 4$). When crossing \eqref{AF_square} 
the transition is simultaneously first {\em and}
second order \cite{JacobsenSaleur06}. Moreover, when moving across $Q=4$ in 
the direction along the integrable curves,
the free energy exhibits an essential singularity \cite{Baxter_book}. There 
are also singularities in the surface and corner free
energies in this case \cite{Jacobsen10,VernierJacobsen12}.

Within the critical regime $0 \le Q \le 4$ much further information can be 
obtained. Along the integrable curves one can compute
various critical exponents, which will typically vary continuously with $Q$. 
These exponents provide crucial information that can
help identifying the CFT describing the continuum limit. This program has 
been carried out for the selfdual curves \eqref{FM_square}, and the 
corresponding CFT is found to be that of a compactified boson, that can be 
effectively described by the Coulomb gas (CG) technique \cite{JacobsenReview}. 
This is arguably the simplest possible CFT with continuously varying exponents.

The coupling constant $g$ of the CG that describes the continuum limit 
along the selfdual curves \eqref{FM_square} is related to the temperature-like
parameter $v$ as: 
\begin{equation}
 v \;=\; -2 \cos(\pi g) \,, \qquad Q \;=\; v^2 \,.
\end{equation}
This CG coupling constant satisfies $\frac{1}{2} \le g \le 1$ for the 
ferromagnetic branch $v = +\sqrt{Q}$, 
while $0 < g \le \frac{1}{2}$ for the other (`unphysical selfdual' 
\cite{Saleur91}) branch $v = -\sqrt{Q}$. The 
thermal operator has critical exponent \cite{Nienhuis84,JacobsenReview}
\begin{equation}
 x_T \;=\;  \frac{3}{2g} - 1
 \label{thermexp}
\end{equation}
and is conjugate to a perturbation in the temperature variable $v$ around the 
critical curve \eqref{FM_square}. In particular, the temperature perturbation
is relevant ($x_T \le 2$) along the upper branch of \eqref{FM_square}, 
which is identified with the ferromagnetic critical curve 
$v_\text{F}(Q)=+\sqrt{Q}$. 
On the other hand, this temperature perturbation is irrelevant ($x_T > 2$) 
along the lower branch of \eqref{FM_square}, which will hence act as a basin 
of attraction for a finite range of $v$-values. 

This basin is delimited by the AF curves \eqref{AF_square} 
and is called the Berker--Kadanoff (BK) phase \cite{Saleur91}.
The BK phase is however unphysical, in the sense that all the scaling levels 
corresponding to the CG description disappear from the spectrum
whenever $Q$ is equal to a Beraha number $B_k$ \eqref{eq.Bk}.
Two distinct representation theoretical mechanisms are responsible for this 
phenomenon. First, the multiplicity (also known as amplitude, or quantum dimension) of certain eigenvalues of the 
corresponding TM vanishes at $Q = B_k$, as can be seen from a 
combinatorial decomposition of the Markov trace \cite{Richard06,Richard07}. 
Second, the representation theory of the quantum group $U_\mathfrak{q}(sl_2)$ 
for $\mathfrak{q}$ a root of unity (with 
$\sqrt{Q} = \mathfrak{q} + \mathfrak{q}^{-1}$) guarantees that other 
eigenvalues are equal in norm at $Q = B_k$ \cite{PasquierSaleur90,Saleur91}, 
and as their combined multiplicity is zero, they vanish from the spectrum as 
well. As a result, for $Q = B_k$ only local observables remain, and these 
can be realised in an equivalent RSOS height model \cite{Pasquier}.

Finally, the continuum limit along the curves $v_\pm$ \eqref{AF_square} 
that bound the BK phase, gives rise to a more exotic CFT 
\cite{JacobsenSaleur06,IkhlefJacobsenSaleur08} with one compact and one 
non-compact boson that couple to form the ${\rm SL}(2,\mathbb{R})/{\rm U}(1)$ 
black hole sigma model \cite{IkhlefJacobsenSaleur12}, 
familiar in the string-theory context \cite{Witten91,DijkgrafVerlinde92}. 

This scenario---the details of which depend somewhat on the boundary 
conditions (free, cylindrical, cyclic, toroidal)---has been expounded in 
detailed studies of partition function zeros 
\cite{SalasSokal01,JacobsenSalas01,JacobsenSalas06,JacobsenSalas07}. 
Indeed, either of the two mechanisms for eigenvalue cancellations outlined above
have an incidence on the accumulation points of partition function zeros, by
the Beraha-Kahane-Weiss theorem.
In particular, the chromatic line $v=-1$ intersects the BK phase
for $0 \le Q \le 3$ on the square lattice, so the accumulation of chromatic 
zeros is confined to this 
region. For $Q > 3$ the chromatic line renormalises to infinite temperature 
($v=0$) and is hence non-critical.

\medskip

\noindent
{\bf Remarks.} 1. The lower branch of \eqref{FM_square} (i.e.,  
$v_\text{BK}=-\sqrt{Q}$) is the analytic continuation of the upper branch 
of \eqref{FM_square}, viz., of the critical ferromagnetic curve $v_\text{F}$.

2. The 3-state Potts antiferromagnet on the square lattice is critical on the
chromatic line $v=-1$, but when the temperature is non-zero (i.e., $-1 < v <0$)
it is disordered; so it renormalises to $v=0$. 

%
%
\subsection{The Berker-Kadanoff phase}

The BK phase actually exists on any lattice and governs a finite part of the 
AF ($v<0$ and $Q > 0$) part of the phase diagram for the 
following reasons, which were succinctly exposed in 
\cite{JacobsenSaleur08bdrychrom}.
First, the ferromagnetic transition curve $v_{\rm F}$---generalising the 
upper branch 
of \eqref{FM_square}---must exist on any lattice, by the universality of the 
CG description with $\frac{1}{2} \le g \le 1$. In the $(Q,v)$-plane this curve 
must have a vertical tangent at the origin, since the corresponding 
symplectic fermion CFT (with central charge $c=-2$) is known to describe 
spanning trees \cite{DuplantierDavid88}.
Indeed, if that tangent was not vertical, we would have instead a 
spanning-forest model \cite{JSS05}, which corresponds to perturbing the 
free fermion CFT by a four-fermion term \cite{CJSSS04}, which would render 
it non-critical \cite{CJSSS04,JacobsenSaleur05}.

This means that the critical curve $Q(v)$ has a vanishing derivative 
$Q'(0) = 0$ on any lattice.%
\footnote{This can indeed be checked explicitly for the square and triangular 
lattices, where the exact critical curve is known; see \eqref{FM_square} 
and \eqref{cubic_tri}.
}
Barring the (unlikely) accident that also $Q''(0) = 0$, we infer
that the critical curve continues analytically from the first 
quadrant $Q,v>0$ into the fourth quadrant $Q>0$, $v<0$. 
Invoking again the universal CG description, now with $g < \frac{1}{2}$, 
we have $x_T > 2$ by \eqref{thermexp}, and a finite range of $v$-values will 
indeed be controlled by the BK critical curve $v_\text{BK}$.
This BK phase is bounded in the $(Q,v)$ plane by two smooth curves $v_\pm$ 
such that for any $Q\in[0,Q_\text{c})$, $0\ge v_{+}(Q) > v_{-}(Q)$. The two
curves merge at some value $Q_\text{c} > 0$, signaling the termination of the 
BK phase. The inequality $Q_\text{c} \le 4$ 
for two-dimensional models 
follows from quantum group results \cite{Saleur91}. 

The upper curve is usually identified with the critical AF 
curve, $v_{+} = v_\text{AF}$, and hits the point $(Q,v)=(0,0)$ at a finite 
negative slope. This value is related to the critical coupling of the 
corresponding spanning-forest model \cite{JSS05}.

Some unusual features of the BK phase have already been pointed out
in the previous section about the square-lattice model, but they hold 
true for any two-dimensional lattice:
At $Q=B_k$, there are vanishing amplitudes and eigenvalue cancellations, so 
that the actual ground state is deeply buried in the spectrum of the 
corresponding TM. We now present another related argument, this time directly 
in the continuum limit, leading to the conclusion that the BK phase
should be dismissed as unphysical.
The critical behaviour of the BK phase can be obtained by analytic 
continuation of the CG results for the ferromagnetic transition 
($g \ge \frac{1}{2}$). It turns out \cite{Saleur90} that some of the critical 
exponents (namely, magnetic and 
watermelon exponents---to be defined in the next section) become negative when 
$g< \frac{1}{2}$. This means that the analytic continuation of the 
ferromagnetic ground state is not anymore the lowest-energy state in
the BK phase, or alternatively, that correlation functions will not decay 
but rather grow with distance. Such features are clearly unphysical and 
non-probabilistic.

The existence of the BK phase has been verified for all Archimedean 
lattices in \cite{JacobsenScullard13,Jacobsen14}, and its extent in 
the $(Q,v)$ plane has been accurately estimated. It was found that in most, 
but not all \cite{JacobsenScullard12}, cases $Q_\text{c} = 4$.

The critical behaviour on the chromatic line $v=-1$ (for general 
$Q \in \mathbb{R}$) then depends in a crucial way on its 
intersection with the BK phase. We have seen above that this intersection
is the interval $0 \le Q \le 3$ for the square lattice. The remainder of the 
interval $[0,4]$, namely $3 < Q \le 4$ in this case, will then have a 
different behaviour. For the square lattice this behaviour is just 
non-critical. The triangular lattice however offers a much more interesting 
alternative, as we shall see presently.

\medskip

\noindent
{\bf Remark.} For non-planar recursive lattices \cite{JacobsenSalas13}, the 
BK phase is also conjectuted to exist, and the value for $Q_\text{c}$ might
be greater than 4. 

%
%
\subsection{The triangular lattice}
\label{sec:pd_tri}

The $(Q,v)$ phase diagram on the triangular lattice appears to be 
considerably more complicated than was the case on the square lattice.
In particular the part close to $Q=4$ possesses several intricate features, 
as witnessed by numerical studies of different types
\cite{JacobsenSalasSokal03,JSS05,JacobsenSalas06,JacobsenSalas07,%
JacobsenSalasScullard15}. 
This model is integrable along the three branches of the 
cubic \cite{BaxterTemperleyAshley78}
\begin{equation}
 v^3 + 3 v^2 \;=\; Q \,.
 \label{cubic_tri}
\end{equation}
Fortunately, the chromatic line 
\begin{equation}
 v \;=\; -1 \,.
 \label{chrom_tri}
\end{equation}
is also integrable \cite{Baxter86,Baxter87} and understanding it in detail 
should provide valuable information that will also be useful for
disentangling the full $(Q,v)$ phase diagram.

\begin{figure}
\begin{center}
\includegraphics[width=200pt]{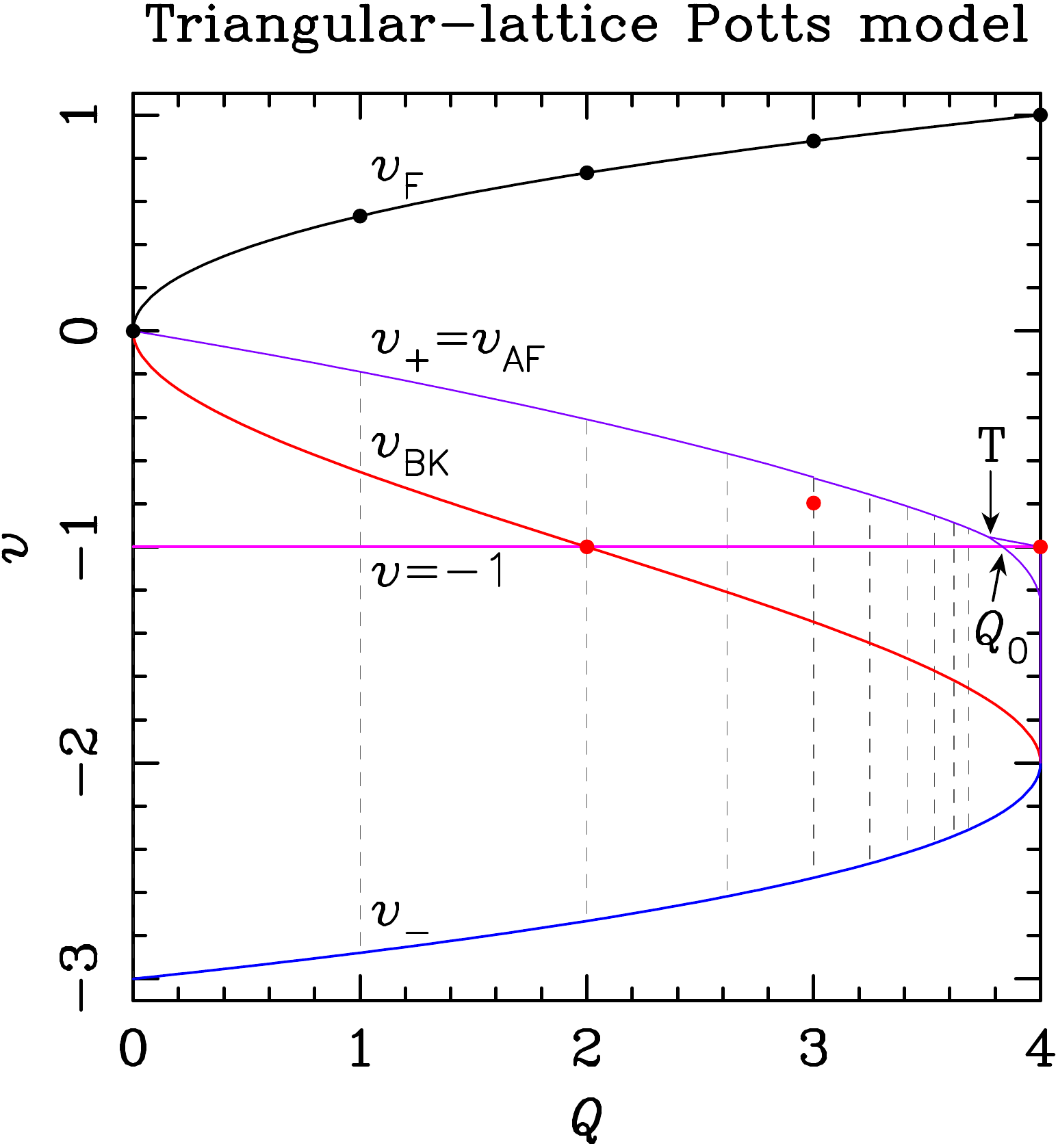}
\end{center}
\caption{Phase diagram of the triangular-lattice $Q$-state Potts model in the 
real $(Q,v)$ plane. The cubic \eqref{cubic_tri} corresponds
to the three thick lines labeled $v_\text{F}$, $v_\text{BK}$, and $v_{-}$, 
respectively. The horizontal line corresponds to $v=-1$ 
\eqref{chrom_tri}. The curve between $v_\text{F}$ and  $v_\text{BK}$ is 
our best numerical estimate of the AF curve $v_+ =v_\text{AF}$
for this lattice \cite{JacobsenSalasScullard15}. This curve has a 
bifurcation point T, leading to two branches: one goes to
the point $(Q_0,-1)$ [cf. \eqref{eq.Q0}], while the other goes to the 
critical point $(4,-1)$. The region defined by these three points and 
$(4,-1.253)$ is governed by the regime~IV that we will consider hereafter, 
while the BK phase corresponds to regime~I. 
(Compare to figure~\ref{fig:trichrompolypd}.)
The vertical dashed lines correspond to the 
Beraha numbers $B_k$ [cf.~\eqref{eq.Bk}], shown here for $3\le k\le 11$. 
The solid dots depict the known critical points for integer values of $Q$. 
}
\label{fig:tripd}
\end{figure}

A schematic phase diagram which is compatible with our current understanding 
is shown in figure~\ref{fig:tripd}. 
Let us parameterise the cubic curve \eqref{cubic_tri} by
\begin{equation}
 v \;=\; -1 + 2 \cos \left( \frac{2\pi(1-g)}{3} \right) \,, \qquad
 Q \;=\; 4 \cos^2 \left(\pi g \right)
\end{equation}
where $-\frac12 \le g \le 1$. The range $0 < g \le 1$ then has then same CG 
interpretation as for the square lattice: 
$\frac{1}{2}\le g\le 1$ corresponds to $v_\text{F}$, and 
$0<g\le \frac{1}{2}$ to $v_\text{BK}$. 
The interval $-\frac12 \le g < 0$ (corresponding to $v_{-}$)
cannot be interpreted within this CG, but has been shown numerically 
\cite{JacobsenSaleur06} to belong to the same universality class 
\cite{IkhlefJacobsenSaleur08,IkhlefJacobsenSaleur12}
as the {\em antiferromagnetic} curve \eqref{AF_square} for the square-lattice 
Potts model (with the same value of $Q$).

This implies that the middle branch $v_\text{BK}$ of \eqref{cubic_tri} 
will control the BK phase, which should extend from the lower branch 
of \eqref{cubic_tri}, that we hence identify with $v_-$, up to some  
AF transition curve $v_\text{AF}=v_{+}$, that unfortunately has not yet been 
determined analytically.
This curve cannot be the chromatic line \eqref{chrom_tri}, since the 
latter is not always above the middle branch of \eqref{cubic_tri}. At the 
origin, $(Q,v) = (0,0)$, the AF curve describes a spanning-forest problem 
and its slope is known numerically \cite{JSS05}. It has be followed 
numerically to higher $Q$ in \cite{JacobsenSalasScullard15}, 
and other pieces of yet unpublished work, and is depicted in 
figure~\ref{fig:tripd}.

Along the chromatic line $v=-1$, Baxter has found \cite{Baxter86} three 
distinct analytic expressions for the TM eigenvalue $g_i$, 
whose logarithm is the bulk free energy (per site): 

\begin{eqnarray}
 g_1(Q) &=& -\frac{1}{x} \prod_{j=1}^\infty
 \frac{(1-x^{6j-3})(1-x^{6j-2})^2(1-x^{6j-1})}
      {(1-x^{6j-5})(1-x^{6j-4})(1-x^{6j})(1-x^{6j+1})} \,, 
      \label{Baxter_g1} \\[2mm]
 \log g_2(Q) &=& \int_{-\infty}^{\infty} {\rm d}k \, 
 \frac{\sinh k \theta}{2 k} \left(
 \frac{\sinh[k(\pi-2\theta)/2]}{(2 \cosh k\theta - 1) 
       \sinh(\pi k/2)} \right. \nonumber \\[2mm]
 & & \qquad \qquad \qquad \qquad \left. 
-\frac{\cosh[k(\pi-2\theta)/2]}{(2 \cosh k\theta +1) 
       \cosh(\pi k/2)} \right) \,, 
 \label{Baxter_g2} \\[2mm]
 \log g_3(Q) &=& \int_{\infty}^\infty {\rm d}k \,
 \frac{\sinh k \theta  \, [\sinh k(\pi-\theta)]}
      {k \sinh \pi k \, [2 \cosh k(\pi-\theta)-1]} \,, \label{Baxter_g3}
\end{eqnarray}

\noindent
where we have parameterised
\begin{equation}
 Q \;=\; 2 - x - x^{-1} \;=\; 2 + 2 \cos \theta \,, 
 \label{eq.def.theta}
\end{equation}
with $|x| < 1$ and $0 < \text{Re}(\theta) < \pi$. 
(See Ref.~\cite{Baxter86} for details.)

In his second paper \cite{Baxter87}, Baxter determined the regions of 
$Q \in \mathbb{C}$ for which each of the expressions
\eqref{Baxter_g1}--\eqref{Baxter_g3} is dominant (see 
figure~\ref{fig:trichrompolypd}). This was done by 
investigating the ratios $g_2/g_1$ and $g_3/g_1$,
which can both the expressed as infinite products, by a combination of 
symmetry arguments and numerical analysis. His final result is:
\begin{enumerate}
 \item $g_1$ is dominant for $Q \in (-\infty,0) \cup (4,\infty)$. 
 \item $g_2$ is dominant for $Q\in (Q_0,4)$. 
 \item $g_3$ is dominant for $Q\in (0,Q_0)$.
\end{enumerate}
where \cite{Baxter87,JacobsenSalasSokal03}
\begin{equation}
 Q_0 = 3.819\,671\,731\,239\,719 \cdots \,.
 \label{eq.Q0}
\end{equation}

Returning now to the phase diagram in the $(Q,v)$-plane, the simplest 
scenario that can account for these findings is that
the AF transition curve must split before $Q_0$ into a branch 
that goes down to $(Q,v) = (Q_0,-1)$ and
delimits the BK phase, and another branch that extends further to the 
4-colouring point $(Q,v) = (4,-1)$. This scenario, which
is shown schematically in figure~\ref{fig:tripd}, is confirmed by studies of 
the limiting curves and the critical polynomial 
\cite{JacobsenSalasScullard15}.
The region $v < -1$ appears to contain further intricate features, which are 
beyond the scope of the present discussion.

The region (i) where $g_1$ dominates is non-critical. In that region, 
exact expressions for the surface and corner free energies are also known
\cite{Jacobsen10}, and the contributions from corners of angle 
$\frac{\pi}{3}$ and $\frac{2\pi}{3}$ can even be determined
separately \cite{VernierJacobsen12}. In the limit $Q \to 4^+$, these corner 
free energies develop essential singularities that have been determined from 
the asymptotic analysis of the exact expressions \cite{VernierJacobsen12}.

For the remainder of this paper we shall restrict the attention to the 
critical regions (ii) and (iii). For reasons that will become clear from 
the mappings in section~\ref{sec:mappings} to a spin-one vertex model,
we shall refer to the critical theory that dominates in region (iii) 
as {\em regime I}, and to the theory dominating (ii) as {\em regime IV}.
(See figure~\ref{fig:trichrompolypd}.)
Our objective is to extend Baxter's analysis to compute the exact critical 
exponents in either of these regimes, and determine the corresponding CFT.

Regime~I is relatively straightforward to deal with, using the mapping to the 
O($n$) model \cite{Nienhuis82} and the BA analysis of the 
associated 7-vertex model \cite{BatchelorBlote}.
Choosing a particular twist of the vertex model to ensure the equivalence 
with the Potts model will enable us to
compute the corresponding critical exponents; see section~\ref{sec:regimeI}.
The end result is that regime~I is {\em also} described by the BK 
universality class. It is remarkable that the triangular-lattice Potts model 
has two distinct integrable curves, \eqref{chrom_tri} and the 
middle branch of \eqref{cubic_tri}, with the same critical behaviour.

Regime~IV is our main interest here. It cannot be obtained within the 
7-vertex model \cite{BatchelorBlote}, and indeed in that reference the 
existence of the region $Q\in (Q_0,4)$ was relegated to a terse footnote. 
However, using further mappings to a 19-vertex model 
(see section~\ref{sec:mappings}),
we shall obtain an efficient handle on regime~IV and derive the corresponding 
critical exponents and CFT interpretation. Achieving this is going to be the 
principal motivation for the remainder of this paper.

\medskip

\noindent
{\bf Remarks.} 1. Although Baxter's analysis of the eigenvalues 
\eqref{Baxter_g1}--\eqref{Baxter_g3} has been subject of some debate 
\cite{JacobsenSalasSokal03}---in particular concerning the
precise numerical evaluation of the infinite products---his conclusions 
for $Q \in \mathbb{R}$ have withstood a number of stringent numerical tests 
\cite{JacobsenSalasScullard15}. 

2. In the square-lattice case, we could say that $Q_0=3$. The corresponding 
chromatic
polynomial does not define an integrable theory. However, since the region 
$0 \le Q < Q_0$
is attracted to $v_{\rm BK}$ under the renormalisation group flow, it can be 
identified with regime~I.
The behaviour right at $Q=Q_0$ is given instead by the AF curve $v_+$.

3. The extension of Baxter's work to the computation of critical exponents, by
means of a BA analysis, is deferred to section~\ref{sec:BA}, where a twist 
should be included in the BA equations. 

%
%
\section{From the chromatic polynomial to a spin-one vertex model}
\label{sec:mappings}
\setcounter{footnote}{1}

Baxter \cite{Baxter86} found an ingenious coordinate BA for the 
triangular-lattice chromatic polynomial,
involving several kinds of particle trajectories (that he represented 
as dotted, full, double-full and wiggly lines).
In view of the subsequent developments in the classification of integrable 
systems, it appears preferable to
instead bring the chromatic polynomial into contact with a (by now) 
well-studied integrable model to which we can apply all the (by now)
standard BA machinery. In this section we therefore reexpress the 
colouring problem in terms of an integrable, 
spin-one vertex model on the square lattice, known as the Izergin--Korepin 
(IK) (or $a_2^{(2)}$) model.

The route that we shall take passes through quite a number of different 
formulations as intermediate steps. We emphasise that although each of the 
mappings
exhibited here is exact, the various models appearing have different domains 
of validity. Some require
$Q \in \mathbb{N}$, while other allow for any $Q \in \mathbb{R}$; sometimes 
the twist (that determines the weight of
non-contractible clusters and loops) is a free parameter, and sometimes 
it is fixed.
In more technical terms, each model provides a distinct representation of the 
underlying Temperley-Lieb
(or dilute TL) algebra, which is not always faithful with respect to the most 
general formulation
(allowing for arbitrary values of $Q$ and the twist). It follows that some 
of the eigenstates---and hence
some of the critical exponents---are absent from some of the models presented, 
and not all of the
correlation functions can be directly interpreted in terms of the original 
colouring problem. To keep track of these subtleties, our discussion puts a
special emphasis on the definition of the corresponding transfer matrices, 
as  well as on the various physical observables of interest. 

Another key point is to keep carefully track of the twist. The bulk free 
energy obtained by Baxter \cite{Baxter86} is
independent on the twist, so he did not need to introduce it. However, 
since we are eventually aiming at exact
expressions for the critical exponents---which are sensitive to the 
boundary condition defined by the twist---we
need to ensure that each model in the series of mappings is identically 
twisted.

It is worth stressing that we will start from the colouring problem defined 
on a strip of the triangular lattice with \emph{cylindrical} boundary 
conditions and a width $L$ multiple of 3. This choice allows for a 3-colouring 
for such strips, with vertices of each of the three sublattices having constant
colour. Strip graphs of widths $L$ not a multiple of 3 correspond to a 
topologically frustrated
ground state, depending on the value of $L \pmod 3$, and
will be considered elsewhere \cite{VernierJacobsenSalas15}.

\subsection{Colouring problem}

We start from the Potts model defined on a strip of the triangular lattice 
$G=(V,E)$, whose partition function we recall to be given by 
\begin{equation}
 Z^{\rm Potts}_G(Q,K) \;=\; \sum_{\{\sigma_i\}} \, 
    \prod_{\langle i,j\rangle} \,
    \mathrm{e}^{K \, \delta_{\sigma_i, \sigma_j}} \,,
\label{eq:ZPottsBis}
\end{equation}
where $Q\in {\mathbb N}$ and $K\in {\mathbb R}$. The colouring problem is 
recovered for $K \to  -\infty$.
Viewing the longitudinal direction of the strip as imaginary time, we can 
build the system row by row acting with the TM, which we represent as follows:
\medskip
$$
\begin{tikzpicture}[scale=2]
   \foreach \x in {0,...,5}
{ 
    \draw (\x-0.5,0) -- (\x+0.5,0);
    \draw (\x,1) -- (\x+1.,1);
    \draw (\x-0.25,0.5)-- (\x,0) -- (\x+0.5,1) -- (\x+0.75,0.5);
    \draw[fill=white] (\x,0) circle[radius=2pt];
    \draw[fill=white] (\x+0.5,1) circle[radius=2pt];
}

\draw[dashed] (-0.5,0) -- (0,1);
\draw[dashed] (5.5,0) -- (6,1);
\node[below] at (0.,-0.2) {$1$};
\node[below] at (1.,-0.2) {$2$};
\node[below] at (2.5,-0.2) {$\ldots$};
\node[below] at (4.,-0.2) {$L-1$};
\node[below] at (5.,-0.2) {$L$};
\end{tikzpicture}
$$
where the dashed lines indicate that the two ends are to be joined to recover 
the cylinder geometry. 

For generic values of $K$, the TM in this spin representation acts 
on the space of all spin configurations of a given row, 
$\{1,2,\ldots,Q\}^L$. Therefore, its dimension is given simply by $Q^L$. 
In the colouring problem ($K=-\infty$) this dimension is reduced
because of the nearest-neighbour constraint $\sigma_i\neq \sigma_j$.
In terms of the corresponding colouring matrix 
$A = \{1 - \delta_{\sigma_i,\sigma_j}\}_{1  \leq i,j\leq Q}$
the dimension reads $d_L = {\rm Tr}\ A^L$. Now $A$ has one eigenvector
$v_1(\sigma_i) = 1$ with eigenvalue $Q-1$, and $Q-1$ eigenvectors
$v_k(\sigma_i) = \delta_{\sigma_i,1}-\delta_{\sigma_i,k}$, for $k=2,3,\ldots,Q$,
each with eigenvalue $-1$. Therefore $d_L = (Q-1)^L + (Q-1) (-1)^L$, which
obviously grows asymptotically as $(Q-1)^L$. Note that $d_L$ can be further
reduced by taking into account the (transverse) translational symmetry 
of the strip. 

The main drawback of the TM method in the spin representation is that for each
value of $Q \in \mathbb{N}$ we have to produce a distinct TM.

We now turn to the physical correlation functions and critical exponents of 
interest in the colouring problem. The first of these is the so-called 
magnetic correlation function, defined between two sites $i,j$ of the 
triangular lattice as
\begin{equation}
G_H (i,j) \;=\; \left\langle \delta_{\sigma_i,\sigma_j} - \frac{1}{Q} 
\right\rangle  \,. 
\label{eq:magneticcorrelation}
\end{equation}
On the critical lines of the phase diagram, the magnetic correlation function 
has an algebraic asymptotic behaviour:
\begin{equation}
G_H(i,j) \;\sim\; |i-j|^{-2 x_H} \,, \qquad \text{for $|i-j|\gg 1$} \,,
\label{eq:magneticexponent}
\end{equation}
where $x_H$ is the \emph{magnetic exponent}. Away from critical 
points, it decays exponentially fast with the distance,  
$G_H(i,j) \sim \rm{e}^{-|i-j|/\xi}$, with a typical length 
scale given by the correlation length $\xi$. The asymptotic behaviour of 
the correlation length in the vicinity of a critical point allows for 
the definition of another critical exponent: 
\begin{equation}
\xi(v) \;\sim\; |v-v_c|^{- \nu} \,,
\end{equation}
which is related to the \emph{thermal exponent} $x_T$ through 
\begin{equation}
\nu \;=\; \frac{1}{2-x_T} \,.
\label{eq:thermalexponent}
\end{equation}
Two-point functions of more general operators, acting on groups of more than 
one spin, have been defined in \cite{VasseurJacobsen14}. They include in 
particular the so-called $k$-cluster `watermelon' operators, which can 
however be formulated more easily in the FK representation to which we 
turn next.

\subsection{FK representation}

The partition function \eqref{eq:ZPottsBis} can be rewritten in the FK
representation as follows \cite{FK1972}:  
\begin{equation}
Z^{\rm FK}_G(Q,v) \;=\; \sum_{A \subseteq E} v^{|A|} \, Q^{k(A)} \,,
\label{eq:ZPottsFK}
\end{equation}
where $v=\mathrm{e}^K -1 = -1$ in the colouring problem, and $k(A)$ is the 
number of connected components of the spanning subgraph $(V,A)$. 
As this function is a polynomial in both $Q$ and $v$, we can extend those 
variables
analytically to arbitrary real or even complex numbers. The original colouring 
model is \emph{in principle} recovered when we specialise (\ref{eq:ZPottsFK}) 
to $v=-1$ and $Q\in {\mathbb N}$; however one must ensure that the free energy 
and the correlation functions obtained in the cluster representation still 
make sense in the colouring problem. Whether this is true or not depends, as 
we will see, on the regime under consideration.

As in the spin representation, we can define the FK representation of the 
TM on the triangular lattice, with the important difference 
that the latter considers all values of 
$Q\in {\mathbb C}$ at the same time. The computation of the TM 
for a triangular-lattice strip with cylindrical boundary conditions is not
straightforward: one needs to consider first a strip of width $L+1$ with free
transverse boundary conditions, and at the end, identify sites $1$ and 
$L+1$ \cite{Baxter87,SalasSokal01}. The final TM acts on the
space of all non-crossing, non-nearest-neighbour partitions of the set
$\{1,2,\ldots,L\}$ on a circle. 
The non-crossing property is due to the planarity of the
graph, and the non-nearest-neighbor property is due to $v=-1$ (i.e., 
neighbour spins cannot be coloured alike) \cite{SalasSokal01,Jacobsen10}.
The number of such
partitions is given by the Riordan number $R_L$ for $L\ge 2$; its asymptotic 
behaviour is $R_L = 3^L L^{-3/2} 3\sqrt{3}/(8 \sqrt{\pi})[1 + O(1/L)]$   
(see e.g. ref.~\cite[section~3.3 and references therein]{SalasSokal01}). 
If we also take into account the translational symmetry along the transverse 
direction, the number of partition classes modulo translations is  
asymptotically equal to $R_L/L$. 

Turning to physical observables, the magnetic correlation function 
(\ref{eq:magneticcorrelation}) gets straightforwardly translated in the 
FK representation as
\begin{equation}
G_H (i,j) \;=\; \frac{Q}{Q-1} \,
   \frac{Z_G^{(i\leftrightarrow j)}(Q,v)}{Z_G^\text{FK}(Q,v)} \,,
\end{equation}
where $Z_G^{(i\leftrightarrow j)}$ is the partition sum of FK configurations 
where sites $i$ and $j$ belong to the same cluster. In the TM picture, 
the points $i$ and $j$ get mapped onto the infinite top and infinite 
bottom of the cylinder, and the exponent $x_H$ can be obtained from the 
scaling of the TM eigenvalues as
\begin{equation}
\log \left|\frac{\Lambda_0^{(L)}}{\Lambda_1^{(L)}}\right| \;=\; 
\frac{2 \pi G \, x_H}{L} + o\left( \frac{1}{L} \right) \,,
\end{equation}
where $G=\frac{\sqrt{3}}{2}$ is a geometrical factor 
\cite{JacobsenSalasSokal03}, 
$\Lambda_0^{(L)}$ is the largest TM eigenvalue, and 
$\Lambda_1^{(L)}$ is the largest eigenvalue in the subsector where one cluster 
(considered `marked') is imposed to propagate along the vertical direction of 
the cylinder, and hence cannot be destroyed by the action of the TM. 
This definition can actually be extended, as we can enlarge our state space by 
imposing the propagation of $k$ distinct marked clusters from the bottom 
to the top of the cylinder, leading to the set of exponents 
\begin{equation}
\log \left|\frac{\Lambda_0^{(L)}}{\Lambda_k^{(L)}}\right| \;=\; 
\frac{2 \pi G \, x_k}{L} + o\left( \frac{1}{L} \right) \,.
\end{equation}
The case $k=1$ actually contains two distinct situations, depending 
whether the propagating cluster is allowed to wrap around the cylinder or not; 
we will come back to this issue in section~\ref{sec:6v}. 

In a similar fashion, we expect that the thermal exponent should be related 
to the relative scaling of the subdominant eigenvalue $\Lambda_0'^{(L)}$ 
in the $k=0$ sector, with respect to the dominant one $\Lambda_0^{(L)}$, 
namely
 \begin{equation}
\log \left| \frac{\Lambda_0^{(L)}}{\Lambda_0'^{(L)}}\right| \;=\; 
\frac{2 \pi G \, x_T}{L} + o\left( \frac{1}{L} \right) \,.
\end{equation}

\medskip

\noindent
{\bf Remark.} 
It is worth mentioning that for $Q= B_p$ [cf.~\eqref{eq.Bk}], with 
integer $p\geq 3$, the Potts-model partition function allows for another 
probabilistic interpretation (with positive weights), namely in terms of a 
restricted-solid-on-solid (RSOS) model \cite{Pasquier,RichardJacobsenSalas}. 
We will discuss this issue in section~\ref{sec:anyonic}.
 
\subsection{Duality transformation} 

\begin{figure}
\centering
\begin{tikzpicture}[scale=2]
\foreach \y in {0,...,2}{

\begin{scope}[shift={(0.5*\y,\y)}]
\foreach \x in {0,...,5}{
 \draw[red] (\x+0.5,0) -- (\x+1.,1);
 \draw[red] (\x+0.5,0) -- (\x,1);
 \draw[red] (\x-0.25,0.5) -- (\x+0.75,0.5);
 \draw[black](\x-0.5,0) -- (\x+0.5,0);
 \draw[black](\x,1) -- (\x+1.,1);
 \draw[black](\x-0.25,0.5)-- (\x,0) -- (\x+0.5,1) -- (\x+0.75,0.5);
 \draw[blue](\x+0.5,0.3) -- (\x,0.7);
 \draw[blue](\x-0.25,0.5) -- (\x,0.7);
 \draw[blue](\x+0.5,0.3) -- (\x+0.75,0.5);
 \draw[blue](\x+0.5,0.3) -- (\x+0.5,0);
 \draw[blue](\x,0.7) -- (\x,1);  
 \fill[blue] (\x,0.7) circle[radius=2.5pt];
 \fill[blue] (\x+0.5,0.3) circle[radius=2.5pt];
 \fill[black] (\x,0) circle[radius=2.5pt];
 \fill[black] (\x+0.5,1) circle[radius=2.5pt];

}
\draw[red] (-0.5,0) -- (0,1);
\draw[dashed] (-0.5,0) -- (0,1);
\draw[dashed] (5.5,0) -- (6,1);
\end{scope}
}

\draw[line width=5pt] (0,0) -- (0.5,1) -- (1.5,1)-- (2,2)-- (1.,2)-- (1.5,3);
\draw[line width=5pt] (2.5,1) -- (3.5,1) -- (3,2) -- (4,2) -- (3.5,1);
\draw[line width=5pt] (6,1)--  (5.5,1) -- (4.5,3);
\draw[line width=5pt] (0,1)--  (0.5,1);
\draw[line width=5pt] (5,2) -- (6,2);
\draw[line width=5pt] (3,0) -- (5,0) -- (4.5,1);

\draw[line width=4pt,blue] (0.5,0.3) -- (1,0.7) -- (1.5,0.3) -- (2,0.7) %
      -- (2.5,0.3) -- (3,0.7) -- (3.5,0.3) -- (4,0.7) -- (4.5,0.3);
\draw[line width=4pt,blue] (1.5,2.3) -- (2,2.7) -- (2.5,2.3) -- (3,2.7) %
      -- (3.5,2.3) -- (4,2.7) -- (4.5,2.3) -- (4.5,1.7) -- (4.,1.3)-- (4.,0.7);
\draw[line width=4pt,blue] (5.75,0.5) -- (5.5,0.3) -- (5,0.7) -- (5,1.3) %
      -- (4.5,1.7);
\draw[line width=4pt,blue] (0.5,0) -- (0.5,0.3);
\draw[line width=4pt,blue] (1.5,0) -- (1.5,0.3);
\draw[line width=4pt,blue] (2.5,0) -- (2.5,0.3);
\draw[line width=4pt,blue] (5.5,0) -- (5.5,0.3);
\draw[line width=4pt,blue] (-0.25,0.5) -- (0.,0.7);
\draw[line width=4pt,blue] (0.25,1.5) -- (0.5,1.7) -- (1,1.3) -- (1.5,1.7);
\draw[line width=4pt,blue] (0.75,2.5) -- (1,2.7) -- (1,3);
\draw[line width=4pt,blue] (0.5,1.7) -- (0.5,2);
\draw[line width=4pt,blue]  (2,2.7) -- (2,3);
\draw[line width=4pt,blue]  (3,2.7) -- (3,3);
\draw[line width=4pt,blue]  (4,2.7) -- (4,3);
\draw[line width=4pt,blue]  (6.75,2.5) --(6.5,2.3) --(6,2.7) --(5.5,2.3) %
      --(5,2.7) -- (5,3);
\draw[line width=4pt,blue]  (6,2.7) -- (6,3);
\draw[line width=4pt,blue]  (6.5,2.) --(6.5,2.3);
\draw[line width=4pt,blue]  (6.25,1.5) --(6.,1.3)--(5.5,1.7);
\draw[line width=4pt,blue]  (2.,0.7) --(2.,1.3) -- (2.5,1.7) -- (3,1.3);
\draw[line width=4pt,blue]  (2.5,1.7) -- (2.5,2.3);

\draw[line width=2pt,red,rounded corners=10pt]  (0.45,0) -- (0.25,0.5) -- (0.75,0.5)--(1,1)--(1.25,0.5)--(1.75,0.5)--(2,1)--(1.75,1.5)--(2.25,1.5)--(2.5,2)--(2.25,2.5)--(1.75,2.5)--(1.5,2)--(1.25,2.5)--(1.75,2.5)--(1.95,3);
\draw[line width=2pt,red,rounded corners=10pt]  (0.55,0) -- (0.75,0.5)--(1.25,0.5) -- (1.45,0);
\draw[line width=2pt,red,rounded corners=10pt]  (1.55,0) -- (1.75,0.5)--(2.25,0.5) -- (2.45,0);
\draw[line width=2pt,red,rounded corners=10pt]  (2.55,0) -- (2.75,0.5)--(3.25,0.5) -- (3.5,0)-- (3.75,0.5)-- (4.25,0.5)-- (4.5,0)-- (4.75,0.5)-- (4.25,0.5)-- (4,1)-- (4.25,1.5)-- (4.75,1.5)-- (5,1)-- (4.75,0.5)-- (5.25,0.5)-- (5.45,0);
\draw[line width=2pt,red,rounded corners=10pt](2.05,3) -- (2.25,2.5)-- (2.75,2.5)-- (2.95,3);
\draw[line width=2pt,red,rounded corners=10pt](3.05,3) -- (3.25,2.5)-- (3.75,2.5)-- (3.95,3);
\draw[line width=2pt,red,rounded corners=10pt](5.05,3) -- (5.25,2.5)-- (5.75,2.5)-- (5.95,3);
\draw[line width=2pt,red,rounded corners=10pt](6.05,3) -- (6.25,2.5)-- (6.75,2.5)-- (6.95,3);
\draw[line width=2pt,red,rounded corners=10pt](2.5,0.5) -- (2.75,0.5)-- (3,1)-- (3.25,0.5)-- (3.75,0.5)-- (4,1)-- (3.75,1.5)-- (4.25,1.5)-- (4.5,2)-- (4.25,2.5)-- (3.75,2.5)-- (3.5,2)-- (3.25,2.5)-- (2.75,2.5)-- (2.5,2)-- (2.75,1.5)-- (3.25,1.5)-- (3,1)-- (2.75,1.5)-- (2.25,1.5)-- (2,1)-- (2.25,0.5)-- (2.5,0.5);
\draw[line width=2pt,red,rounded corners=10pt](3.5,1.5)-- (3.75,1.5)-- (3.5,2)-- (3.25,1.5)-- (3.5,1.5);
\draw[line width=2pt,red,rounded corners=10pt](4.05,3)-- (4.25,2.5)-- (4.75,2.5)-- (4.5,2)-- (4.75,1.5)-- (5.25,1.5)-- (5,1.)-- (5.25,0.5)-- (5.75,0.5) -- (5.95,0.9);
\draw[line width=2pt,red,rounded corners=10pt] (-0.2,0.6) -- (0,1) -- (0.25,0.5) -- (-0.25,0.5) -- (-0.5,0);
\draw[line width=2pt,red,rounded corners=10pt] (6.2,1.4) -- (6,1) -- (5.75,1.5) -- (5.25,1.5)-- (5.5,2)-- (5.75,1.5)-- (6.25,1.5)-- (6.5,2)-- (6.25,2.5)-- (5.75,2.5)-- (5.5,2)-- (5.25,2.5)-- (4.75,2.5)-- (4.95,3);
\draw[line width=2pt,red,rounded corners=10pt] (1.05,3) -- (1.25,2.5)-- (0.75,2.5)-- (0.5,2.)-- (0.75,1.5)-- (1.25,1.5)-- (1.5,2.)-- (1.75,1.5)-- (1.25,1.5)-- (1.,1.)-- (0.75,1.5)-- (0.25,1.5)-- (0.05,1.1);
\end{tikzpicture}
\caption{
Configuration of FK clusters on the triangular lattice $G$ (black), 
and the associated configuration of dual FK clusters on the hexagonal lattice 
$G^*$ (blue). In red, we have represented the surrounding loops on the medial, 
kagome lattice.  
}            
\label{fig:FKclustersloops}                    
\end{figure}
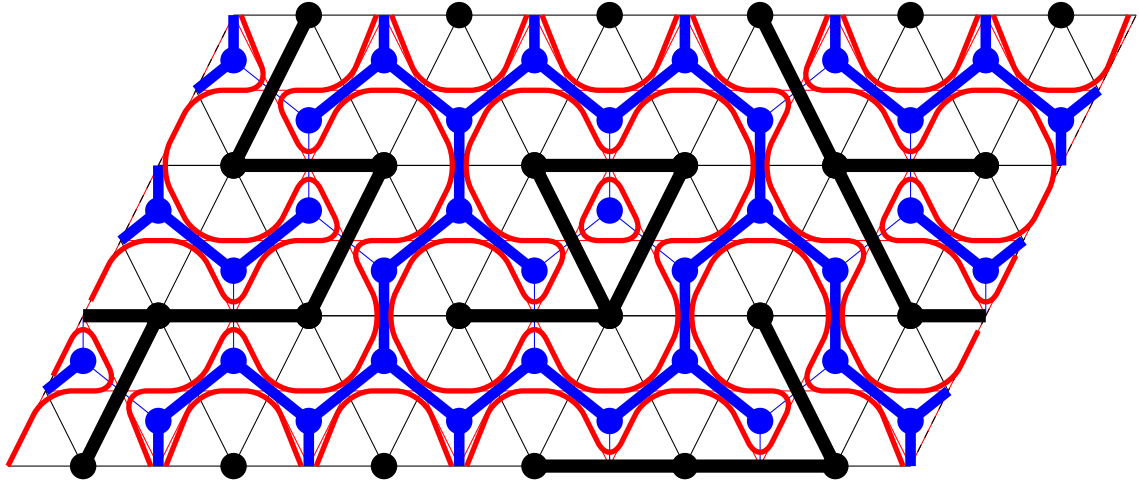

Given a $Q$-state Potts model with parameter $v$ in the FK representation 
(\ref{eq:ZPottsFK}) and defined on a triangular lattice $G=(V,E)$, a
duality transformation \cite{Wu82} allows us to obtain an equivalent 
$Q$-state Potts model 
on the dual, hexagonal lattice $G^*=(V^*,E^*)$, by putting a FK link 
on each edge of $E^*$ not crossed by a FK link on $E$ 
(see figure~\ref{fig:FKclustersloops}). 

The corresponding dual parameter is given by
\begin{equation}
v^* \;=\; \frac{Q}{v} \;=\; -Q  \,,
\end{equation}
and we can represent the corresponding TM as 
\medskip
$$
\begin{tikzpicture}[scale=2]
   \foreach \x in {0,...,5}
{ 
    \draw (\x-0.5,0) -- (\x+0.5,0);
    \draw (\x,1) -- (\x+1.,1);
    \draw (\x-0.25,0.5)-- (\x,0) -- (\x+0.5,1) -- (\x+0.75,0.5);
    \draw[fill=white] (\x,0) circle[radius=2pt];
    \draw[fill=white] (\x+0.5,1) circle[radius=2pt];
    \draw[blue] (\x+0.5,0.3) -- (\x,0.7);
    \draw[blue] (\x-0.25,0.5) -- (\x,0.7);
    \draw[blue] (\x+0.5,0.3) -- (\x+0.75,0.5);
    \draw[blue] (\x+0.5,0.3) -- (\x+0.5,0);
 \draw[blue] (\x,0.7) -- (\x,1);
}
   \foreach \x in {0,...,5}
{ 
    \draw[blue,fill=white] (\x,0.7) circle[radius=2pt];
    \draw[blue,fill=white] (\x+0.5,0.3) circle[radius=2pt];
}
\draw[dashed] (-0.5,0) -- (0,1);
\draw[dashed] (5.5,0) -- (6,1);
\node[below] at (0.,-0.2) {$1$};
\node[below] at (1.,-0.2) {$2$};
\node[below] at (2.5,-0.2) {$\ldots$};
\node[below] at (4.,-0.2) {$L-1$};
\node[below] at (5.,-0.2) {$L$};
\end{tikzpicture}
$$
where the dual, hexagonal lattice has been represented in blue. The space 
of states in terms of a set of connectivities follows from a construction 
analogous
to that used on the triangular lattice, and we point out that imposing a number 
$k$ of propagating clusters in the 
triangular-lattice Potts model is equivalent to imposing the same 
number of propagating clusters in the dual, hexagonal-lattice model, so 
that the definition of critical exponents sketched in the previous section 
translates straightforwardly to the dual model.  

Note that under duality the chromatic polynomial $\chi_G$ of a planar graph 
$G$ transforms into the flow polynomial $\Phi_{G^*}$ on the dual graph $G^*$. 
To be more precise, $\chi_{G^*}(Q) = Q \, \Phi_G(Q)$. 

\subsection{Six-vertex model on the kagome lattice}
\label{sec:6v}

The Potts model on the dual, hexagonal lattice $G^*$, can be further mapped 
mapped onto a model of loops living on the edges on the medial, kagome lattice,
and corresponding to the surrounding contours of the FK clusters in the 
spanning subgraph $(V^*,A)$ (see figure~\ref{fig:FKclustersloops}). 
This is achieved through the following identity \cite{BaxterKellandWu76}
\begin{equation}
Z^{\rm FK}_{G^*}(Q,v^*) \;=\; Q^{|V^*|/2}\, 
    \sum\limits_{A \subseteq E^*} (x^*)^{|A|} \, Q^{\ell(A)/2} \,,
\label{eq:ZPottsloops}
\end{equation}
where $\ell(A)$ counts the number of loops in the spanning subgraph 
$(V^*,A)$, and 
\begin{equation} 
x^* \;=\;  \frac{v^*}{\sqrt{Q}} = - \sqrt{Q}  \,.
\end{equation}

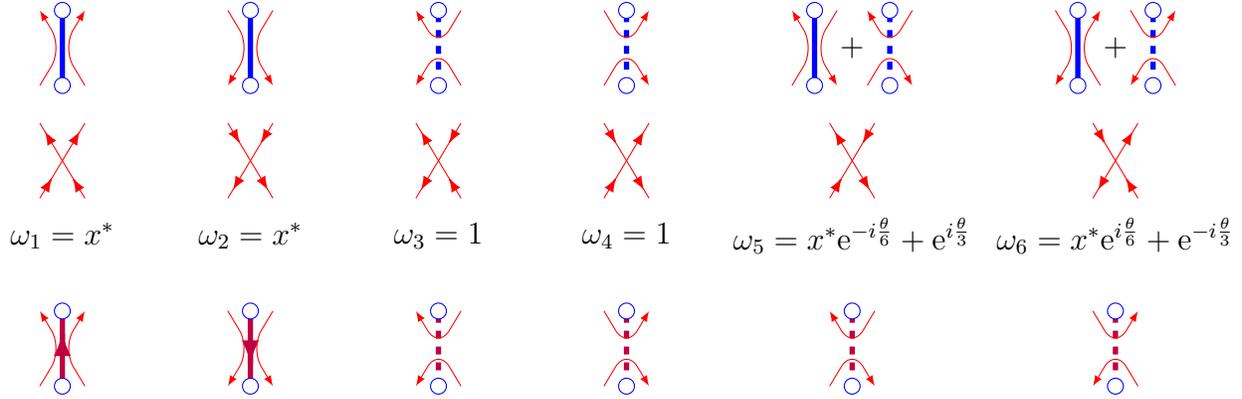
\begin{figure}
\centering
\begin{tikzpicture}
\begin{scope}
  \draw[blue,line width=2pt] (0,0) -- (0,1);
  \draw[red,rounded corners=8pt,->,>=latex] (-0.3,0) -- (0,0.5) -- (-0.3,1);
  \draw[red,rounded corners=8pt,->,>=latex] (0.3,0) -- (0,0.5) -- (0.3,1);
  \draw[blue,fill=white](0,0) circle [radius=3pt];
  \draw[blue,fill=white](0,1) circle [radius=3pt];
  \begin{scope}[shift={(0,-1.5)}]
     \draw[red] (-0.3,0) -- (0.3,1);
     \draw[red] (0.3,0) -- (-0.3,1);
     \draw[red,thick,->,>=latex] (-0.15,0.25) -- (-0.14,0.267);
     \draw[red,thick,->,>=latex] (0.15,0.25) -- (0.14,0.267);
     \draw[red,thick,-<,>=latex] (-0.15,0.75) -- (-0.14,0.733);
     \draw[red,thick,-<,>=latex] (0.15,0.75) -- (0.14,0.733);
  \end{scope}
  \node at (0,-2) {$\omega_1 = x^*$};
  \begin{scope}[shift={(0,-4)}]
     \draw[purple,line width=2pt] (0,0) -- (0,1);
     \draw[purple,line width=2pt,->,>=latex] (0,0.7) -- (0,0.71);
     \draw[red,rounded corners=8pt,->,>=latex] (-0.3,0) -- (0,0.5) -- (-0.3,1);
     \draw[red,rounded corners=8pt,->,>=latex] (0.3,0) -- (0,0.5) -- (0.3,1);
     \draw[blue,fill=white](0,0) circle [radius=3pt];
     \draw[blue,fill=white](0,1) circle [radius=3pt];
  \end{scope}
\end{scope}
\begin{scope}[shift={(2.5,0)}]
  \draw[blue,line width=2pt] (0,0) -- (0,1);
  \draw[red,rounded corners=8pt,<-,>=latex] (-0.3,0) -- (0,0.5) -- (-0.3,1);
  \draw[red,rounded corners=8pt,<-,>=latex] (0.3,0) -- (0,0.5) -- (0.3,1);
  \draw[blue,fill=white](0,0) circle [radius=3pt];
  \draw[blue,fill=white](0,1) circle [radius=3pt];
  \begin{scope}[shift={(0,-1.5)}]
     \draw[red] (-0.3,0) -- (0.3,1);
     \draw[red] (0.3,0) -- (-0.3,1);
     \draw[red,thick,-<,>=latex] (-0.15,0.25) -- (-0.14,0.267);
     \draw[red,thick,-<,>=latex] (0.15,0.25) -- (0.14,0.267);
     \draw[red,thick,->,>=latex] (-0.15,0.75) -- (-0.14,0.733);
     \draw[red,thick,->,>=latex] (0.15,0.75) -- (0.14,0.733);
  \end{scope}
 \node at (0,-2) {$\omega_2 = x^*$};
 \begin{scope}[shift={(0,-4)}]
     \draw[purple,line width=2pt] (0,0) -- (0,1);
     \draw[purple,line width=2pt,<-,>=latex] (0,0.3) -- (0,0.31);
     \draw[red,rounded corners=8pt,<-,>=latex] (-0.3,0) -- (0,0.5) -- (-0.3,1);
     \draw[red,rounded corners=8pt,<-,>=latex] (0.3,0) -- (0,0.5) -- (0.3,1);
     \draw[blue,fill=white](0,0) circle [radius=3pt];
     \draw[blue,fill=white](0,1) circle [radius=3pt];
 \end{scope}
\end{scope}
\begin{scope}[shift={(5,0)}]
  \draw[blue,dashed,line width=2pt] (0,0) -- (0,1);
  \draw[red,rounded corners=8pt,<-,>=latex] (-0.3,0) -- (0,0.5) -- (0.3,0);
  \draw[red,rounded corners=8pt,<-,>=latex] (-0.3,1) -- (0,0.5) -- (0.3,1);
  \draw[blue,fill=white](0,0) circle [radius=3pt];
  \draw[blue,fill=white](0,1) circle [radius=3pt];
  \begin{scope}[shift={(0,-1.5)}]
     \draw[red] (-0.3,0) -- (0.3,1);
     \draw[red] (0.3,0) -- (-0.3,1);
     \draw[red,thick,-<,>=latex] (-0.15,0.25) -- (-0.14,0.267);
     \draw[red,thick,->,>=latex] (0.15,0.25) -- (0.14,0.267);
     \draw[red,thick,-<,>=latex] (-0.15,0.75) -- (-0.14,0.733);
     \draw[red,thick,->,>=latex] (0.15,0.75) -- (0.14,0.733);
  \end{scope}
  \node at (0,-2) {$\omega_3 = 1$};
  \begin{scope}[shift={(0,-4)}]
     \draw[purple,dashed,line width=2pt] (0,0) -- (0,1);
     \draw[red,rounded corners=8pt,<-,>=latex] (-0.3,0) -- (0,0.5) -- (0.3,0);
     \draw[red,rounded corners=8pt,<-,>=latex] (-0.3,1) -- (0,0.5) -- (0.3,1);
     \draw[blue,fill=white](0,0) circle [radius=3pt];
     \draw[blue,fill=white](0,1) circle [radius=3pt];
  \end{scope}
\end{scope}
\begin{scope}[shift={(7.5,0)}]
  \draw[blue,dashed,line width=2pt] (0,0) -- (0,1);
  \draw[red,rounded corners=8pt,->,>=latex] (-0.3,0) -- (0,0.5) -- (0.3,0);
  \draw[red,rounded corners=8pt,->,>=latex] (-0.3,1) -- (0,0.5) -- (0.3,1);
  \draw[blue,fill=white](0,0) circle [radius=3pt];
  \draw[blue,fill=white](0,1) circle [radius=3pt];
  \begin{scope}[shift={(0,-1.5)}]
     \draw[red] (-0.3,0) -- (0.3,1);
     \draw[red] (0.3,0) -- (-0.3,1);
     \draw[red,thick,->,>=latex] (-0.15,0.25) -- (-0.14,0.267);
     \draw[red,thick,-<,>=latex] (0.15,0.25) -- (0.14,0.267);
     \draw[red,thick,->,>=latex] (-0.15,0.75) -- (-0.14,0.733);
     \draw[red,thick,-<,>=latex] (0.15,0.75) -- (0.14,0.733);
  \end{scope}
  \node at (0,-2) {$\omega_4 = 1$};
  \begin{scope}[shift={(0,-4)}]
     \draw[purple,dashed,line width=2pt] (0,0) -- (0,1);
     \draw[red,rounded corners=8pt,->,>=latex] (-0.3,0) -- (0,0.5) -- (0.3,0);
     \draw[red,rounded corners=8pt,->,>=latex] (-0.3,1) -- (0,0.5) -- (0.3,1);
     \draw[blue,fill=white](0,0) circle [radius=3pt];
     \draw[blue,fill=white](0,1) circle [radius=3pt];
  \end{scope}
\end{scope}
\begin{scope}[shift={(10.5,0)}]
  \begin{scope}[shift={(-0.5,0)}]
     \draw[blue,line width=2pt] (0,0) -- (0,1);
     \draw[red,rounded corners=8pt,->,>=latex] (-0.3,0) -- (0,0.5) -- (-0.3,1);
     \draw[red,rounded corners=8pt,<-,>=latex] (0.3,0) -- (0,0.5) -- (0.3,1);
     \draw[blue,fill=white](0,0) circle [radius=3pt];
     \draw[blue,fill=white](0,1) circle [radius=3pt];
  \end{scope}
  \node at (0,0.5) {$+$};
  \begin{scope}[shift={(0.5,0)}]
     \draw[blue,dashed,line width=2pt] (0,0) -- (0,1);
     \draw[red,rounded corners=8pt,->,>=latex] (-0.3,0) -- (0,0.5) -- (0.3,0);
     \draw[red,rounded corners=8pt,<-,>=latex] (-0.3,1) -- (0,0.5) -- (0.3,1);
     \draw[blue,fill=white](0,0) circle [radius=3pt];
     \draw[blue,fill=white](0,1) circle [radius=3pt];
  \end{scope}
  \begin{scope}[shift={(0,-1.5)}]
     \draw[red] (-0.3,0) -- (0.3,1);
     \draw[red] (0.3,0) -- (-0.3,1);
     \draw[red,thick,->,>=latex] (-0.15,0.25) -- (-0.14,0.267);
     \draw[red,thick,-<,>=latex] (0.15,0.25) -- (0.14,0.267);
     \draw[red,thick,-<,>=latex] (-0.15,0.75) -- (-0.14,0.733);
     \draw[red,thick,->,>=latex] (0.15,0.75) -- (0.14,0.733);
  \end{scope}
  \node at (0,-2) {$\omega_5 = x^* %
        \mathrm{e}^{-i \frac{\theta}{6}}+\mathrm{e}^{i \frac{\theta}{3}}$};
  \begin{scope}[shift={(0,-4)}]
     \draw[purple,dashed,line width=2pt] (0,0) -- (0,1);
     \draw[red,rounded corners=8pt,->,>=latex] (-0.3,0) -- (0,0.5) -- (0.3,0);
     \draw[red,rounded corners=8pt,<-,>=latex] (-0.3,1) -- (0,0.5) -- (0.3,1);
     \draw[blue,fill=white](0,0) circle [radius=3pt];
     \draw[blue,fill=white](0,1) circle [radius=3pt];
  \end{scope}
\end{scope}
\begin{scope}[shift={(14.,0)}]
  \begin{scope}[shift={(-0.5,0)}]
     \draw[blue,line width=2pt] (0,0) -- (0,1);
     \draw[red,rounded corners=8pt,->,>=latex] (-0.3,1) -- (0,0.5) -- (-0.3,0);
     \draw[red,rounded corners=8pt,->,>=latex] (0.3,0) -- (0,0.5) -- (0.3,1);
     \draw[blue,fill=white](0,0) circle [radius=3pt];
     \draw[blue,fill=white](0,1) circle [radius=3pt];
  \end{scope}
  \node at (0,0.5) {$+$};
  \begin{scope}[shift={(0.5,0)}]
     \draw[blue,dashed,line width=2pt] (0,0) -- (0,1);
     \draw[red,rounded corners=8pt,->,>=latex] (0.3,0) -- (0,0.5) -- (-0.3,0);
     \draw[red,rounded corners=8pt,<-,>=latex] (0.3,1) -- (0,0.5) -- (-0.3,1);
     \draw[blue,fill=white](0,0) circle [radius=3pt];
     \draw[blue,fill=white](0,1) circle [radius=3pt];
  \end{scope}
  \begin{scope}[shift={(0,-1.5)}]
     \draw[red] (-0.3,0) -- (0.3,1);
     \draw[red] (0.3,0) -- (-0.3,1);
     \draw[red,thick,-<,>=latex] (-0.15,0.25) -- (-0.14,0.267);
     \draw[red,thick,->,>=latex] (0.15,0.25) -- (0.14,0.267);
     \draw[red,thick,->,>=latex] (-0.15,0.75) -- (-0.14,0.733);
     \draw[red,thick,-<,>=latex] (0.15,0.75) -- (0.14,0.733);
  \end{scope}
  \node at (0,-2) {$\omega_6 = x^* %
        \mathrm{e}^{i \frac{\theta}{6}}+\mathrm{e}^{-i \frac{\theta}{3}}$};
  \begin{scope}[shift={(0,-4)}]
     \draw[purple,dashed,line width=2pt] (0,0) -- (0,1);
     \draw[red,rounded corners=8pt,->,>=latex] (-0.3,1) -- (0,0.5) -- (0.3,1);
     \draw[red,rounded corners=8pt,<-,>=latex] (-0.3,0) -- (0,0.5) -- (0.3,0);
     \draw[blue,fill=white](0,0) circle [radius=3pt];
     \draw[blue,fill=white](0,1) circle [radius=3pt];
  \end{scope}
\end{scope}
\end{tikzpicture}
\caption{
Mapping from the Potts model (\ref{eq:ZPottsloops}) on the hexagonal lattice 
(first line) onto a six-vertex model on the kagome lattice (second line), and
then onto a model of dilute loops on the hexagonal lattice (third line).
}                                 
\label{fig:Pottshexato6vtoOnhexa}
\end{figure}

The configurations of this loop model can now be mapped onto those of a 
six-vertex model on the kagome lattice \cite{BaxterTemperleyAshley78}.
This is done by reexpressing the factor $Q^{\ell(A)/2}$ in terms of local 
factors by setting [cf. (\ref{eq.def.theta})]:  
\begin{equation}
\sqrt{Q} \;=\; 2 \cos \frac{\theta}{2} \,,
\end{equation}
and summing over the orientations of loops, as represented in the first 
two lines of figure~\ref{fig:Pottshexato6vtoOnhexa}. Notice that in this   
figure we represent only one of the possible three types of vertices
on the kagome lattice (related by $2\pi/3$ rotations); however, contrary 
to what happens for the
square lattice, we can always see which weights are which because the two
angles in the kagome lattice are distinct: $\pi/3$ and $2\pi/3$, respectively. 

The TM of the resulting six-vertex model can be represented 
as follows (the vertices and edges of the medial kagome lattice are 
depicted as red dots and lines, respectively):
\medskip
$$
\begin{tikzpicture}[scale=2]
\foreach \x in {0,...,5}{
 \draw[red] (\x+0.5,0) -- (\x+1.,1);
 \draw[red] (\x+0.5,0) -- (\x,1);
 \draw[red] (\x-0.25,0.5) -- (\x+0.75,0.5);
 \draw[black](\x-0.5,0) -- (\x+0.5,0);
 \draw[black](\x,1) -- (\x+1.,1);
 \draw[black](\x-0.25,0.5)-- (\x,0) -- (\x+0.5,1) -- (\x+0.75,0.5);
 \draw[blue](\x+0.5,0.3) -- (\x,0.7);
 \draw[blue](\x-0.25,0.5) -- (\x,0.7);
 \draw[blue](\x+0.5,0.3) -- (\x+0.75,0.5);
 \draw[blue](\x+0.5,0.3) -- (\x+0.5,0);
 \draw[blue](\x,0.7) -- (\x,1);  
}
\draw[red] (-0.5,0) -- (0,1);
\draw[dashed] (-0.5,0) -- (0,1);
\draw[dashed] (5.5,0) -- (6,1);
\foreach \x in {0,...,5}{
\draw[fill=white] (\x,0) circle[radius=2pt];
\draw[fill=white] (\x+0.5,1) circle[radius=2pt];
\draw[red,fill=white](\x+0.5,0) circle[radius=2pt];
 \draw[red,fill=white](\x,1) circle[radius=2pt];
\draw[red,fill=white](\x+1,1) circle[radius=2pt];
\draw[red,fill=white](\x+0.75,0.5) circle[radius=2pt];
 \draw[red,fill=white](\x+0.25,0.5) circle[radius=2pt];
 \draw[red,fill=white](\x-0.25,0.5) circle[radius=2pt];
 \draw[blue,fill=white] (\x,0.7) circle[radius=2pt];
 \draw[blue,fill=white] (\x+0.5,0.3) circle[radius=2pt];
}
\draw[red,fill=white](-0.5,0) circle[radius=2pt];
\node[red] at (5.2,-0.25) {$\mathrm{e}^{\pm i \frac{\varphi}{2}}$};
\draw[red,decorate,decoration={snake,amplitude=1pt, segment length=4pt}]%
     (5.2,-0.05) -- (5.8,1.05);
\end{tikzpicture}
$$
 
The degrees of freedom are now the orientations on each edge of the kagome 
lattice, so the TM acts on each row on the space of $2L$ edges oriented 
either upwards and downwards, therefore with a total dimension $2^{2L} = 4^L$.

As suggested by the figure above, a twist factor 
$\mathrm{e}^{\pm i \frac{\varphi}{2}}$ has to be introduced in the 
six-vertex model in order to ensure the correspondence with the original 
periodic Potts model. It is 
represented by a red wavy line, which contributes a factor 
$\mathrm{e}^{\pm i \frac{\varphi}{2}}$ each time it is crossed by a 
rightwards (resp.\/ leftwards) arrow of the six-vertex model.  
The twist parameter ${\varphi}$ must be chosen as follows: 
\begin{itemize}
\item In the sector with no propagating clusters ($k=0$), there may be closed 
loops on the kagome lattice, that wind around the {transverse}
periodic boundary condition. To recover the original model, these loops must 
be given the same weight $\sqrt{Q} = 2 \cos \frac{\theta}{2}$ as the
one given to contractible loops. Choosing the twist
$\frac{\varphi}{2} = \frac{\theta}{2}$ (or, as we shall see, 
$\frac{\varphi}{2} = \frac{2\pi-\theta}{2}$, depending on the regime), 
since each loop has to be summed over 
the two possible orientations, it is given a weight 
\begin{equation}
\mathrm{e}^{i \frac{\varphi}{2}} + \mathrm{e}^{-i \frac{\varphi}{2}} 
\;=\; 2 \cos \frac{\theta}{2} \,,
\end{equation}
as required. 
Therefore we expect the central charge of the Potts model to be given in terms
of the free energy (density per site) by  \cite{Cardy86,Affleck86}
\begin{equation}
f_L(\varphi) \;=\; 
\frac{1}{L}\, \log \left|{\Lambda_{0}^{(L)}}(\varphi)\right| 
\;=\; f_\text{bulk} + \frac{\pi G \, c}{6 L^2} + o\left( \frac{1}{L^2}\right) 
\,,
\label{eq:CFT_fL}
\end{equation}
where $\Lambda_{0}^{(L)}(\varphi)$ is the largest eigenvalue of the 
six-vertex TM, appropriately twisted, in the sector with $\ell=0$ 
`through-lines' (Potts loops propagating along the vertical direction 
of the cylinder).

\item In the sector with a single propagating FK cluster ($k=1$) we have two 
  distinct situations: 
  \begin{itemize}
     \item If the propagating FK cluster is allowed to wrap around the 
           cylinder, then this translates in the six-vertex language 
           to no loop propagating in the vertical direction, 
           but no closed loop allowed to wrap around 
           the cylinder. Forbidding loops on the kagome lattice to wrap 
           around the cylinder amounts to giving such loops a weight $0$, 
           therefore to fixing the twist to $\frac{\varphi}{2} = \frac{\pi}{2}$.
           The magnetic exponent can thus be obtained as 
\begin{equation}
\log \left| \frac{\Lambda_{0}^{(L)}(\varphi)}
                 {\Lambda_{0}^{(L)}(\pi)}\right| \;=\;
   \frac{2 \pi G\, x_H}{L} + o\left( \frac{1}{L} \right) \,.
\end{equation}
           
     \item If the propagating FK cluster is \emph{not} allowed to wrap 
           around the cylinder, then it translates in the Potts 
           loops/six-vertex language into 
           the presence of two loops propagating in the vertical direction 
           (`through-lines') 
           delimiting the vertical borders of the propagating FK cluster. 
           The situation is then similar to the case $k\ge 2$, described 
           in the following paragraph. 
  \end{itemize}   

\item In the sectors with $k\ge 2$ propagating clusters (or $k=1$ cluster not 
      allowed to wrap around the cylinder, see above), there are no 
      non-contractible loops anymore. Each propagating cluster is associated 
      with two Potts through-lines, and one can 
      take each of the $\ell =2k$ through-lines around the axis of the 
      cylinder without affecting the Boltzmann weights, if every such line 
      acquires in doing so a 
      phase which is a $\ell^{\rm th}$ root of unity. This means that the 
      eigenvalues of the original model are obtained in the twisted vertex 
      model with a twist 
      being of the form $\frac{2\pi p}{\ell}$, with any integer $p$ satisfying 
      $\gcd(p,\ell)=1$. 
\end{itemize}

\subsection{O$(n)$ model on the hexagonal lattice}

As shown in the last line of figure~\ref{fig:Pottshexato6vtoOnhexa}, the 
configurations of the six-vertex model can in turn be mapped onto those of 
dilute loops living on the edges on the hexagonal lattice. 
In this mapping, the thin spin-$\frac12$ loops run around the inside and 
outside of the thick spin-$1$ loops, therefore a dilute loop on the hexagonal 
lattice is a double Potts loop on the kagome lattice. 

The resulting gas of dilute loops is nothing but the O$(n)$ model on the 
hexagonal lattice \cite{Nienhuis82}, where the weight of each closed loop is  
\begin{equation}
n \;=\; - 2 \cos 2\theta 
\end{equation}
and the fugacity of each loop segment is 
\begin{equation}
K \;=\; \frac{1}{2 \sin\frac{\theta}{2}}
\label{eq:fugacityK}
\end{equation}

The corresponding TM can be represented as follows: 
\medskip
$$
\begin{tikzpicture}[scale=2]
\draw[red] (-0.5,0) -- (0,1);
\foreach \x in {0,...,5}{
   \draw[red] (\x+0.5,0) -- (\x+1.,1);
   \draw[red] (\x+0.5,0) -- (\x,1);
   \draw[red] (\x-0.25,0.5) -- (\x+0.75,0.5);
   \draw[blue];
   \draw (\x-0.5,0) -- (\x+0.5,0);
   \draw(\x,1) -- (\x+1.,1);
   \draw (\x-0.25,0.5)-- (\x,0) -- (\x+0.5,1) -- (\x+0.75,0.5);
   \draw[fill=white] (\x,0) circle[radius=2pt];
   \draw[fill=white] (\x+0.5,1) circle[radius=2pt];
   \draw[fill=white] (\x,0.7) circle[radius=2pt];
   \draw[fill=white] (\x+0.5,0.3) circle[radius=2pt];
}
\draw[dashed] (-0.5,0) -- (0,1);
\draw[dashed] (5.5,0) -- (6,1);
\foreach \x in {0,...,5}{
   \draw[purple, line width=5pt] (\x+0.5,0.3) -- (\x,0.7);
   \draw[purple, line width=5pt] (\x-0.25,0.5) -- (\x,0.7);
   \draw[purple, line width=5pt] (\x+0.5,0.3) -- (\x+0.75,0.5);
   \draw[purple, line width=5pt] (\x+0.5,0.3) -- (\x+0.5,0);
   \draw[purple, line width=5pt] (\x,0.7) -- (\x,1);  
}
\node[purple] at (5.2,-0.25) {$\mathrm{e}^{\pm i {\varphi}}$};
\draw[purple,decorate,decoration={snake,amplitude=1pt, segment length=4pt}]%
     (5.2,-0.05) -- (5.8,1.05);
\end{tikzpicture}
$$
  
One needs to be careful about what becomes in this model of the twist 
parameter introduced in the six-vertex model: 
each loop segment of the O$(n)$ model crossing the vertical seam has to be 
summed over two possible orientations.  Now, from 
figure~\ref{fig:Pottshexato6vtoOnhexa}, one sees that each oriented O$(n)$ 
loop crossing the vertical seam is equivalent to two edges of the six-vertex 
model crossing the seam with the same orientation, hence a twist 
$\varphi$ rather than $\frac{\varphi}{2}$ (where $\varphi$ has to be 
conveniently chosen depending on the sector we are considering, as explained 
in the previous section). 

It is important to stress that making the right choice for the twist in 
order to ensure the correspondence with the original periodic Potts model 
results in a `wrong' choice of the weight for non-contractible loops in the 
O$(n)$ model. For instance, in the $k=0$ sector, if we choose 
$\frac{\varphi}{2} = \frac{\theta}{2}$, we get the right weight for the 
winding FK clusters $\sqrt{Q} = 2 \cos \frac{\theta}{2}$; but the wrong one 
for the winding O($n$) loops $\widetilde{n} = 2\cos \theta$. Similarly, if we 
choose $\frac{\varphi}{2} = \theta-\frac{\pi}{2}$, we get the right weight for 
a winding O($n$) loop $n=-2\cos 2\theta$, but the wrong one for a winding FK 
cluster $\widetilde{Q}^{1/2}=2\sin\theta$. 
In other words, the periodic Potts model is equivalent to a O$(n)$ model 
on the hexagonal lattice with a particular choice of twisted 
boundary conditions.

The TM acts on the set of non-crossing dilute matchings of the set 
$\{1,2,\ldots,L\}$.
By dilute (in opposition to perfect) we here mean that any given point may be
empty and hence not paired with another point. The number of such matchings
is given by the Motzkin number $M_L$ for $L \ge 1$; its asymptotic behaviour
is $3^L L^{-3/2} 3 \sqrt{3} / (2 \sqrt{\pi}) [1 + O(1/L)]$.

\subsection{Spin-one vertex model on the square lattice}

\begin{figure}
\centering
\begin{tikzpicture}[scale=1]
\begin{scope}[rotate=45]
\draw [black, dashed,line width=0.2] (-0.5,-0.5) -- (0.5,0.5); 
\draw [black, dashed,line width=0.2] (0.5,-0.5) -- (-0.5,0.5); 
\end{scope}
\node at (0,-1.) {$\rho_1$};
\begin{scope}[shift={(0,-2.5)},rotate=45,scale=0.35]
\draw [line width=0.4, dashed] (-1,-1) -- (0,0) -- (-1,1); 
\draw [line width=0.4, dashed] (0,0) -- (1.5,0);
\draw [line width=0.4, dashed] (2.5,-1) -- (1.5,0) -- (2.5,1); 
\end{scope}
\begin{scope}[shift={(1.75,0)}]
\begin{scope}[rotate=45]
\draw [black, dashed,line width=0.2] (-0.5,-0.5) -- (0.5,0.5); 
\draw [black, dashed,line width=0.2] (0.5,-0.5) -- (-0.5,0.5); 
\draw[purple,line width=2,rounded corners=10pt]%
   (-0.5,-0.5) -- (0,0) -- (-0.5,0.5);
 \end{scope}
\node at (0,-1.) {$\rho_2$};
 \begin{scope}[shift={(0,-2.5)},rotate=45,scale=0.35]
\draw [line width=0.4, dashed] (-1,-1) -- (0,0) -- (-1,1); 
\draw [line width=0.4, dashed] (0,0) -- (1.5,0);
\draw [line width=0.4, dashed] (2.5,-1) -- (1.5,0) -- (2.5,1); 
\draw[purple,line width=2,rounded corners=10pt] (-1,-1) -- (0,0) -- (-1,1);
\end{scope}
\end{scope}
\begin{scope}[shift={(3.5,0)}]
\begin{scope}[rotate=45]
\draw [black, dashed,line width=0.2] (-0.5,-0.5) -- (0.5,0.5); 
\draw [black, dashed,line width=0.2] (0.5,-0.5) -- (-0.5,0.5); 
\draw[purple,line width=2,rounded corners=10pt]%
   (0.5,-0.5) -- (0,0) -- (0.5,0.5);
\end{scope}
\node at (0,-1.) {$\rho_3$};
\begin{scope}[shift={(0,-2.5)},rotate=45,scale=0.35]
\draw [line width=0.4, dashed] (-1,-1) -- (0,0) -- (-1,1); 
\draw [line width=0.4, dashed] (0,0) -- (1.5,0);
\draw [line width=0.4, dashed] (2.5,-1) -- (1.5,0) -- (2.5,1);
\draw[purple,line width=2,rounded corners=10pt] (2.5,1) -- (1.5,0) -- (2.5,-1);
 
\end{scope}
\end{scope}
\begin{scope}[shift={(5.25,0)}]
\begin{scope}[rotate=45]
\draw [black, dashed,line width=0.2] (-0.5,-0.5) -- (0.5,0.5); 
\draw [black, dashed,line width=0.2] (0.5,-0.5) -- (-0.5,0.5); 
\draw[purple,line width=2,rounded corners=10pt]%
   (-0.5,-0.5) -- (0,0) -- (0.5,-0.5);
\end{scope}
\node at (0,-1.) {$\rho_4$};
\begin{scope}[shift={(0,-2.5)},rotate=45,scale=0.35]
\draw [line width=0.4, dashed] (-1,-1) -- (0,0) -- (-1,1); 
\draw [line width=0.4, dashed] (0,0) -- (1.5,0);
\draw [line width=0.4, dashed] (2.5,-1) -- (1.5,0) -- (2.5,1); 
\draw[purple,line width=2,rounded corners=5pt] (-1,-1) --(0,0)-- (1.5,0) -- (2.5,-1);
\end{scope}
\end{scope}
\begin{scope}[shift={(7,0)}]
\begin{scope}[rotate=45]
\draw [black, dashed,line width=0.2] (-0.5,-0.5) -- (0.5,0.5); 
\draw [black, dashed,line width=0.2] (0.5,-0.5) -- (-0.5,0.5); 
\draw[purple,line width=2,rounded corners=10pt]%
   (-0.5,0.5) -- (0,0) -- (0.5,0.5);
\end{scope}
\node at (0,-1.) {$\rho_5$};
\begin{scope}[shift={(0,-2.5)},rotate=45,scale=0.35]
\draw [line width=0.4, dashed] (-1,-1) -- (0,0) -- (-1,1); 
\draw [line width=0.4, dashed] (0,0) -- (1.5,0);
\draw [line width=0.4, dashed] (2.5,-1) -- (1.5,0) -- (2.5,1); 
\draw[purple,line width=2,rounded corners=5pt] (-1,1) --(0,0)-- (1.5,0) -- (2.5,1);
\end{scope}
\end{scope}
\begin{scope}[shift={(8.75,0)}]
\begin{scope}[rotate=45]
\draw [black, dashed,line width=0.2] (-0.5,-0.5) -- (0.5,0.5); 
\draw [black, dashed,line width=0.2] (0.5,-0.5) -- (-0.5,0.5); 
\draw[purple,line width=2,rounded corners=10pt]%
   (-0.5,-0.5) -- (0,0) -- (0.5,0.5);
   \end{scope}
\node at (0,-1.) {$\rho_6$};
\begin{scope}[shift={(0,-2.5)},rotate=45,scale=0.35]
\draw [line width=0.4, dashed] (-1,-1) -- (0,0) -- (-1,1); 
\draw [line width=0.4, dashed] (0,0) -- (1.5,0);
\draw [line width=0.4, dashed] (2.5,-1) -- (1.5,0) -- (2.5,1); 
\draw[purple,line width=2,rounded corners=5pt] (-1,-1) --(0,0)-- (1.5,0) -- (2.5,1);
\end{scope}
\end{scope}
\begin{scope}[shift={(10.5,0)}]
\begin{scope}[rotate=45]
\draw [black, dashed,line width=0.2] (-0.5,-0.5) -- (0.5,0.5); 
\draw [black, dashed,line width=0.2] (0.5,-0.5) -- (-0.5,0.5); 
\draw[purple,line width=2,rounded corners=10pt]%
   (0.5,-0.5) -- (0,0) -- (-0.5,0.5);
   \end{scope}
\node at (0,-1.) {$\rho_7$};
\begin{scope}[shift={(0,-2.5)},rotate=45,scale=0.35]
\draw [line width=0.4, dashed] (-1,-1) -- (0,0) -- (-1,1); 
\draw [line width=0.4, dashed] (0,0) -- (1.5,0);
\draw [line width=0.4, dashed] (2.5,-1) -- (1.5,0) -- (2.5,1); 
\draw[purple,line width=2,rounded corners=5pt] (-1,1) --(0,0)-- (1.5,0) -- (2.5,-1);
\end{scope}
\end{scope}
\begin{scope}[shift={(12.25,0)}]
\begin{scope}[rotate=45]
\draw [black, dashed,line width=0.2] (-0.5,-0.5) -- (0.5,0.5); 
\draw [black, dashed,line width=0.2] (0.5,-0.5) -- (-0.5,0.5); 
\draw[purple,line width=2,rounded corners=10pt]%
   (-0.5,-0.5) -- (0,0) -- (-0.5,0.5);
\draw[purple,line width=2,rounded corners=10pt]%
   (0.5,-0.5) -- (0,0) -- (0.5,0.5);
 \end{scope}  
\node at (0,-1.) {$\rho_8$};
\begin{scope}[shift={(0,-2.5)},rotate=45,scale=0.35]
\draw [line width=0.4, dashed] (-1,-1) -- (0,0) -- (-1,1); 
\draw [line width=0.4, dashed] (0,0) -- (1.5,0);
\draw [line width=0.4, dashed] (2.5,-1) -- (1.5,0) -- (2.5,1); 
\draw[purple,line width=2,rounded corners=10pt] (-1,1) -- (0,0) -- (-1,-1);
\draw[purple,line width=2,rounded corners=10pt] (2.5,1) -- (1.5,0) -- (2.5,-1);
\end{scope}
\end{scope}
\begin{scope}[shift={(14,0)}]
\begin{scope}[rotate=45]
\draw [black, dashed,line width=0.2] (-0.5,-0.5) -- (0.5,0.5); 
\draw [black, dashed,line width=0.2] (0.5,-0.5) -- (-0.5,0.5); 
\draw[purple,line width=2,rounded corners=10pt]%
   (-0.5,-0.5) -- (0,0) -- (0.5,-0.5);
\draw[purple,line width=2,rounded corners=10pt]%
   (-0.5,0.5) -- (0,0) -- (0.5,0.5);
   \end{scope}
\node at (0,-1.) {$\rho_9$};
\begin{scope}[shift={(0,-2.5)},rotate=45,scale=0.35]
\node at (0,0) {$0$};
\end{scope}
\end{scope}
\end{tikzpicture}
\caption{Vertices of the square lattice O$(n)$ loop model (first row), 
together with the corresponding Boltzmann weights, and their mapping onto 
the vertices of the hexagonal lattice O$(n)$ model when the weight 
$\rho_9$ is zero.}
\label{fig:Onsquaretohexa}
\end{figure}
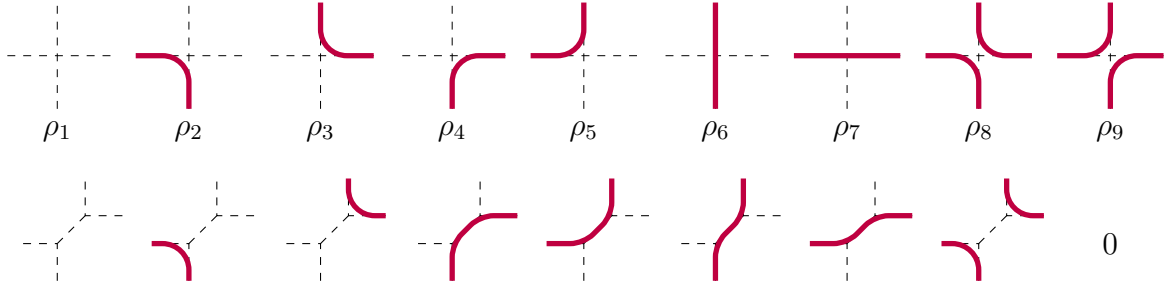

The O$(n)$ model on the hexagonal lattice is a particular case of the O$(n)$
model on the square lattice, whose vertex weights are parameterised as shown
in the first line of figure \ref{fig:Onsquaretohexa}.
Indeed, in the particular case $\rho_9=0$, the vertices can be stretched 
horizontally by adding one additional horizontal edge as shown in the 
second line of the figure.

An integrable choice of the Boltzmann weights is well known 
\cite{BloteNienhuis89} (we use here the notations of \cite{VJS:a22}):
\begin{subequations}
\label{eq:IKweights}
\begin{align}
\rho_1(u) &\;=\; 
        1 + \frac{2 \cos(u+\frac{3\theta}{2})\sin u 
       \sin\frac{\theta}{2}}{(2\cos\theta-1)\sin^2\theta}\,, \\
\rho_2(u) &\;=\;\rho_3(u) \;=\; \frac{\cos(u+\frac{3\theta}{2})}
                            {\cos\frac{3\theta}{2}}\,,\\
\rho_4(u) &\;=\;\rho_5(u) \;=\; -\frac{\sin u}
                             {\cos\frac{3\theta}{2}}\,,\\
\rho_6(u) &\;=\;\rho_7(u) \;=\; \frac{2 \cos(u+\frac{3\theta}{2})\sin u
       \sin\frac{\theta}{2}}{(2\cos\theta-1)\sin^2\theta}\,, \\
\rho_8(u) &\;=\; \frac{\sin(2u+\frac{5\theta}{2})-
       \sin\frac{\theta}{2}}{2(\cos\theta+\cos2\theta)}\,, \\
\rho_9(u) &\;=\; \frac{\cos(u+\frac{\theta}{2})\sin u}
           {2 \cos^2\frac{\theta}{2}\sin \frac{\theta}{2}(2\cos\theta-1)} \,.
\end{align}
\end{subequations}
In all the following we rescale $\rho_4$ and $\rho_5$ by a factor $-1$, 
which is just a change of basis. This model has several symmetries:
\begin{itemize}
\item The transformation $u \to \frac{3\pi}{2} - \frac{3\theta}{2} - u$ 
 for fixed $\theta$ is equivalent to rotating the vertices of
 figure~\ref{fig:Onsquaretohexa}  through $\frac{\pi}{2}$
 (crossing symmetry).
\item Up to some gauge symmetry, the weights are also invariant under 
 $\pi$ shifts of $u$, as well as under $2\pi$ shifts of $\theta$.
\end{itemize}
For a given $\theta$, there is only one non-trivial value of the spectral 
parameter $u$ such that $\rho_9=0$, namely 
$u=\frac{\pi}{2} - \frac{\theta}{2}$.
The corresponding weights, which we rescale such that $\rho_1 =1$, are the 
following [cf. (\ref{eq:IKweights})]:
\begin{subequations}
\begin{align}
 \rho_1 &\;=\; 1\,, \\
 \rho_2 &\;=\; \rho_3 \;=\; \frac{1}{2 \sin \frac{\theta}{2}}\,, \\
 \rho_4 &\;=\; \ldots \;=\; \rho_8 \;=\; 
                            \frac{1}{4 \sin^2 \frac{\theta}{2}}\,, \\
 \rho_9 &\;=\; 0 \,, 
\end{align}
\end{subequations}
so they recover precisely the fugacity $K$ given by 
eq.~(\ref{eq:fugacityK}). 

Now, it is well-known (see \cite{VJS:a22} for a review) that the integrable 
O$(n)$ model on the square lattice can be reformulated as a spin-one vertex 
model, the so-called IK (or $a_2^{(2)}$) model. 
The equivalence 
is worked out by once again giving orientations to the loops (to be summed 
over), and turning the loop weight $n$ into local, angular contributions. Each 
occupied edge is associated with a state $\pm 1$ (depending on the orientation 
of the corresponding loop), while empty edges are associated with the state 
$0$. 

We can now bend the lattice, and represent the corresponding TM as follows: 
\medskip
$$
\begin{tikzpicture}[scale=2]
\foreach \x in {0,...,5}{ 
\draw[purple, line width=5pt] (\x+0.5,-0.5) -- (\x+0.5,0.5);
}
\draw[purple, line width=5pt] (0,0) -- (6.5,0);
\node[purple] at (6,-0.75) {$\mathrm{e}^{\pm i {\varphi}}$};
\draw[purple,decorate,decoration={snake,amplitude=1pt, segment length=4pt}]%
   (6.,-0.5) -- (6.,0.5);
\draw[dashed] (0,-0.7) -- (0,0.5);
\draw[dashed] (6.5,-0.7) -- (6.5,0.5);
\node[below] at (0.5,-0.7) {$1$};
\node[below] at (1.5,-0.7) {$2$};
\node[below] at (3,-0.7) {$\ldots$};
\node[below] at (4.5,-0.7) {$L-1$};
\node[below] at (5.5,-0.7) {$L$};
\end{tikzpicture} 
$$
where each edge carries a $S^{(z)}=-1,0,1$ state, the interactions are 
encoded by local vertex weights depending on the 19 possible states of the
adjacent edges, and the vertical seam (red wavy line) 
corresponds to a boundary twist 
$\mathrm{e}^{-i \varphi}, 1, \mathrm{e}^{i \varphi}$. The fact that the 
vertical seam now only crosses one horizontal edge of each row (whereas it 
also crosses the last ``vertical'' edge in the six-vertex formulation on the
kagome lattice) is irrelevant,  since the vertical crossing can be eliminated
by a change of basis. 

The TM of the model commutes with the total magnetization 
\begin{equation}
m \;=\; \sum_{i=1}^{L} S^{(z)}_i  \,,
\label{eq:magnetizationOndef}
\end{equation}
which can naturally be associated to the number of through-lines in the 
O$(n)$ loop formulation.%
\footnote{Here and below, the word through-line refers to the spin-one
O($n$) loop model. Each such through-line corresponds to a pair of 
through-lines
in the spin-$1/2$ loop model; see figure~\ref{fig:Pottshexato6vtoOnhexa}.}
Therefore each sector of fixed $m$ can be considered independently. 

Rather than checking now explicitly that the eigenvalues dominating the 
spectrum of the TM for the different sectors of the colouring 
problem are indeed recovered by the spin-one vertex model, we defer this 
discussion to the next section, where we will deal with the BA 
construction of these different eigenvalues. 

The total dimension of the spin-one vertex model TM, summed over all
magnetisation sectors $m$, is obviously $3^L$. In the $m=0$ sector
the dimension is given by the central trinomial coefficient, that is, the
largest coefficient in the expansion of $(1+x+x^2)^L$. This has asymptotic
behaviour $3^L L^{-1/2} \sqrt{3/ 4\pi} [1 + O(1/L)]$.
Similarly, the dimension in sector $m=1$ is given by the
next-to-central trinomial coefficient, and more generally, that of the
sector with magnetisation $m$ by the trinomial coefficient that is
$m$ places away from the centre.

We stress that these dimensions are larger than those of the O$(n)$ loop model,
for the reason that
the TM of the latter does not distinguish between contractible and
non-contractible loops (it gives a weight $n$ to either type of loop).
It is however possible to work with an `augmented' loop model TM in
which each arc connecting two points is distinguished by the parity
(even or odd) of the number of times it has crossed the vertical seam.
When closing an even (resp.\ odd) arc we obtain a contractible
(resp.\ non-contractible) loop. It is easy to see that states of this
augmented loop model are in bijection with those of the spin-one
vertex model. It follows by the above equivalences that the spectra
of the respective TM are identical.

%
%
\section{Bethe Ansatz solution}
\label{sec:BA}
\setcounter{footnote}{1}

After all the mappings of section~\ref{sec:mappings}, we have managed to 
reformulate the original colouring problem as
an integrable spin-one model, namely the $a_2^{(2)}$ model with a particular 
value of the twist, related to the weight of
non-contractible FK clusters. This means that we are now in a good position 
to apply the toolbox of the BA method.

The goal of the present section is to set up the general framework, and 
define properly the regimes~I and IV that were introduced in
section~\ref{sec:pd_tri}, as well as the possible excitations of the 
ground state configuration at finite size $L$.
With these ingredients at hand, we can then in 
section~\ref{sec:criticalcontent} take the continuum limit and work out 
the corresponding critical exponents.

\subsection{Bethe Ansatz equations}

It is well-known how to construct the eigenstates of the integrable spin-one 
model through the BA. Each eigenstate is parameterised by a set of 
Bethe roots, solution of the Bethe Ansatz equations (BAE). In our particular 
case, the BAE for the square-lattice O($n$) model (which is related to
the IK model) are given by:
\begin{equation}
\left( \frac{\sinh\left( \lambda_j - \mathrm{i}\frac{\theta}{2}\right)}
            {\sinh\left( \lambda_j + \mathrm{i}\frac{\theta}{2}\right)} 
\right)^L \;=\; 
 \mathrm{e}^{\mathrm{i} \varphi} \; \prod\limits_{i(\neq j)} 
\frac{ \sinh\left( \lambda_j-\lambda_i - \mathrm{i}\theta \right) 
       \cosh\left(\lambda_j-\lambda_i + \mathrm{i}\frac{\theta}{2}\right) }
 {\sinh\left( \lambda_j-\lambda_i + \mathrm{i}\theta \right) 
  \cosh\left(\lambda_j-\lambda_i - \mathrm{i}\frac{\theta}{2}\right)} \,. 
\label{eq.BAE1}
\end{equation}

From there, we look for the leading TM eigenstates of the 
colouring problem in terms of Bethe roots for an appropriate twist. 
The following conclusions have been known since the paper of Baxter 
\cite{Baxter87}: 

\begin{itemize}
\item The range $0<Q<Q_0$ is described by the so-called regime~I in the 
      notations of \cite{Nienhuis}. Namely the ground state is made of roots 
      with imaginary part $\frac{\pi}{2}$.

\item The $Q_0 < Q < 4$ regime is dominated by another configuration of roots
      \cite{Baxter86}, namely $\frac{2L}{3}$ real roots, and $\frac{L}{3}$ 
      roots with imaginary 
      part $\frac{\pi}{2}$. It is worth noting that the corresponding 
      eigenstates are absent from the loop formulations, both on the square and
      hexagonal lattice. Also, it never dominates at one of the isotropic 
      values of the spectral parameter studied in \cite{Nienhuis,VJS:a22}, 
      nor does it dominate the spectrum of the corresponding Hamiltonians, 
      obtained by differentiating the TM with respect to the 
      spectral parameter $u$, at the value $u=0$. As a consequence, this 
      regime has escaped the attention of previous works on the IK 
      vertex model, in which three regimes labelled I, II and III were 
      identified and studied \cite{Nienhuis,VJS:a22}. Extending this 
      nomenclature, we shall call the regime dominated by this root 
      configuration {\em regime~IV}.  
\end{itemize}

Before going further in the study of the regimes, let us compare the BAE 
(\ref{eq.BAE1}) with those of \cite{Baxter86,BatchelorBlote}, where a 
7-vertex formulation of the hexagonal-lattice O$(n)$ loop model is solved 
without any reference to the IK 19-vertex model, which, as we have seen, 
is related to the more general square-lattice O($n$) loop model.
The BAE are written in terms of a set of complex Bethe roots $\{z_j\}$, 
and read 
\begin{equation}
z_j^L \;=\; (-1)^{n-1} \, \mathrm{e}^{i \epsilon} \, \prod_{i (\neq j)}
            \frac{S(z_i,z_j)}{S(z_j,z_i)} \,,
\label{eq:BAEBB}
\end{equation}
where the scattering phases are
\begin{equation}
S(z,w) \;=\; \left(1-z-w +z w+t^2 z \right)\left(1-2w +z w+t^2 w \right) \,,
\end{equation}
and $t^2 = 2 - 2\cos \theta$. In (\ref{eq:BAEBB}), we use the parameter  
$n=L+n_{-}-n_{+}$ equal to the number of lines in Baxter's formulation 
\cite{Baxter86}, with $n_{+}$ (resp.\/ $n_{-}$) being the number of upward 
(resp.\/ downward) pointing vertical arrows in each row 
\cite{BatchelorBlote}. Our sector $k=0$ corresponds to 
$n_{-}=n_{+}$, hence $n=L$.  
Following \cite{BatchelorBlote}, we set $z_j = \frac{1+rw_j}{r+w_j}$, 
where $r=\mathrm{e}^{i \theta}$; however, while these authors 
\cite{BatchelorBlote}
further define $w_j = \mathrm{e}^{2 \lambda_j}$, we choose instead to set 
$w_j = -\mathrm{e}^{2 \lambda_j}$. From there, the BAE (\ref{eq:BAEBB}) 
are easily seen to take the form (\ref{eq.BAE1}), where the correspondence 
between the twist parameters is simply $\varphi =\epsilon$. In other words, 
the BAE underlying the construction of eigenstates for the two models 
(the 7-vertex model of \cite{BatchelorBlote} and the IK 19-vertex model) are 
the {\em }same, even though the space of states of the two models 
are not equivalent.%
\footnote{In particular, we expect that either model is able to reproduce 
both regimes~I and~IV. Indeed, the BAE construction from the 7-vertex model 
is the one originally used by Baxter \cite{Baxter86}, which gave the first 
evidence for the existence of regime~IV.}

\subsection{Regime I: Summary of its properties}
\label{sec:regimeI}

Regime~I, originally investigated in the context of the square lattice 
\cite{Nienhuis},  was studied in great detail for the O($n$) loop model on 
the hexagonal lattice in \cite{BatchelorBlote}. The CFT is that of a single 
compactified bosonic field, and in particular, the dependence 
of the central charge $c$ on the twist $\varphi$ was exactly found to be%
\footnote{
Note that there is a typo in eq.~(2.62) of \cite{BatchelorBlote}. 
We give the right expression here.
} 
\begin{equation}
c(\varphi) \;=\; 1 - \frac{3\varphi^2}{\pi \theta} \,,
\label{eq.c_vs_twist_BB}
\end{equation} 
while the set of exponents with respect to the untwisted central charge $c=1$ 
were found as
\begin{equation}
\Delta_{m,w} \;=\; \frac{m^2 \theta}{4 \pi} + w^2\frac{\pi}{\theta} \,,
\end{equation}
and therefore, with respect to the twisted theory (\ref{eq.c_vs_twist_BB}), 
\begin{equation}
x_{m,w} \;=\;\frac{m^2 \theta}{4 \pi} + w^2\frac{\pi}{\theta}- 
             \frac{\varphi^2}{4\pi \theta}  \,,
\label{eq:RI:xmw}
\end{equation}
where $m$ the total magnetization introduced in (\ref{eq:magnetizationOndef}), 
and $w$ can be thought of as related to the electric charge in the CG 
picture. 

It turns out that the ground state of the O$(n)$ model with twist 
$\varphi = \theta$ is not observed in the Potts-model spectrum, and the central 
charge of the Potts model is recovered by choosing the
twist in this regime as $\varphi = 2\pi - \theta$.%
\footnote{We have already remarked (in section~\ref{sec:6v}) that both 
possibilities for the twist parameter would give the correct weight to 
non-contractible loops, but the comparison with the
spectrum of the Potts model here imposes the latter choice as the correct one.}
Then (\ref{eq.c_vs_twist_BB}) reduces to
\begin{equation}
c_\text{Potts} \;=\; 1 - \frac{3 (2\pi-\theta)^2}{\pi \theta} 
               \;=\; 1 - \frac{6 (p-1)^2}{p} \,,
\end{equation}  
where we have parameterised $\theta = 2\pi/p$. This is precisely the central
charge for the BK phase found by Saleur \cite{Saleur90,Saleur91}, matching
the CG results for coupling $g = \frac{1}{p} \in (0,\frac12)$. The $\ell$-leg
watermelon exponents are straightforwardly obtained from (\ref{eq:RI:xmw}), 
namely 
\begin{equation}
x_{\frac{\ell}{2},0} \;=\; \frac{\ell^2 \theta}{16 \pi}- 
                           \frac{(2\pi-\theta)^2}{4\pi \theta} 
                     \;=\; \frac{\ell^2}{8p}- \frac{(p-1)^2}{2p} \,,
\end{equation}
and coincide with the known values for bulk watermelon exponents in the BK 
phase (see for instance \cite{DuplantierSaleur87,Ikhlef11}). 
As explained in section \ref{sec:6v}, the magnetic exponent is associated to 
the scaling of the TM eigenvalue with twist $\frac{\varphi}{2}=\frac{\pi}{2}$, 
namely
\begin{equation}
x_H \;=\; \frac{1}{12}\left(c_\text{Potts}-c(\pi)\right) \;=\; 
          \frac{p}{8} - \frac{(p-1)^2}{2p} \,,
\end{equation}
which once again coincides with the known value in the BK phase. 
The thermal exponent is obtained from the scaling of the subdominant 
eigenvalues in the ground state ($m=0, w=0$) sector. These are obtained by 
taking values of the twist parameter shifted by $2\pi$, namely 
$\varphi=2\pi(1+j)-\theta$, resulting in 
\begin{equation}
x_T^{(j)} \;=\; \frac{1}{12}\left(c_\text{Potts}-c(2\pi(1+j)-\theta)\right) 
          \;=\; \frac{j}{2}\left[(j+2)p-2\right] \,.
\end{equation}
The two most relevant of these exponents (excluding the case $j=-1$, not 
observed in the spectrum) correspond to $j=1$ and $j=-2$, hence respectively
\begin{equation}
x_T  \;=\;  x_T^{(1)}  \;=\; \frac{3 p}{2} -1 \;\geq\; 2 \,, \qquad 
x_T' \;=\;  x_T^{(-2)} \;=\; 2 \,.
\end{equation}
Recalling that $p = \frac{1}{g}$, the first of these coincides with 
eq.~(\ref{thermexp}).

\subsection{Regime IV: structure of the leading eigenlevels}
\label{sec:RegimeIV}

We now study in detail the so-called regime~IV. 
Let us set notations: 
For sizes $L \in 3 \mathbb{N}$, the ground state is described by 
$m_1=\frac{2L}{3}$ real roots $x_i$, and $m_2 = \frac{L}{3}$ roots with 
imaginary part $\frac{\pi}{2}$ that we denote 
$y_k + \mathrm{i}\frac{\pi}{2}$ with $y_k \in \mathbb{R}$. 

The Bethe equations (\ref{eq.BAE1}) can be recast as:  
\begin{eqnarray}
\left( \frac{\sinh\left( x_j - \mathrm{i}\frac{\theta}{2}\right)}
       {\sinh\left(x_j + \mathrm{i}\frac{\theta}{2}\right)} \right)^L &=& 
 \mathrm{e}^{\mathrm{i} \varphi}\,  
\prod_{i(\neq j)} 
\frac{\sinh\left( x_j-x_i - \mathrm{i}\theta \right) 
      \cosh\left(x_j-x_i + \mathrm{i}\frac{\theta}{2}\right) }
     {\sinh\left( x_j-x_i + \mathrm{i}\theta \right) 
      \cosh\left(x_j-x_i - \mathrm{i}\frac{\theta}{2}\right)} \nonumber \\[2mm]
  & & \quad \times \, 
\prod_{k} 
\frac{\cosh\left( x_j-y_k - \mathrm{i}\theta \right) 
      \sinh\left(x_j-y_k + \mathrm{i}\frac{\theta}{2}\right) }
     {\cosh\left( x_j-y_k + \mathrm{i}\theta \right) 
      \sinh\left(x_j-y_k - \mathrm{i}\frac{\theta}{2}\right)} \,, \\[2mm]
\left( 
\frac{\cosh\left( y_k - \mathrm{i}\frac{\theta}{2}\right)}
     {\cosh\left( y_k + \mathrm{i}\frac{\theta}{2}\right)} \right)^L &=& 
 \mathrm{e}^{\mathrm{i} \varphi} \,  
\prod_{l(\neq k)} 
\frac{\sinh\left( y_k-y_l- \mathrm{i}\theta \right) 
      \cosh\left(y_k-y_l + \mathrm{i}\frac{\theta}{2}\right) }
     {\sinh\left( y_k-y_l + \mathrm{i}\theta \right) 
      \cosh\left(y_k-y_l- \mathrm{i}\frac{\theta}{2}\right)} \nonumber \\[2mm]
 & & \quad \times \, 
\prod_{j} 
\frac{\cosh\left( y_k-x_j- \mathrm{i}\theta \right) 
      \sinh\left(y_k-x_j + \mathrm{i}\frac{\theta}{2}\right) }
     {\cosh\left( y_k-x_j + \mathrm{i}\theta \right) 
      \sinh\left(y_k-x_j- \mathrm{i}\frac{\theta}{2}\right)} \,. 
\end{eqnarray}
Introducing the real function 
$\phi(\lambda,\alpha) = \frac{i}{2} 
\log \frac{\sinh \left(\lambda - i \alpha\right)}
          {\sinh \left(\lambda + i \alpha\right)}$, and taking the logarithm 
of these equations, they can be written as a set of coupled non-linear 
\emph{real} equations in the variables $\{x_j,y_k\}$: 
\begin{eqnarray} 
L \phi\left(x_j,\frac{\theta}{2}\right) &=&
\frac{\varphi}{2} - \pi I_j + 
\sum_{i(\neq j)} \left[\phi\left(x_j-x_i,\theta \right) +
                       \phi\left(x_j-x_i,\frac{\pi-\theta}{2}
\right)\right]  \nonumber \\[2mm] 
& & \quad +
\sum_{k} \left[\phi\left(x_j-y_k, \theta+\frac{\pi}{2} \right) +
               \phi\left(x_j-y_k,-\frac{\theta}{2}\right)\right] \,, \\[2mm]
L \phi\left(y_k,\frac{\pi+\theta}{2}\right) &=&
  \frac{\varphi}{2} - \pi J_k +  
\sum_{l(\neq k)} \left[ \phi\left(y_k-y_l,\theta \right) +
                        \phi\left(y_k-y_l,\frac{\pi-\theta}{2}\right) 
\right] \nonumber \\[2mm]
& & +
\sum_{j} \left[\phi\left(y_k-x_j, \theta+\frac{\pi}{2} \right)+
               \phi\left(y_k-x_j,-\frac{\theta}{2}\right)\right]\,,  
\label{eq:BAE:RIV:logform}
\end{eqnarray}
where the $I_j$ and $J_k$ are distinct integers (or half-integers), 
the so-called Bethe integers. In the ground state configuration, these are 
maximally and symmetrically packed, namely, for $L$ odd
\medskip
$$
\begin{tikzpicture}
\draw(-4,0) -- (4,0);
\draw(0,-1) -- (0,2);
\draw[dashed] (-4,1.5) -- (4,1.5);
\node at (5,0) {${\rm Im} = {0}$};
\node at (5,1.5) {${\rm Im} = \frac{\pi}{2}$};
\draw[fill=white](-2.4,1.5) circle [radius=3pt];
\draw[fill=white](-1.6,1.5) circle [radius=3pt];
\draw[fill=white](-0.8,1.5) circle [radius=3pt];
\draw[fill=white](0,1.5) circle [radius=3pt];
\draw[fill=white](0.8,1.5) circle [radius=3pt];
\draw[fill=white](1.6,1.5) circle [radius=3pt];
\draw[fill=white](2.4,1.5) circle [radius=3pt];
\draw[fill=black](-0.4,0) circle [radius=3pt];
\draw[fill=black](-1.2,0) circle [radius=3pt];
\draw[fill=black](-2.,0) circle [radius=3pt];
\draw[fill=black](-2.8,0) circle [radius=3pt];
\draw[fill=black](-3.6,0) circle [radius=3pt];
\draw[fill=black](0.4,0) circle [radius=3pt];
\draw[fill=black](1.2,0) circle [radius=3pt];
\draw[fill=black](2.,0) circle [radius=3pt];
\draw[fill=black](2.8,0) circle [radius=3pt];
\draw[fill=black](3.6,0) circle [radius=3pt];
\node at (2.6,1.1) {$\frac{L}{6}-\frac{1}{2}$};
\node at (1.6,1.1) {$\ldots$};
\node at (0.8,1.1) {$1$};
\node at (0.2,1.1) {$0$};
\node at (-0.8,1.1) {$-1$};
\node at (-1.4,1.1) {$\ldots$};
\node at (-2.4,1.1) {$-\frac{L}{6}+\frac{1}{2}$};
\node at (-3.6,0.4) {$-\frac{L}{3}+ \frac{1}{2}$};
\node at (2.2,0.4) {$\ldots$};
\node at (-1.2,0.4) {$-\frac{3}{2}$};
\node at (-0.4,0.4) {$-\frac{1}{2}$};
\node at (3.5,0.4) {$\frac{L}{3}- \frac{1}{2}$};
\node at (-2.2,0.4) {$\ldots$};
\node at (1.2,0.4) {$\frac{3}{2}$};
\node at (0.4,0.4) {$\frac{1}{2}$};
\end{tikzpicture}
$$
while for $L$ even 
\medskip
$$
\begin{tikzpicture}
\draw(-4,0) -- (4,0);
\draw(0,-1) -- (0,2);
\draw[dashed] (-4,1.5) -- (4,1.5);
\node at (5,0) {${\rm Im} = {0}$};
\node at (5,1.5) {${\rm Im} = \frac{\pi}{2}$};
\draw[fill=white](-2.0,1.5) circle [radius=3pt];
\draw[fill=white](-0.4,1.5) circle [radius=3pt];
\draw[fill=white](0.4,1.5) circle [radius=3pt];
\draw[fill=white](2.0,1.5) circle [radius=3pt];
\draw[fill=black](-0.4,0) circle [radius=3pt];
\draw[fill=black](-1.2,0) circle [radius=3pt];
\draw[fill=black](-2.,0) circle [radius=3pt];
\draw[fill=black](-3.,0) circle [radius=3pt];
\draw[fill=black](0.4,0) circle [radius=3pt];
\draw[fill=black](1.2,0) circle [radius=3pt];
\draw[fill=black](2.,0) circle [radius=3pt];
\draw[fill=black](3.,0) circle [radius=3pt];
\node at (2.2,1.1) {$\frac{L}{6}-\frac{1}{2}$};
\node at (1.1,1.1) {$\ldots$};
\node at (0.4,1.1) {$\frac{1}{2}$};
\node at (-0.4,1.1) {$-\frac{1}{2}$};
\node at (-1.,1.1) {$\ldots$};
\node at (-2.,1.1) {$-\frac{L}{6}+\frac{1}{2}$};
\node at (-3.,0.4) {$-\frac{L}{3}+\frac{1}{2}$};
\node at (-2.,0.4) {$\ldots$};
\node at (-1.2,0.4) {$-\frac{3}{2}$};
\node at (-0.4,0.4) {$-\frac{1}{2}$};
\node at (3.2,0.4) {$\frac{L}{3}-\frac{1}{2}$};
\node at (2.,0.4) {$\ldots$};
\node at (1.2,0.4) {$\frac{3}{2}$};
\node at (0.4,0.4) {$\frac{1}{2}$};
\end{tikzpicture}
$$
From this ground state solution several elementary excitations can be 
considered: 
\begin{itemize}
\item Magnetic excitations, corresponding to removing 
$\delta m_1, \delta m_2$ roots of each Fermi sea, together with a rearranging 
of the Bethe integers such that these remain symmetric and maximally packed.  

\item Global backscatterings, corresponding to shifting all the Bethe 
integers of either Fermi sea by an amount $2 w_1$, or respectively $2 w_2$ 
(where $w_i$ are either integers of half-integers).

\item Particle-hole excitations of either of the Fermi seas, which correspond 
to shifting $\delta_1^\pm$ (resp.\/ $\delta_2^\pm$) of the Bethe roots to a 
higher/lower Bethe integer. This can be represented as follows 
(for the plus sign):
\medskip
$$
\begin{tikzpicture}
\draw(-4,0) -- (4,0);
\draw[fill=black](-0.4,0) circle [radius=3pt];
\draw[fill=black](-1.2,0) circle [radius=3pt];
\draw[fill=black](-2.,0) circle [radius=3pt];
\draw[fill=black](-3.,0) circle [radius=3pt];
\draw[fill=black](0.4,0) circle [radius=3pt];
\draw[fill=black](1.2,0) circle [radius=3pt];
\draw[fill=black](2.,0) circle [radius=3pt];
\draw[fill=white](3.,0) circle [radius=3pt];
\draw[thick,->,rounded corners=10pt,>=latex] (3,0.2) -- (3.4,0.5) -- (3.8,0.2);
\draw[fill=black](3.8,0) circle [radius=3pt];
\end{tikzpicture}
$$
\end{itemize}

Other excitations can also be considered, namely roots can be rearranged 
in complex conjugate pairs called {\em 2-strings}: $x_i \pm \mathrm{i} \phi_i$. 
However, such configurations can be considered as `artefacts', in the 
following sense: not only these configurations are highly unstable under 
changes of $\theta$ or $\varphi$ and their analytical treatment (if any) 
escapes the range of the methods used in this paper, but the associated 
conformal weights can actually be obtained from the understanding of states 
with no 2-strings. 
How this comes about shall be made clearer in section \ref{sec:k0spectrum}, 
where we will examine in detail the structure of the eigenstates of the 
colouring problem in the $k=0$ sector. 

\subsection{Regime IV: analysis of the Bethe-Ansatz equations}
\label{sec:regIV.analysis.BAE}

In the continuum limit, the logarithmic BAE (\ref{eq:BAE:RIV:logform}) turn 
into equations for the densities of real parts $\rho(x), \sigma(y)$. In 
Fourier space (where the densities are denoted by 
$\widetilde{\rho}(\omega), \widetilde{\sigma}(\omega)$), these equations 
are easily rewritten as 
\begin{eqnarray}
\frac{\sinh\omega\left(\frac{\pi}{2}-
\frac{\theta}{2}\right)}{\sinh\frac{\omega\pi}{2}}
&=&
\widetilde{\rho} + 
\frac{\sinh\omega\left( \frac{\pi}{2}-{\theta}\right) + 
      \sinh\frac{\omega\theta}{2}}{\sinh\frac{\omega\pi}{2}} \,  
      \widetilde{\rho} \nonumber \\ 
 & & \qquad - 
\frac{\sinh\omega\left( \frac{\pi}{2}-\frac{\theta}{2}\right) + 
      \sinh\omega\theta}{\sinh\frac{\omega\pi}{2}} \,  
      \widetilde{\sigma} \quad 
\\
-\frac{\sinh\frac{\omega\theta}{2}}{\sinh\frac{\omega\pi}{2}}
&=&
\widetilde{\sigma} + 
\frac{\sinh\omega\left( \frac{\pi}{2}-{\theta}\right) + 
      \sinh\frac{\omega\theta}{2}}{\sinh\frac{\omega\pi}{2}}\,  
\widetilde{\sigma} \nonumber  \\
 & & \qquad - 
\frac{\sinh\omega\left( \frac{\pi}{2}-\frac{\theta}{2}\right) + 
      \sinh \omega\theta}{\sinh\frac{\omega\pi}{2}} \,  
\widetilde{\rho}
\end{eqnarray}

These equations are easily solved, and we obtain the densities in the 
thermodynamic limit of the real roots in Fourier space:
\begin{equation}
\widetilde{\rho} \;=\; 
\frac{2 \cosh \frac{\omega\theta}{2}}{1+2\cosh\omega\theta} \,, \quad
\widetilde{\sigma} \;=\; \frac{1}{1+2\cosh\omega\theta} \,. 
\end{equation}
So, back to real space, 
\begin{equation}
 \rho(x)  \;=\; \frac{1}{\theta \sqrt{3}} 
          \frac{\sinh \frac{2 \pi x}{3 \theta}+
                \sinh \frac{4 \pi x}{3 \theta}}{\sinh \frac{2 \pi x}{\theta}}
\,, \quad
 \sigma(y)  \;=\; \frac{1}{\theta \sqrt{3}} 
           \frac{\sinh \frac{\pi y}{3 \theta}}{\sinh \frac{\pi y}{\theta}}\,.
\end{equation}

\begin{figure}
\centering
\includegraphics[width=200pt]{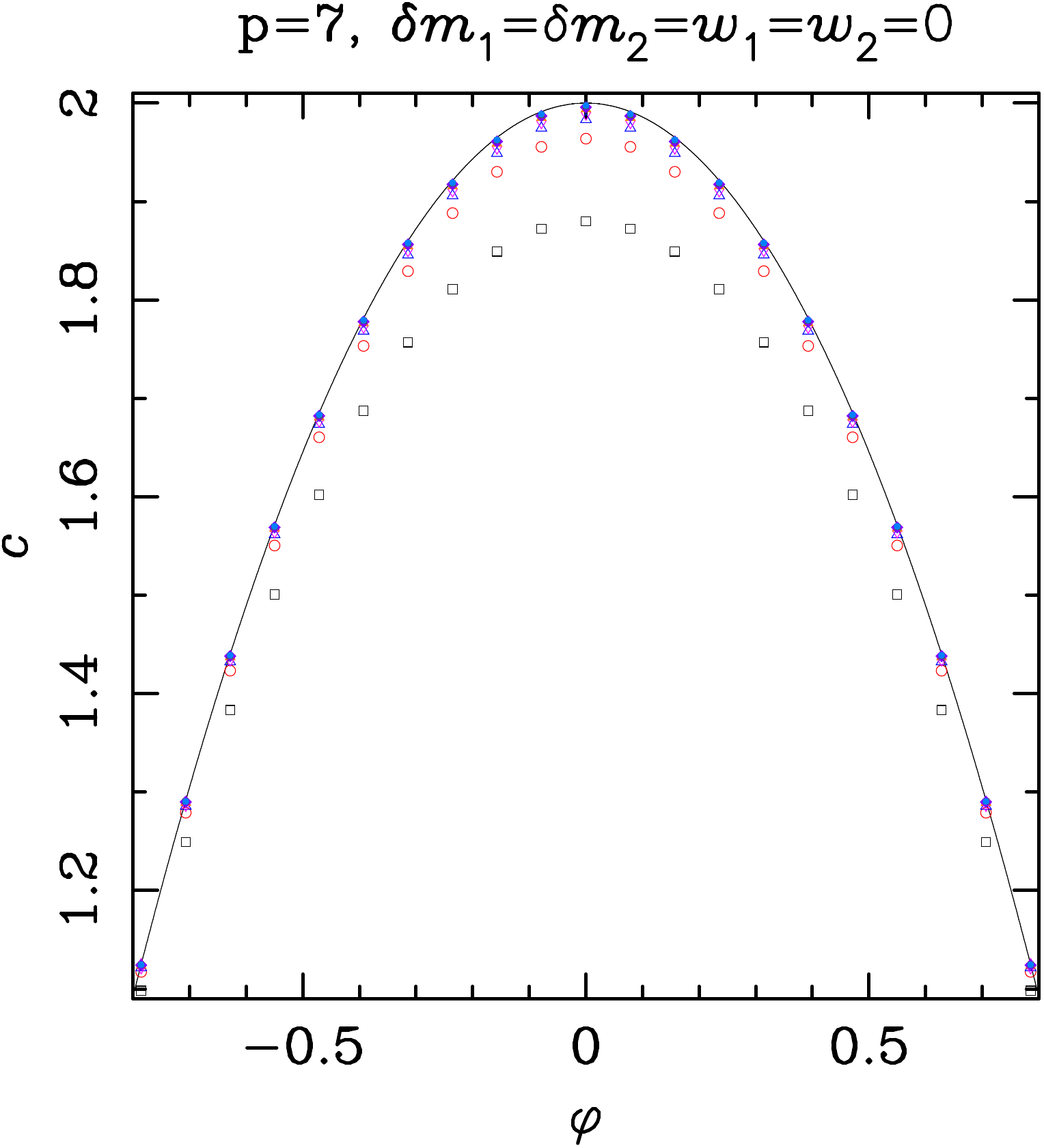}
\caption{Estimates of the central charge for the ground state (with 
root configuration given by (\ref{eq:quantumnumbersgroundstate})) for 
$p=7$ ($Q \simeq 3.2469796$) as a function of the twist $\varphi$. 
We show the numerical solutions of the BAE
for sizes $L=6$ (black $\square$), $L=12$ (red $\circ$), $L=18$ 
(navy blue $\triangle$), $L=24$ (pink $\lozenge$), $L=30$ (orange $\times$),
$L=36$ (violet $\blacklozenge$), $L=42$ (dark gray $\blacksquare$), 
and $L=48$ (blue $\bullet$).   
The convergence towards the analytic formula 
$c = 2 - \frac{6p}{p-4} \left(\frac{\varphi}{\pi}\right)^2$ [cf. 
(\ref{eq:BAE:RIV:ctwisted})], depicted as a black curve, is excellent.}
\label{fig:p7cphi}
\end{figure}

As discussed in the previous section, the excitations above the ground state 
are obtained by creating `holes' and `backscatterings' in the Fermi seas of 
$x$ and $y$ (we shall not consider here the excitation obtained by forming 
2-strings). Introducing the densities of holes $\rho^h$ and $\sigma^h$ (which 
play an analogous role as the densities of roots $\rho$ and $\sigma$), we can 
write the scattering equations in a matrix form 
\begin{equation}
 \left(  \begin{array}{c}  
          \rho + \rho^h   \\
          \sigma + \sigma^h
         \end{array}
 \right) 
 \;=\;   s + K   \left(  \begin{array}{c}  
          \rho    \\
          \sigma 
         \end{array}
 \right)  \,.
\end{equation}
The conformal weights of excitations indexed by the holes numbers 
$(\delta m_1,\delta m_2)$ and backscatterings $(w_1,w_2)$ are obtained 
(at zero twist) as \cite{korepin1997quantum}
\begin{equation}
  \Delta_\alpha + \bar{\Delta}_\alpha \;=\;  
  \frac{1}{4} (\delta m_1,\delta m_2) \left(1-K\right) 
  \left(  \begin{array}{c}  
          \delta m_1  \\
          \delta m_2 
         \end{array}   \right) 
         + 
         (w_1,w_2) \left(1-K\right)^{-1} \left(  \begin{array}{c}  
          w_1   \\
          w_2 
         \end{array}   \right)  \,, 
\end{equation}
where the label $\alpha$ denotes for the set of quantum numbers of a given 
excitation, and in $1-K$ the zero-frequency limit $\omega \to 0$ has to 
be taken, so 
\begin{equation}
 1-K \;=\; \left(    \begin{array}{rr}  
          2- \frac{\theta}{\pi}  & -1-\frac{\theta}{\pi}  \\
          -1-\frac{\theta}{\pi}  & 2- \frac{\theta}{\pi}
         \end{array}  \right) \,.
\end{equation}
Therefore,  
\begin{eqnarray}
  \Delta_\alpha + \bar{\Delta}_\alpha &=&  
  \frac{1}{2}
  \left[ \left(1-\frac{1}{p}\right) \left({\delta m_1}^2 + {\delta m_1}^2 
      \right) - \left(1+\frac{2}{p}\right) {\delta m_1} {\delta m_2} \right]
  \nonumber \\
 & & \qquad \qquad + 
 \frac{2}{3}\left({w_1}^2+{w_1} {w_2} + {w_2}^2\right) + 
 \frac{2}{p-4}\left({w_1}+{w_2}\right)^2 \,,
 \label{eq:BAE:RIV:Deltaanal}
\end{eqnarray}
which can readily be checked numerically. 
Further, we also check numerically that introducing particle-hole excitations 
results in the following conformal weights with respect to the ground state 
of the untwisted $c=2$ theory
\begin{eqnarray}
  \Delta_\alpha + \bar{\Delta}_\alpha &=& \frac{1}{2}
  \left[ \left(1-\frac{1}{p}\right) \left({\delta m_1}^2 + 
  {\delta m_1}^2 \right) - 
  \left(1+\frac{2}{p}\right) {\delta m_1} {\delta m_2} \right] \nonumber\\
 & & \qquad + 
 \frac{2}{3}\left({w_1}^2+{w_1} {w_2} + {w_2}^2\right) + 
 \frac{2}{p-4}\left(w_1+w_2\right)^2 \nonumber\\
 & & \qquad + \delta_1^{+}  + \delta_1^{-} + \delta_2^{+} + \delta_2^{-} 
 \,.
 \label{eq:BAE:RIV:Deltaanalbis}
\end{eqnarray}

It is also of interest to study what happens to these excited states when a 
twist $\varphi$ is turned on. We therefore measure numerically the 
corresponding effective central charges, 
$c_{\alpha} = 2-12 \left( \Delta_\alpha + \bar{\Delta}_\alpha \right)$, 
as a function of the twist. 
Solving the corresponding BAE for sizes up to $L \sim 100$, we find the 
corresponding formula 
\begin{eqnarray}
c_\alpha(\varphi) &=& 2 -
6\left[ \left(1-\frac{1}{p}\right) \left({\delta m_1}^2 + {\delta m_1}^2 
        \right) - 
  \left(1+\frac{2}{p}\right) {\delta m_1} {\delta m_2} \right] \nonumber \\
 & & \qquad - 8\left({w_1}^2+{w_1} {w_2} + {w_2}^2\right) - 
     \frac{24}{p-4}\left({w_1}+{w_2}\right)^2  \nonumber \\
 & & \qquad - \frac{6 p}{p-4}  \frac{\varphi}{\pi}\left( 
     \frac{\varphi}{\pi} + 2(w_1 + w_2)  \right) - 
    12\left(\delta_1^{+}  + \delta_1^{-} + \delta_2^{+} + \delta_2^{-}\right)
\,.
\label{eq:BAE:RIV:ctwisted}
\end{eqnarray}
An example is shown in figure~\ref{fig:p7cphi}, where we measure numerically 
the central charge associated with the ground state 
(\ref{eq:quantumnumbersgroundstate}) 
as a function of the twist, and which shows an excellent agreement 
with the formula (\ref{eq:BAE:RIV:ctwisted}).

The exponents $\Delta_\alpha, \bar{\Delta}_\alpha$ are measured with respect 
to the untwisted model with central charge $c=2$. Anticipating on the results 
of section~\ref{sec:ccharge}, the exponents of the Potts model should 
instead be measured 
with the twisted central charge (\ref{eq:BAE:RIV:cPotts}). We shall therefore 
consider the shifted exponents $h_\alpha, \bar{h}_\alpha$, 
\begin{equation}
 h_\alpha + \bar{h}_\alpha \;=\; \Delta_\alpha + \bar{\Delta}_\alpha 
- \frac{2}{p(p-4)} \,. 
\end{equation}
Or, more precisely, 
\begin{eqnarray}
h_\alpha + \bar{h}_\alpha &=& 
\frac{1}{2}\left[ \left(1-\frac{1}{p}\right) \left({\delta m_1}^2 + 
      {\delta m_1}^2 \right) - 
 \left(1+\frac{2}{p}\right) {\delta m_1} {\delta m_2} \right] \nonumber\\
 & & \qquad + 
 \frac{2}{3}\left({w_1}^2+{w_1} {w_2} + {w_2}^2\right) + 
 \frac{2}{p-4}\left({w_1}+{w_2}\right)^2  \nonumber\\
 & & \qquad + \delta_1^{+} + \delta_1^{-} + \delta_2^{+} + \delta_2^{-} 
 \nonumber\\
 & & \qquad + \frac{p}{2(p-4)}  \frac{\varphi}{\pi} 
    \left( \frac{\varphi}{\pi} + 2(w_1+w_2)  \right) - \frac{2}{p(p-4)}\,.
\label{eq:BAE:RIV:hhbar}
\end{eqnarray}

To disentangle $h_\alpha$ from $\bar{h}_\alpha$, we can further use the 
translational invariance of the system, implying that the TM commutes with the 
momentum operator. The latter can be written as a sum over Bethe integers, 
\begin{equation} 
P \;=\; \frac{2\pi}{L} \left( \sum_{i=1}^{m_1}2I_j +
                              \sum_{k=1}^{m_2}2J_k \right) \,.
\end{equation}
So we readily have, for the states of interest: 
\begin{eqnarray}
P_\alpha &=& \frac{2\pi}{L} \left[  
   \left( \frac{2L}{3}-\delta m_1\right)w_1 + 
   \left( \frac{L}{3}-\delta m_2\right)w_2  + 
   \delta_1^+ + \delta_2^+ -\delta_1^- - \delta_2^- \right] \nonumber \\
&=& 2\pi \left( \frac{2}{3}w_1+ \frac{1}{3}w_2 \right) 
+ \frac{2 \pi}{L} \left(   \delta_1^+ + \delta_2^+ -\delta_1^- - \delta_2^- 
- \delta m_1 w_1 - \delta m_2 w_2 \right) \,.
\end{eqnarray}
In the continuum limit, the $\frac{1}{L}$-dependent part is related to 
the conformal spin $h_\alpha - \bar{h}_\alpha $, as 
\begin{equation}
h_\alpha - \bar{h}_\alpha \;=\; \delta_1^+ + \delta_2^+ -
    \delta_1^- - \delta_2^- - \delta m_1 w_1 - \delta m_2 w_2 \,.
\label{eq:hhbarmomentum}
\end{equation}

%
%
\section{Critical content of the colouring problem}
\label{sec:criticalcontent}
\setcounter{footnote}{1}

We will now restrict to the specific states observed in the Potts spectrum. 
This will lead to analytic expressions for the central charge, the thermal 
and magnetic exponents, and the watermelon exponents describing the
propagation of several FK clusters. Combining this information will 
finally lead to the identification of the CFT describing vertex colourings of
the triangular lattice.

\subsection{Ground state ($k=0$) sector} 
 \label{sec:k0spectrum}
 
\subsubsection{Central charge.} 
\label{sec:ccharge}

As already explained, the ground state of the Potts model (namely, in the 
$k=0$ sector), can be found as an eigenstate in the zero magnetisation sector 
of the spin-one vertex model, at twist $\varphi =\theta$. In terms of Bethe 
roots, it simply corresponds to 
\begin{equation}
(\delta m_1, \delta m_2) \;=\; (0,0)\,, \quad 
(w_1,w_2) \;=\; (0,0)\,, \quad 
\varphi\;=\; \theta \,. 
\label{eq:quantumnumbersgroundstate}
\end{equation}
 
From equation (\ref{eq:BAE:RIV:ctwisted}), we therefore directly read the 
corresponding central charge, 
\begin{equation}
c \;=\; 2 - \frac{24}{p(p-4)} \,.
 \label{eq:BAE:RIV:cPotts}
\end{equation}
We have made an extensive numerical test of this formula by 
\begin{itemize}
  \item Direct diagonalisation of the TM for small values of $L=6,9,12$. In
        this case we were able to cover the whole interval $Q\in (2,4)$ in 
        steps of size $10^{-3}$. 
  \item Numerically solving the BAE for sizes $L=6,9,12,24,48,96$ with the 
        right twist $\varphi=\theta$. Except for $L=6,9$, we could solve 
        these equations in the smaller interval $Q \in [2.469,4)$, again 
        in steps of size $10^{-3}$. Note that for $Q\in(2,B_5)$, the root 
        configurations contain one or more 2-strings. Furthermore, the higher 
        the value of $L$, the larger the number of 2-strings that might appear 
        in the configuration. 
        This technical complication prevented us from studying the 
        interval $(2,2.469)$ for $L\ge 12$ by solving the BAE. Indeed, for
        $L=6,9,12$ the results obtained from this method agree
        perfectly well with those obtained from the TM diagonalisation. 
\end{itemize} 
In particular, for each value of $Q \in [2.469,4)$, we computed the 
free energy $f_L(Q)$ for $L=6,12,24,48,96$. ($f_9(Q)$ was computed mainly for
testing our methods.)
We then performed the standard three-parameter fit based 
[cf.~(\ref{eq:CFT_fL})] on the CFT Ansatz
$f_L(Q) = f_\text{bulk}(Q) + \frac{\pi G}{6L^2} c(Q) + \frac{A}{L^4}$, using
three consecutive values of the width, 
$L=L_\text{min}, 2L_\text{min}, 4L_\text{min}$,
to obtain the unknowns $f_\text{bulk}(Q)$, $c(Q)$ and $A$.
As a precaution against finite-size effects,
we repeated the fits for three different values of $L_\text{min}$.

The values obtained for $f_\text{bulk}$ agree very well with those
given by Baxter's exact expression for regime~IV (\ref{Baxter_g2}), 
even though we are also computing the former values for $Q$ much
less than $Q_0$ [cf.~(\ref{eq.Q0})], where the ground state of regime~I 
is dominant. This vividly illustrates the fact---discussed at the end of 
section~\ref{sec:pd_tri}---that regimes~I and~IV coexist, in the sense that
the scaling levels of either of them are well-defined by analytic 
continuation on the other side of $Q_0$.
 
\begin{figure}
\centering
\begin{tabular}{cc}
\includegraphics[width=200pt]{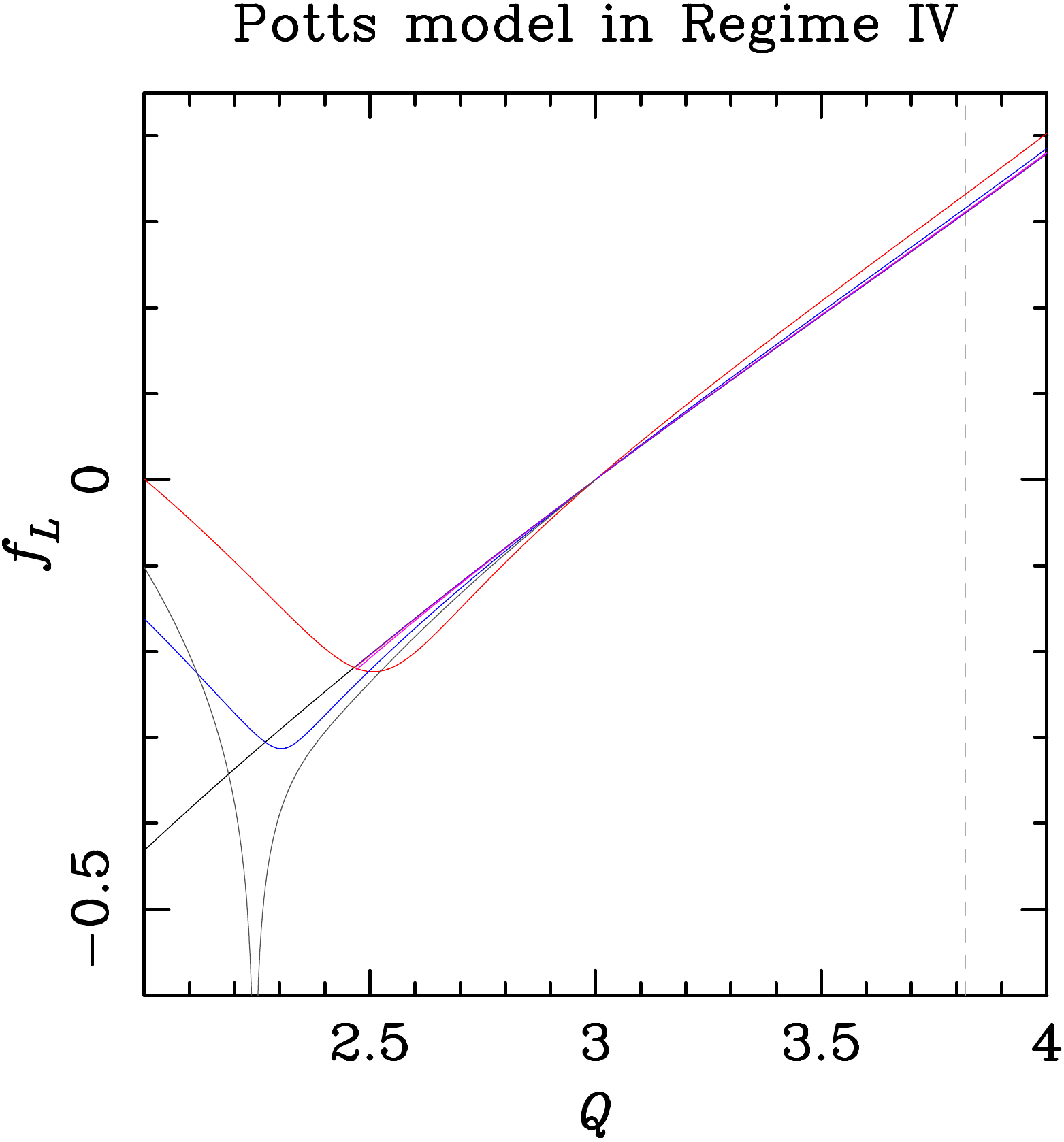} & 
\includegraphics[width=200pt]{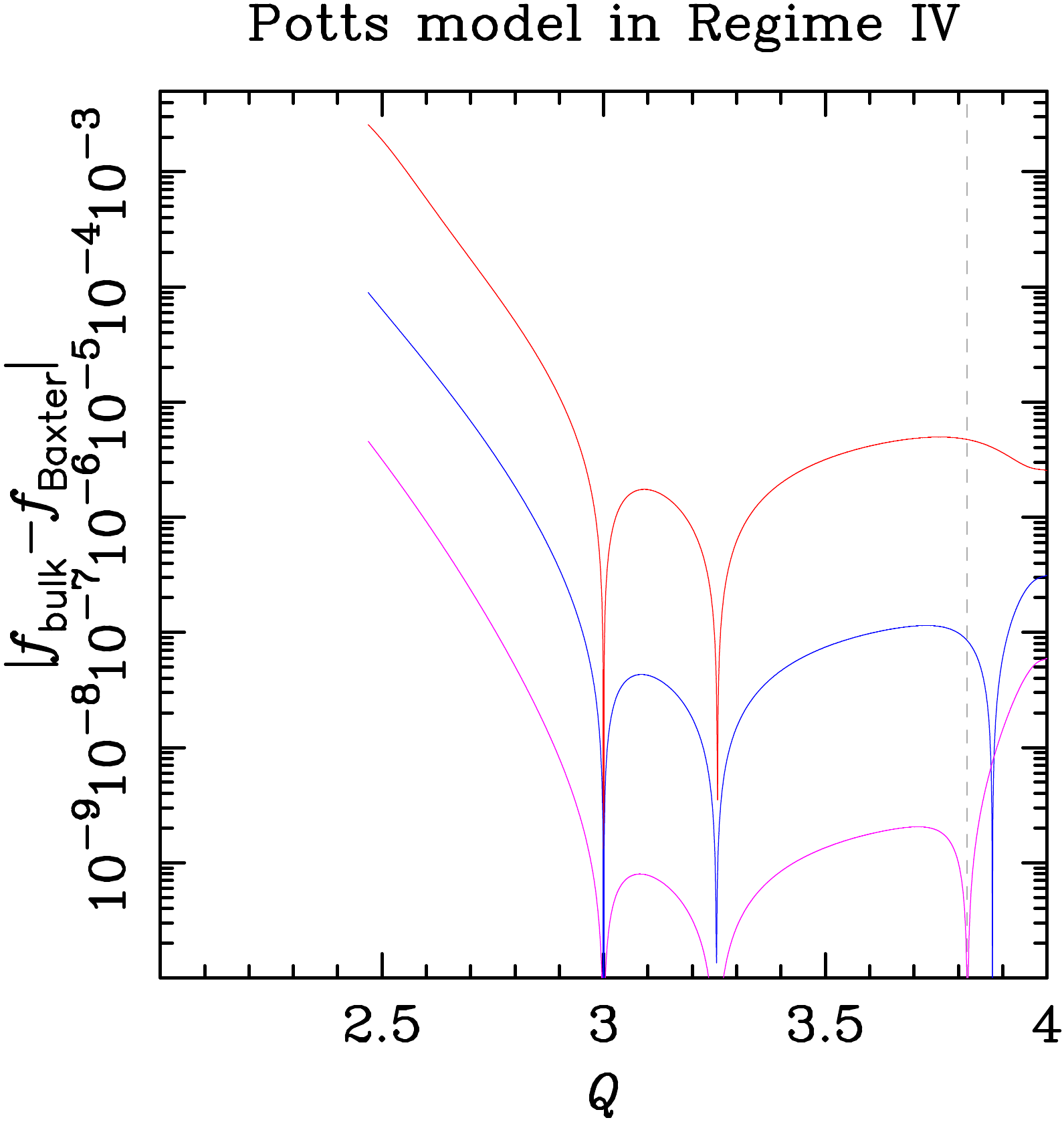} \\[2mm] 
\qquad  (a) &\quad\quad\; (b) \\
\end{tabular}
\caption{(a) Values of the free energies $f_L$ as a function of $Q$ for
several values of the width: $L=6$ (red), $L=12$ (blue), $L=24$ (pink),
$L=48$ (orange), and $L=96$ (violet).
The dark-gray curve diverging at $Q\approx 2.224$ corresponds to $L=9$.
The black curve corresponds to Baxter's exact result (\ref{Baxter_g2}). 
(b) Absolute difference
$|f_\text{bulk}-f_\text{Baxter}|$ for the bulk free energy obtained from 
the three-parameter fit 
$f_L(Q) = f_\text{bulk}(Q) + \frac{\pi G}{6L^2}c(Q) +  \frac{A}{L^4}$ 
with $L_\text{min}=6$ (red), $L_\text{min}=12$ (blue), and 
$L_\text{min}=24$ (pink). 
}
\label{fig:free}
\end{figure}

In figure~\ref{fig:free}(a) we depict the raw data for $f_L(Q)$, as well as 
Baxter's free energy (\ref{Baxter_g2}), and in panel (b) we show the 
discrepancies between $f_\text{bulk}$ and $f_\text{Baxter}$ (\ref{Baxter_g2})
in semi-logarithmic scale. The agreement is very good. This result 
shows that the ground state of regime~IV can be analytically continued 
deep inside the regime~I. 
We believe that this agreement can be extended to the range $2 < Q \le 4$,
and we have made extensive tests to verify this claim. Indeed we have followed
the free-energy level $f_L$ which is dominant at $Q=4$ (i.e., the ground 
state of regime~IV) by analytic continuation to smaller $Q$---using both direct 
diagonalisation of the TM for small $L$, and numerical BA 
computations for larger $L$---and found it to have the following features:
\begin{itemize}
 \item If $L\in 6{\mathbb N}$, the level $f_L$ follows Baxter's $g_2$ closely 
       down 
       to some value $Q_{\rm min}(L) > 2$ where $f_L(Q)$ has a minimum. For 
       $Q < Q_{\rm min}(L)$, $f_L$ breaks away from $g_2$. However, for 
       $L=9$ (see figure~\ref{fig:free}(a)), although the  
       behaviour is similar, we find a divergence at 
       $Q_{\rm div}(9) \approx 2.244$, rather than a minimum. Therefore we 
       expect parity effects, so in our main fits we will only include even 
       values of $L$.
 \item There is good evidence that $Q_{\rm min}(L) \to 2$ as $L \to \infty$ 
       for $L\in 6{\mathbb N}$. We also conjecture that 
       $Q_{\rm div}(L)\to 2$ as $L \to \infty$ for $L$ being 
       an odd multiple of 3. 
 \item Right at $Q=2$, we find that $f_L$ takes a finite value that, 
       at least for $L=6,9,12$ decreases as $L$ increases. If we fit 
       the corresponding $f_L(2)$ values to the standard CFT 3-parameter 
       Ansatz, we obtain $f_\text{bulk}(2) \approx -0.2570$, which is still 
       larger than Baxter's value $g_2(2) \approx -0.4315231$. 
\end{itemize}
These observations indicate that $f_L \to g_2$ as $L \to \infty$ on the 
interval $Q \in (2,4]$, but the convergence is not uniform. 
This feature is responsible of the `largest' discrepancies 
observed for $Q \lessapprox B_5$. Indeed, for 
$Q \to 2^+$ the central charge (\ref{eq:BAE:RIV:cPotts}) diverges, signaling 
the break-down of regime~IV. On the other hand, we have found no evidence 
of regime~IV levels whose analytic continuation converge to $g_2$ for 
$0 \le Q < 2$. Since $g_2$ is analytic on the whole interval $0 \le Q \le 4$, 
it is possible that it serves as the ground state of yet another regime (and 
a different CFT) for $Q \in [0,2)$, but further examination of
this issue is beyond the scope of the present paper.

A further noteworthy feature concerns the value of the analytic continuation 
of regime~IV's ground state to $Q=3$. We find that $f_L(3)=0$ exactly for any 
$L$. This should be compared with the simple fact that a triangular-lattice 
strip with cylindrical boundary conditions and width $L$ a multiple of 3 is 
uniquely 3-colourable modulo global colour permutations (i.e. $Z_G(3,-1)=3!$). 
Therefore regime~IV captures an essential feature of the colouring problem 
at $Q=3$.

Finally, in figure~\ref{fig:free}(b) we observe that the convergence of 
$f_L$ to $g_2$ is noticeable slower close to $Q=4$. This is probably due 
to the usual logarithmic corrections at $Q=4$, well-known in the Potts-model 
context.

The reader might worry if all of this does not contradict the simple 
observation that $Z=0$ for $Q=2$, since a triangular lattice manifestly does 
not admit a proper vertex 2-colouring. As a matter of fact, there is no 
problem, because at $Q=2$ massive cancellations of TM eigenvalues will occur,%
\footnote{
The same happens for other special values of $Q$, which are the Beraha numbers 
for cylindrical boundary conditions \cite{JacobsenSalasSokal03}, 
and $Q \in \mathbb{N}$ for toroidal boundary conditions 
\cite{JacobsenSalas07}.}
due to three distinct mechanisms:
\begin{enumerate}
 \item Some of the eigenvalues are simply zero;
 \item Other eigenvalues have a zero amplitude in the decomposition of the 
       Markov trace \cite{Richard07};
 \item Some eigenvalues are equal up to a sign, and their combined amplitude 
       summed over the various sectors of the TM is zero.
\end{enumerate}
The end result is that $Z=0$ indeed, despite the existence of finite 
eigenvalues, and we have checked this explicitly for small $L$.

\begin{figure}
\centering
\begin{tabular}{cc}
\includegraphics[width=200pt]{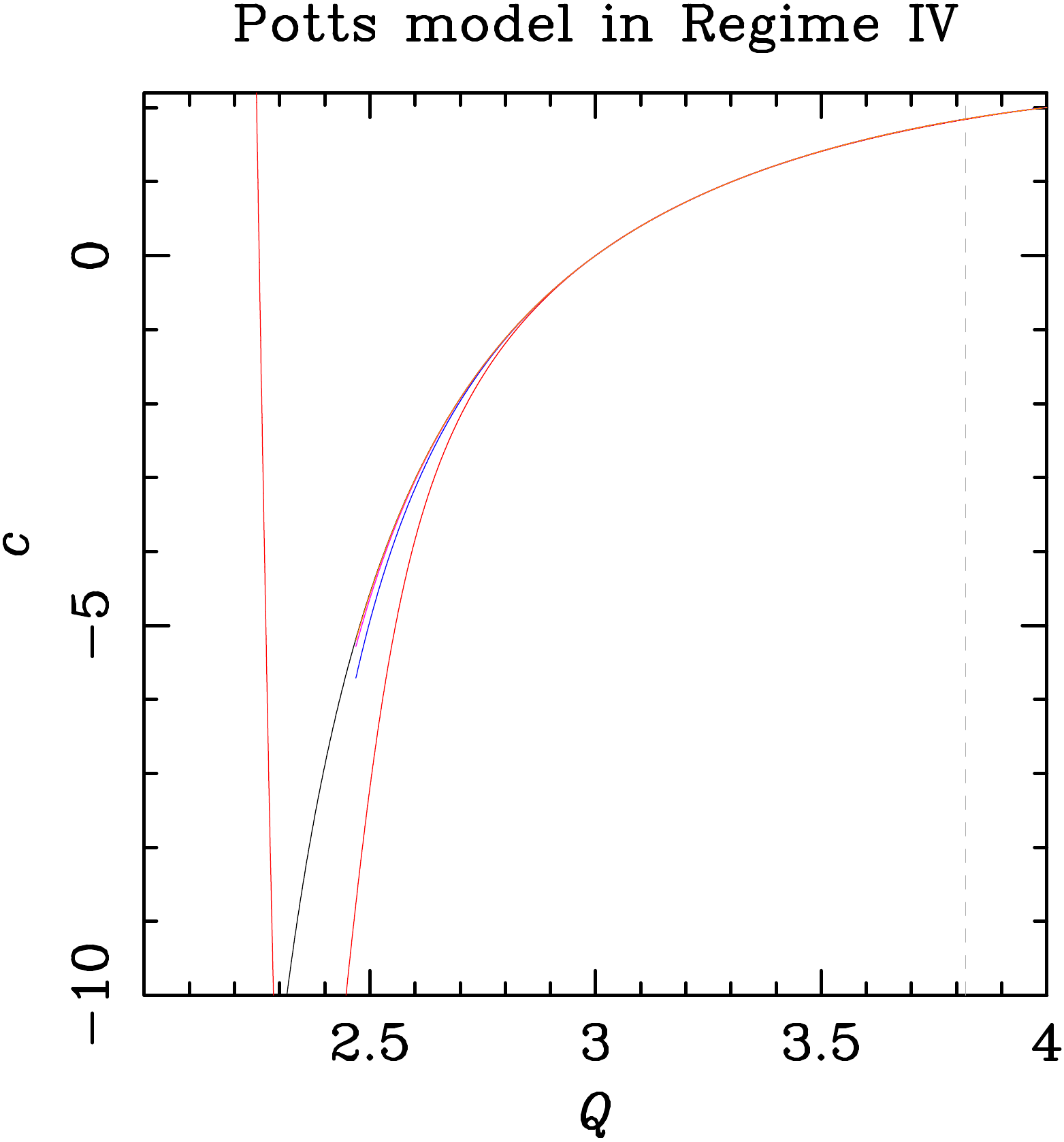} &
\includegraphics[width=200pt]{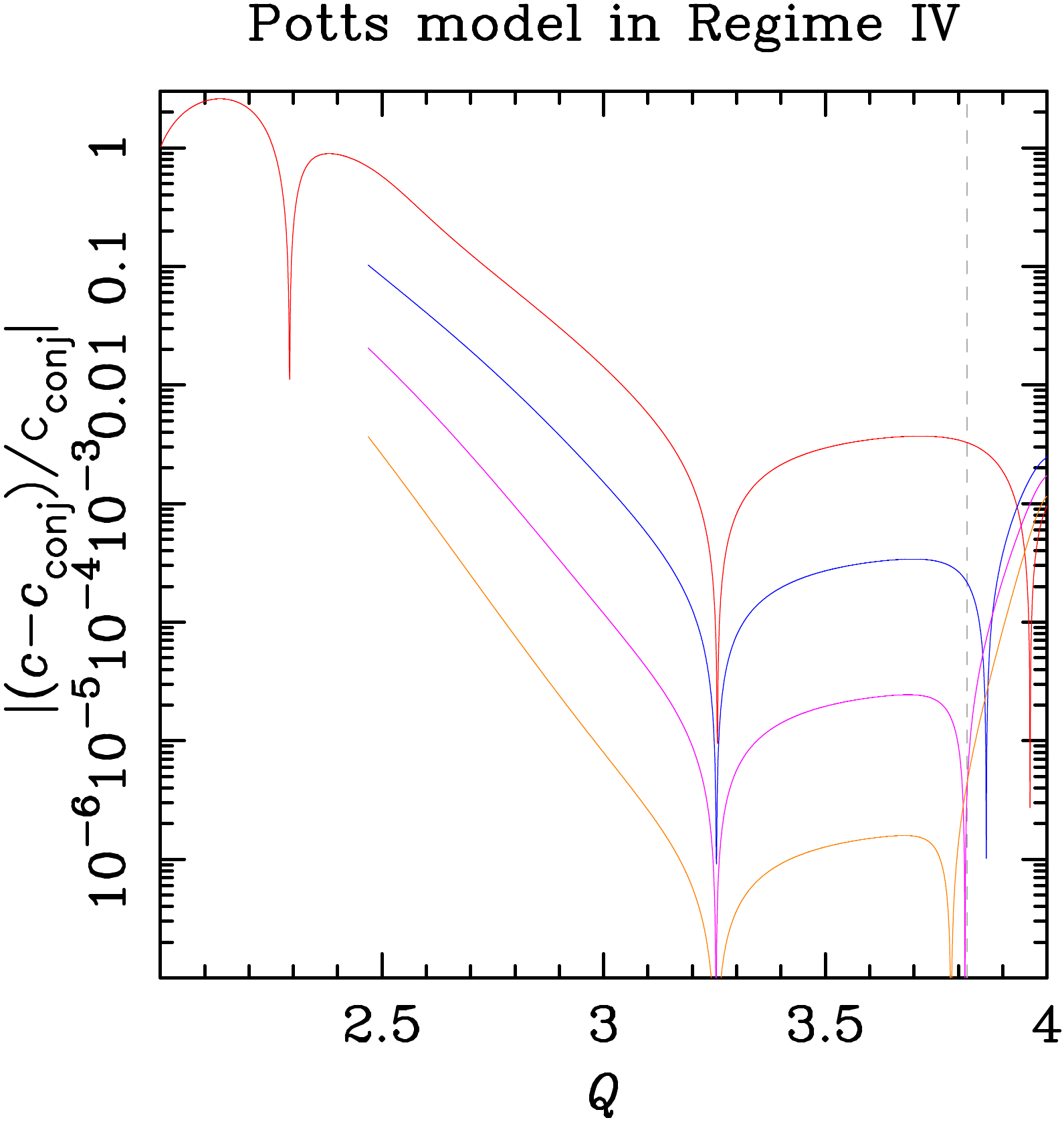} \\[2mm]
\qquad  (a) &\quad\quad\; (b) \\
\end{tabular}
\caption{(a) Values of the central charge $c$ obtained by performing the
2-parameter fit 
$f_L(Q) - f_\text{Baxter}(Q) = \frac{\pi G}{6L^2}c(Q) +  \frac{A}{L^4}$ 
for $L_\text{min}=6$ (red), $L_\text{min}=12$ (blue), 
$L_\text{min}=24$ (pink), and $L_\text{min}=48$ (orange). The black curve
shows the conjectured value (\ref{eq:BAE:RIV:cPotts}). Note
that the fits cover the interval $(2,4)$ for $L_\text{min}=6$, while they
cover the smaller interval $[2.469,4)$ for the other values.  
(b) Relative difference between the estimated 
central charge and the conjecture (\ref{eq:BAE:RIV:cPotts}) in 
semi-logarithmic scale. 
}
\label{fig:c}
\end{figure}
 
Now that the agreement of $f_\text{bulk}$ with $f_\text{Baxter}$ has been 
firmly established, we can obtain more precise estimates for the central 
charge $c$ by performing fits with only two parameters, $c(Q)$ and $A$, using 
the CFT Ansatz 
$f_L(Q) - f_\text{Baxter}(Q) = \frac{\pi G}{6L^2} c(Q) +  \frac{A}{L^4}$ and
the exact result (\ref{Baxter_g2}). 
Accordingly, we fitted our data for two consecutive values of $L$, namely 
$L=L_\text{min},2L_\text{min}$, for different values of $L_\text{min}$. 
The results are depicted in 
figure~\ref{fig:c}. The agreement between the numerical estimates for
the central charge and conjecture (\ref{eq:BAE:RIV:cPotts}) is excellent,
including for $Q < Q_0$. There is indeed good evidence that $c$
converges to (\ref{eq:BAE:RIV:cPotts}) for all $Q \in (2,4]$, 
but once again the convergence is not uniform; note in particular that 
$c \to -\infty$ as $Q \to 2^+$, whereas $f_\text{Baxter}$ stays finite
at $Q=2$.

At $Q=3$, we find that $c=0$ exactly for any $L$,
as expected from the total absence of finite-size effects noticed above.
This value $c=0$ agrees with the obvious non-critical nature of the 3-colouring
problem. On the other hand, close to $Q=4$, the observed larger discrepancies 
are again due to logarithmic corrections.

\subsubsection{Thermal exponent.} 
\label{sec:RIV:thermal}

It is now interesting to further consider the excited states in the $k=0$ 
sector, and in particular the first of these, which in the continuum limit 
corresponds to the \emph{thermal field}. Although the corresponding root 
configuration is highly unstable under variations of the twist, we can 
nevertheless describe its analytic continuation at zero twist, which is 
made of 
\begin{itemize}
\item $\frac{2L}{3}-2$ real roots. 
\item $\frac{L}{3}-2$ roots with imaginary part $\frac{\pi}{2}$.
\item Two degenerate 2-strings, with zero real part and the same imaginary part.
\end{itemize}

\medskip
\newcommand{\nOOtwistO}{
\begin{tikzpicture}[scale=0.5]
\draw [black, line width=0.1] (0,-0.8)  -- (0,2); 
\draw [black, line width=0.2] (-3,0) -- (3,0); 
\draw [black, dashed, line width=0.2] (-2.8,1.5) -- (3,1.5);
\draw[fill] (-3.2,1.6) node[scale=1]{$\frac{\pi}{2}$};
\draw[fill] (-3.3,0) node[scale=1]{$0$};
\draw[fill=white] (-2.,1.5)  circle[radius=3pt];
\draw[fill=white] (-1.3,1.5) circle[radius=3pt];
\draw[fill=white] (-0.5,1.5) circle[radius=3pt];
\draw[fill=white] (0.5,1.5) circle[radius=3pt];
\draw[fill=white] (1.3,1.5) circle[radius=3pt];
\draw[fill=white] (2.,1.5) circle[radius=3pt];
\draw [decorate,decoration={brace,amplitude=5pt},xshift=0pt,yshift=-3pt]
(2.5,-0.8) -- (-2.5,-0.8)node [black,midway,yshift=-15pt]{\footnotesize{$2L/3$}};
\draw[fill=black] (-2.7,0)  circle[radius=3pt];
\draw[fill=black] (-2.,0)  circle[radius=3pt];
\draw[fill=black] (-1.5,0)  circle[radius=3pt];
\draw[fill=black] (-1,0)  circle[radius=3pt];
\draw[fill=black] (-0.6,0)  circle[radius=3pt];
\draw[fill=black] (-0.2,0)  circle[radius=3pt];
\draw[fill=black] (2.7,0)  circle[radius=3pt];
\draw[fill=black] (2.,0)  circle[radius=3pt];
\draw[fill=black] (1.5,0)  circle[radius=3pt];
\draw[fill=black] (1,0)  circle[radius=3pt];
\draw[fill=black] (0.6,0)  circle[radius=3pt];
\draw[fill=black] (0.2,0)  circle[radius=3pt];
\draw [decorate,decoration={brace,amplitude=5pt},xshift=0pt,yshift=10pt]
(-2,1.5) -- (2,1.5)node [black,midway,yshift=15pt] {\footnotesize{$L/3$}};
\end{tikzpicture}}

\newcommand{\nOOtwistpim}{
\begin{tikzpicture}[scale=0.5]
\draw [black, line width=0.1] (-0.18,-0.8)  -- (-0.18,2); 
\draw [black, line width=0.2] (-3,0) -- (3,0); 
\draw [black, dashed, line width=0.2] (-2.8,1.5) -- (3,1.5);
\draw[fill=white] (-2.,1.5)  circle[radius=3pt];
\draw[fill=white] (-1.3,1.5) circle[radius=3pt];
\draw[fill=white] (-0.5,1.5) circle[radius=3pt];
\draw[fill=white] (0.5,1.5) circle[radius=3pt];
\draw[fill=white] (1.3,1.5) circle[radius=3pt];
\draw[fill=white] (2.5,1.5) circle[radius=3pt];
\node[above] at(2.5,1.5) {\scriptsize{$\to \infty$}};

\draw[fill=black] (-2.7,0)  circle[radius=3pt];
\draw[fill=black] (-2.,0)  circle[radius=3pt];
\draw[fill=black] (-1.5,0)  circle[radius=3pt];
\draw[fill=black] (-1,0)  circle[radius=3pt];
\draw[fill=black] (-0.6,0)  circle[radius=3pt];
\draw[fill=black] (-0.2,0)  circle[radius=3pt];
\draw[fill=black] (2.7,0)  circle[radius=3pt];
\draw[fill=black] (2.,0)  circle[radius=3pt];
\draw[fill=black] (1.5,0)  circle[radius=3pt];
\draw[fill=black] (1,0)  circle[radius=3pt];
\draw[fill=black] (0.6,0)  circle[radius=3pt];
\draw[fill=black] (0.2,0)  circle[radius=3pt];
\node[above] at(2.7,0) {\scriptsize{$\to \infty$}};
\end{tikzpicture}}

\newcommand{\nOOtwistpip}{
\begin{tikzpicture}[scale=0.5]
\draw [black, line width=0.1] (-0.2,-0.8)  -- (-0.2,2); 
\draw [black, line width=0.2] (-3,0) -- (3,0); 
\draw [black, dashed, line width=0.2] (-2.8,1.5) -- (3,1.5);
\draw[fill=white] (-1.6,1.5)  circle[radius=3pt];
\draw[fill=white] (-0.9,1.5) circle[radius=3pt];
\draw[fill=white] (-0.2,1.5) circle[radius=3pt];
\draw[fill=white] (0.5,1.5) circle[radius=3pt];
\draw[fill=white] (1.2,1.5) circle[radius=3pt];
\draw[fill=black] (-2.2,0)  circle[radius=3pt];
\draw[fill=black] (-1.8,0)  circle[radius=3pt];
\draw[fill=black] (-1.4,0)  circle[radius=3pt];
\draw[fill=black] (-1,0)  circle[radius=3pt];
\draw[fill=black] (-0.6,0)  circle[radius=3pt];
\draw[fill=black] (-0.2,0)  circle[radius=3pt];
\draw[fill=black] (0.2,0)  circle[radius=3pt];
\draw[fill=black] (0.6,0)  circle[radius=3pt];
\draw[fill=black] (1,0)  circle[radius=3pt];
\draw[fill=black] (1.4,0)  circle[radius=3pt];
\draw[fill=black] (1.8,0)  circle[radius=3pt];
\draw[fill=black] (-0.3,0.5)  rectangle +(0.2,0.2);
\draw[fill=black] (-0.3,-0.7) rectangle +(0.2,0.2);
\end{tikzpicture}}

\newcommand{\nOOtwisttwopim}{
\begin{tikzpicture}[scale=0.5]
\draw [black, line width=0.1] (-0.5,-0.8)  -- (-0.5,2); 
\draw [black, line width=0.2] (-3,0) -- (3,0); 
\draw [black, dashed, line width=0.2] (-2.8,1.5) -- (3,1.5);
\draw[fill=white] (-1.5,1.5)  circle[radius=3pt];
\draw[fill=white] (-0.8,1.5) circle[radius=3pt];
\draw[fill=white] (-0.1,1.5) circle[radius=3pt];
\draw[fill=white] (0.4,1.5) circle[radius=3pt];
\draw[fill=white] (2.3,1.5) circle[radius=3pt];
\node[above] at(2.3,1.5) {\scriptsize{$\to \infty$}};
\node[above] at(2.5,0) {\scriptsize{$\to \infty$}};
\draw[fill=black] (-2.4,0)  circle[radius=3pt];
\draw[fill=black] (-2.0,0)  circle[radius=3pt];
\draw[fill=black] (-1.6,0)  circle[radius=3pt];
\draw[fill=black] (-1.2,0)  circle[radius=3pt];
\draw[fill=black] (-0.8,0)  circle[radius=3pt];
\draw[fill=black] (-0.4,0)  circle[radius=3pt];
\draw[fill=black] (0.0,0)  circle[radius=3pt];
\draw[fill=black] (0.4,0)  circle[radius=3pt];
\draw[fill=black] (0.8,0)  circle[radius=3pt];
\draw[fill=black] (1.2,0)  circle[radius=3pt];
\draw[fill=black] (2.55,0)  circle[radius=3pt];
\draw[fill=black] (-0.6,0.5)  rectangle +(0.2,0.2);
\draw[fill=black] (-0.6,-0.7)  rectangle +(0.2,0.2);
\end{tikzpicture}}

\newcommand{\nOOtwisttwopip}{
\begin{tikzpicture}[scale=0.5]
\draw [black, line width=0.1] (-0.5,-0.8)  -- (-0.5,2); 
\draw [black, line width=0.2] (-3,0) -- (2,0); 
\draw [black, dashed, line width=0.2] (-2.8,1.5) -- (2,1.5);
\draw[fill=white] (-1.7,1.5)  circle[radius=3pt];
\draw[fill=white] (-0.9,1.5) circle[radius=3pt];
\draw[fill=white] (-0.1,1.5) circle[radius=3pt];
\draw[fill=white] (0.7,1.5) circle[radius=3pt];
\draw[fill=black] (-2.3,0)  circle[radius=3pt];
\draw[fill=black] (-1.9,0)  circle[radius=3pt];
\draw[fill=black] (-1.5,0)  circle[radius=3pt];
\draw[fill=black] (-1.1,0)  circle[radius=3pt];
\draw[fill=black] (-0.7,0)  circle[radius=3pt];
\draw[fill=black] (-0.3,0)  circle[radius=3pt];
\draw[fill=black] (0.1,0)  circle[radius=3pt];
\draw[fill=black] (0.5,0)  circle[radius=3pt];
\draw[fill=black] (0.9,0)  circle[radius=3pt];
\draw[fill=black] (1.3,0)  circle[radius=3pt];
\draw[fill=black] (-0.6,0.5)  rectangle +(0.2,0.2);
\draw[fill=black] (-0.6,-0.7)  rectangle +(0.2,0.2);
\draw[fill=black] (-0.6,0.20)  rectangle +(0.2,0.2);
\draw[fill=black] (-0.6,-0.40)  rectangle +(0.2,0.2);
\end{tikzpicture}}

\begin{figure}
\centering
\begin{tikzpicture}[scale=0.85]
\node at (0,0)  {\nOOtwistO};
\node at (4.3,0.16){\nOOtwistpim};
\node at (8,0.06)  {\nOOtwistpip};
\node at (12,0.15) {\nOOtwisttwopim};
\node at (16,0.07) {\nOOtwisttwopip};
\draw[line width=1pt,->,>=latex] (-1,-2.6) -- (17,-2.6) node[right] {$\varphi$};
\draw  (0,-2.6)--(0,-2.4)  node[above] {$0$};
\draw  (8,-2.6)--(8,-2.4)  node[above] {$\pi$};
\draw  (16,-2.6)--(16,-2.4) node[above] {$2\pi$};
\end{tikzpicture}
\caption{Configurations of the Bethe roots associated with the ground state 
in regime~IV, as a function of the twist parameter. For $\varphi < \pi$, it 
is described by a Fermi sea of $\frac{2 L}{3}$ real roots (black dots) and 
a Fermi sea of $\frac{L}{3}$ roots with imaginary part $\frac{\pi}{2}$ 
(white dots). As $\varphi \to \pi^-$, the real parts of the rightmost roots 
of each Fermi sea go to $+\infty$. As $\varphi$ crosses $\pi$, these two roots 
form one 2-string (black squares). The same process occurs once again as 
$\varphi \to 2\pi^-$, resulting in a configuration with two degenerate 
2-strings at $\varphi = 2\pi$. Although on the picture these are distinct, it 
should be kept in mind that the corresponding roots actually coincide 
two by two.
}
\label{fig:Betherootstwist}
\end{figure}
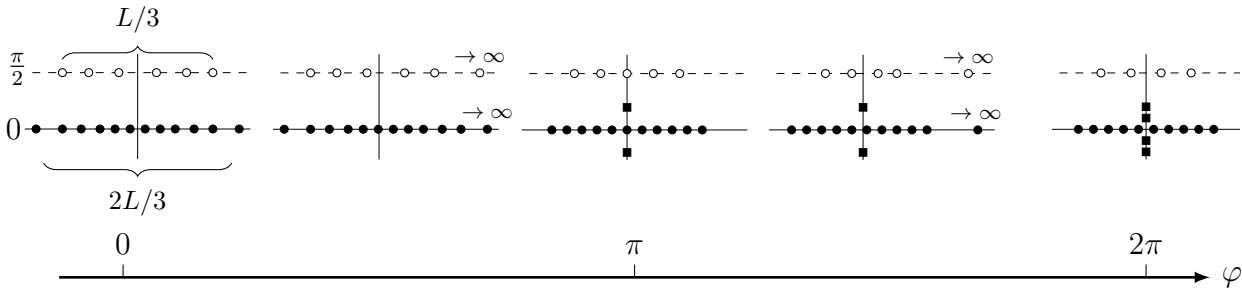

Let us now go back to the ground state configuration 
(\ref{eq:quantumnumbersgroundstate}), and try to follow it continuously as 
the twist is increased. The evolution of the Bethe roots 
from $\varphi=0$ to $\varphi=2\pi$ is depicted schematically in 
figure~\ref{fig:Betherootstwist}. It is seen that the root 
configurations encounters qualitative changes, namely, the formation of 
2-strings, as $\varphi$ crosses multiples of $\pi$. More precisely, the root 
configuration associated with the thermal exponent at $\varphi=0$ actually 
corresponds to the continuation of the ground state configuration to 
$\varphi = 2\pi$. The corresponding conformal weight can therefore be 
obtained straightforwardly by analytic continuation of the conformal weight 
for the twisted ground state (this is after all exactly the same as what was 
observed in regime~I, see section~\ref{sec:regimeI})~\footnote{
More generally, we expect that all root configurations with 2-strings can 
be understood as the continuation of `standard' root configurations to 
larger values of the twist parameter. This explains a posteriori our comment 
at the end of section~\ref{sec:RegimeIV}, and we shall not comment anymore 
on the presence of such configurations in the other Potts sectors. 
}. 

\begin{figure}
\centering
\begin{tabular}{cc}
\includegraphics[width=200pt]{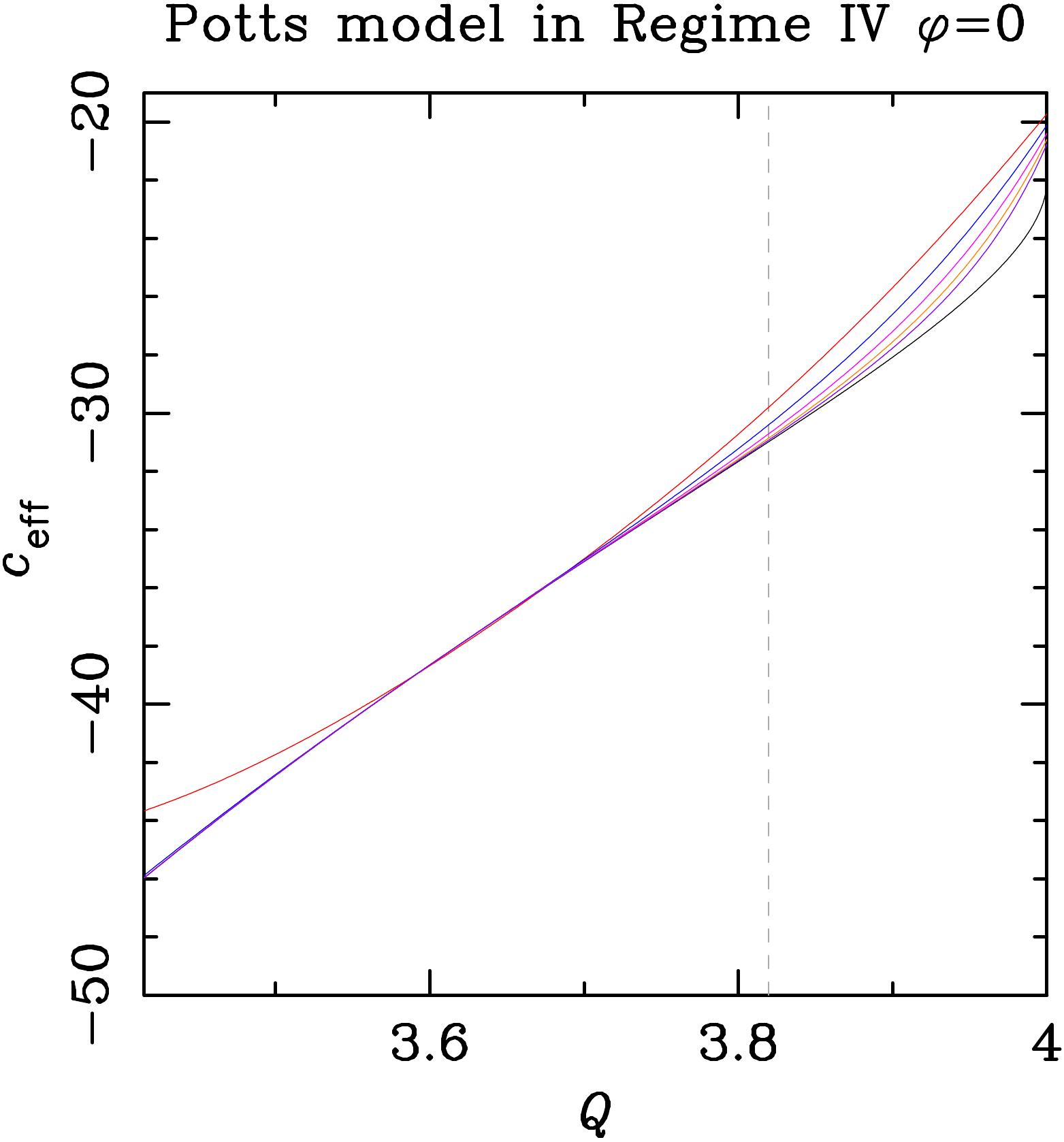} &
\includegraphics[width=200pt]{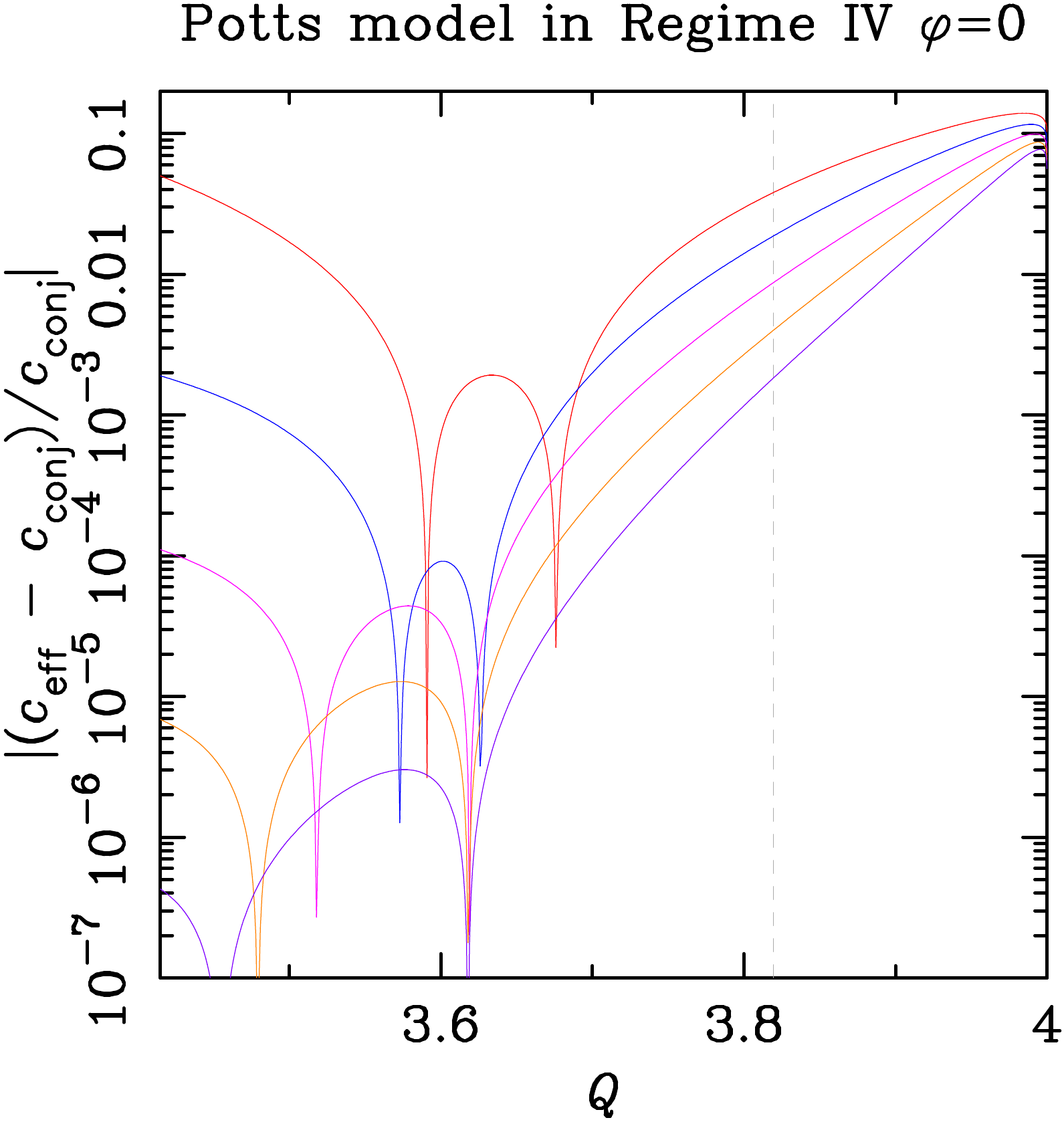}\\[2mm]
\qquad  (a) &\quad\quad\; (b) \\
\end{tabular}
\caption{Fits for the subdominant free energy in regime~IV at zero twist.
(a) Values of the effective central charge $c_\text{eff}$ 
obtained by performing the 2-parameter fit 
$f_{2,L}(Q) - f_\text{Baxter}(Q) = 
\frac{\pi G}{6L^2}c_\text{eff}(Q)+\frac{A}{L^4}$ 
for $L_\text{min}=12$ (red), $L_\text{min}=24$ (blue), 
$L_\text{min}=48$ (pink), $L_\text{min}=96$ (orange), and $L=192$ (violet). 
The black curve shows the conjectured value (\ref{eq:BAE:RIV:cPottsT}).  
(b) Relative difference
between the estimated central charge and the conjecture 
(\ref{eq:BAE:RIV:cPottsT}) in semi-logarithmic scale. 
}
\label{fig:ceff}
\end{figure}

To check this assumption, we have solved numerically the zero-twist BAE 
associated with the {\em first excited state} continued to $\varphi=0$ 
(remember that the true level present in the Potts model, namely at a twist 
$\varphi = \theta$, corresponds to a highly unstable configuration of roots, 
whose numerical study does not appear feasible) for sizes up to $L=384$ 
in the interval $Q\in (B_8,4)$. 
For each value of $Q$, we computed the corresponding free energy $f_{2,L}(Q)$ 
for $L=12,24,48,96,192$, and $384$. We then performed the standard 
three-parameter fit $f_{2,L}(Q) = f_\text{bulk}(Q) + 
\frac{\pi G}{6L^2} c_\text{eff}(Q) + \frac{A}{L^4}$ for three  
consecutive values of the width $L=L_\text{min}, 2L_\text{min}, 4L_\text{min}$.
As for the leading eigenvalue, the values obtained for $f_\text{bulk}$ 
agree very well with those given by Baxter's exact expression for regime~IV 
(\ref{Baxter_g2}). A better estimate for $c_\text{eff}$ is 
obtained by using the 2-parameter CFT Ansatz
$f_{2,L}(Q) - f_\text{Baxter}(Q) = 
\frac{\pi G}{6L^2} c_\text{eff}(Q) +  \frac{A}{L^4}$, and two  
consecutive values of $L=L_\text{min},2L_\text{min}$. 
The results are depicted in figure~\ref{fig:ceff}, and agree with the 
conjecture 
\begin{equation}
c_\text{eff} \;=\; -22 - \frac{96}{p-4} \,,
\label{eq:BAE:RIV:cPottsT}
\end{equation}
that is, precisely the expression obtained from (\ref{eq:BAE:RIV:ctwisted}) 
at $\delta m_1 = \delta m_2 = w_1 = w_2 = \delta_1^\pm = \delta_2^\pm = 0$ 
for a twist $\varphi = 2 \pi$. 
The agreement between the numerical estimates for
the effective central charge and conjecture (\ref{eq:BAE:RIV:cPottsT}) 
is excellent, especially for values of $Q$ far away from $4$. Close to $Q=4$,
we should expect 
logarithmic corrections as those shown in figure~\ref{fig:c}, implying 
a slower convergence towards (\ref{eq:BAE:RIV:cPottsT}).  

From there we assume, as in regime I, that the subleading eigenlevels in the 
$k=0$ sector are obtained from the ground state level by taking twists of the 
form $\varphi = \theta + 2 \pi j$, leading to effective central charges 
\begin{equation}
c_{\text{eff},j} \;=\; 2 - \frac{24}{p(p-4)}\left(1+jp\right)^2 \,, 
\end{equation}
and therefore to conformal weights of the form 
\begin{equation}
h_j + \bar{h}_j \;=\; \frac{1}{12}\left(c- c_{\text{eff},j}\right)  
                \;=\; \frac{2j}{p-4}\left(2+jp\right) \,.
\label{eq:RIV:thermalj}
\end{equation}
Among these, the two least irrelevant correspond to $j=1$ and $j=-1$, namely 
\begin{equation}
x_T  \;=\; h_{-1}+\bar{h}_{-1} \;=\; 2 + \frac{4}{p-4} \,, \qquad 
x'_T \;=\; h_{1}+\bar{h}_{1}   \;=\; 2 + \frac{12}{p-4} \,.
\label{eq:RIV:thermexps}
\end{equation}

\subsubsection{Magnetic exponent.}
\label{sec:RIV:magnetic}

As explained in section~\ref{sec:6v}, the magnetic exponent is associated to 
the scaling of the leading TM eigenvalue in sector $k=0$, with twist 
$\frac{\varphi}{2}=\frac{\pi}{2}$. From our observations in 
section~\ref{sec:k1spectrum}, we know that this eigenvalue can be associated 
with the quantum numbers 
\begin{equation}
(\delta m_1, \delta m_2)\;=\; (0,0)\,, \quad 
(w_1,w_2) \;=\; \left(-\frac{1}{2},\frac{1}{2}\right)\,, \quad 
\varphi\;=\;0 \,,
\end{equation}
and therefore with an effective central charge
\begin{equation}
c_\text{eff} = 0 \,.
\end{equation}
From there, we directly read off the magnetic exponent, 
\begin{equation}
x_H \;=\; \frac{1}{12} \left( c- c_\text{eff}\right) 
    \;=\; \frac{1}{6} - \frac{2}{p(p-4)} \,.
\label{eq:RIV:XH}
\end{equation}

\subsection{Even $k$ sectors.}
\label{sec:kevenspectrum}
\label{sec:RIV:even}

For instance we study the sector $k=2$, for $Q=3.99$, and $L=9$. 
The eigenenergies given are of the form 
$f_i = \frac{1}{L}\log \Lambda_i$, where $\Lambda_i$ are the TM eigenvalues: 
$$
\begin{array}{cccl}
f_0 &=& 0.356989 &\quad (\delta m_1, \delta m_2)=(1,1)\,, 
        (w_1,w_2)=(0,0)\,,  \varphi=0\\
f_1 = f_2 &=& 0.343760 & \quad (\delta m_1, \delta m_2)=(1,1)\,, 
        (w_1,w_2)=(\pm 1,0)\,, \varphi=\mp\pi \\
f_3 = f_4 &=& 0.321961 & \quad (\delta m_1, \delta m_2)=(1,1)\,, 
        (w_1,w_2)=(\pm 1,0)\,, \varphi=0 \\
f_5 = f_6 &=& 0.316346 & \quad (\delta m_1, \delta m_2)=(2,2)\,     
               \text{+ one 2-string} \\
f_7 = f_8 &=& 0.314850 & \quad (\delta m_1, \delta m_2)=(1,1)\,, 
        (w_1,w_2)=(0,0)\,, \delta_1^\pm= 1\,, \varphi= 0\\
f_9 = f_{10} &=& 0.309532 & \quad (\delta m_1, \delta m_2)=(1,1)\,, 
        (w_1,w_2)=(\pm 1,\pm 1)\,, \varphi=0\\
f_{11} = f_{12} &=& 0.304716 & \quad (\delta m_1, \delta m_2)=(1,1)\,, 
        (w_1,w_2)=(\pm 2,0)\,, \varphi=  \mp 2 \pi\\
f_{13} = f_{14} &=& 0.302223 & \quad (\delta m_1, \delta m_2)=(1,1)\,, 
        (w_1,w_2)=(\pm 1,0)\,, \delta_1^\mp= 1\,, \varphi=  0\\
&\ldots& & \\
\end{array}  
$$
(when no mention is made of the $\delta_1, \delta_2$ numbers, these are 
equal to zero).

From this example as well as further examination of the $k=4,6,8$ cases, we 
see that the 
least excited states making up the even $k\geq 2$ sectors are associated with 
\begin{equation}
\delta m_1 \;=\; \delta m_2 \;=\; \frac{k}{2}  \,, 
\quad w_1 , w_2 \in \mathbb{Z}  \,, \quad \varphi \;=\; { n \pi} \,, \quad  
n \in \mathbb{Z} \,.
\end{equation}

From (\ref{eq:BAE:RIV:hhbar})/(\ref{eq:hhbarmomentum}), we find the 
corresponding conformal spins 
\begin{equation}
h - \bar{h} = -\frac{1}{2}k \left(w_1 + w_2\right)\,,
\end{equation}
which indeed assume integer values, while the conformal weights are found to be 
\begin{eqnarray}
h &=&  \frac{\left[ k(p-4)- (2n+2(w_1+w_2))p \right]^2-16}{16 p(p-4)}
 + 
\frac{1}{12}\left(3 k n + \left(w_2-w_1\right)^2\right)
\label{eq:hevenk}
\\
\bar{h} &=& \frac{\left[ k(p-4)- (2n+2(w_1+w_2))p \right]^2-16}{16 p(p-4)}
 \nonumber \\ & &  \qquad \qquad + 
\frac{1}{12}\left(3 k (n+2w_1+2w_2) + \left(w_2-w_1\right)^2\right) \,.
\label{eq:hbarevenk}
\end{eqnarray}

\subsection{Odd $k\geq 3$ sectors.}
 \label{sec:koddspectrum}
 \label{sec:RIV:odd}
 
We here specialize to odd sectors with $k \geq 3$, as the $k=1$ case is a bit 
particular, as we shall see in the next section. 
Similarly, for $k=3$, $Q=3.99$, and $L=9$, the lowest-lying levels are the 
following: 
$$
\begin{array}{cccl}
f_1 =f_2 &=&  0.316285 & \quad (\delta m_1, \delta m_2)=(2,1)\,, 
              (w_1,w_2)=(\pm\frac{1}{2},0)\,, \varphi=0\\
f_3 = f_4 &=& 0.313692 & \quad (\delta m_1, \delta m_2)=(2,1)\,, 
              (w_1,w_2)=(\pm\frac{1}{2},0)\,, \varphi=\mp\frac{2 \pi}{3}\\
f_5 = f_6 &=& 0.290629 & \quad (\delta m_1, \delta m_2)=(2,1)\,, 
               (w_1,w_2)=(\pm \frac{1}{2},0)\,, \varphi=0\,, 
               \delta_1^\mp=1\\
f_7 = f_8 &=& 0.288732 & \quad (\delta m_1, \delta m_2)=(2,1)\,, 
              (w_1,w_2)=(\pm \frac{1}{2},\mp 1)\,, \varphi= \pm\frac{4 \pi}{3}\\
&\ldots& & \\
\end{array}  
$$
For $k=5$, $Q=3.99$, and $L=9$, we find
$$
\begin{array}{cccl}
f_1 =f_2 &=&  0.239272 & \quad (\delta m_1, \delta m_2)=(3,2)\,, 
              (w_1,w_2)=(0,\pm\frac{1}{2})\,, \varphi=\mp\frac{4\pi}{5}\\
f_3 = f_4 &=& 0.229800 & \quad (\delta m_1, \delta m_2)=(3,2)\,, 
              (w_1,w_2)=(0,\pm\frac{1}{2})\,, \varphi=\mp\frac{2 \pi}{5}\\
f_5 = f_6 &=& 0.224099 & \quad (\delta m_1, \delta m_2)=(3,2)\,, 
               (w_1,w_2)=(\mp 1,\pm\frac{1}{2})\,, \varphi=\mp\frac{2\pi}{5}\\
f_7 = f_8 &=& 0.222247, & \quad (\delta m_1, \delta m_2)=(3,2)\,, 
              (w_1,w_2)=(0,\pm\frac{1}{2}) \,, \varphi= \mp\frac{4 \pi}{5}\,, 
               \delta_1^\pm=1\\
f_9 = f_{10} &=&0.220503 & \quad (\delta m_1, \delta m_2)=(3,2)\,, 
              (w_1,w_2)=(\pm 1,\pm\frac{1}{2})\,, \varphi= \mp\frac{2 \pi}{5}\,, 
               \delta_1^\mp=1\\
&\ldots& & \\
\end{array}  
$$
In the latter case states corresponding to $\varphi=0$, for instance 
$(\delta m_1, \delta m_2)=(3,2)\,, (w_1,w_2)=(0,\pm\frac{1}{2})\,, \varphi=0$ 
are also found, but lie deeper in the Potts spectrum. 

From these examples, as well as further examination of the $k=7$ and $k=9$ 
cases, we see that 
the least excited states making up the odd $k>1$ sectors are associated with 
\begin{equation}
\delta m_1  = \frac{k+1}{2} \,, \quad \delta m_2 = \frac{k-1}{2} \,, 
\quad w_1 = v_1+\frac{k-1}{4} \,, \quad  w_2= v_2 + \frac{k+1}{4}  \,, 
\quad \varphi = \frac{2 n \pi}{k} \,,
\end{equation}
where $v_1$, $v_2$ and $n$ take integer values.

From (\ref{eq:BAE:RIV:hhbar})/(\ref{eq:hhbarmomentum}), we find the 
corresponding conformal spins 
\begin{equation}
h - \bar{h} = \frac{1}{4}\left(1-k^2 + 2(v_2-v_1)-2k(v_1+v_2)\right) \,,
\end{equation}
which indeed assumes integer values, while the conformal weights are found 
to be 
\begin{eqnarray}
h &=&  \frac{\left[ k(p-4)- (k+4\frac{n}{k}+2(v_1+v_2))p \right]^2-16}
            {16 p(p-4)}
\nonumber  \\ & & + 
\frac{1}{12}\left(6 n + \left(v_2-v_1+2\right)^2\right)
\label{eq:hoddk} \\
\bar{h} &=& \frac{\left[ k(p-4)- (k+4\frac{n}{k}+2(v_1+v_2))p \right]^2-16}
                 {16 p(p-4)}
\nonumber  \\ & & \qquad \qquad + 
\frac{1}{12}\left( 3 k^2 + 6 (n+k v_1+k v_2) + \left(v_2-v_1-1\right)^2\right) 
\label{eq:hbaroddk}
\end{eqnarray}

\subsection{$k = 1$ sectors.}

\label{sec:k1spectrum}

As we mentioned in section~\ref{sec:6v}, the $k=1$ case is a bit particular, 
in that there are two 
ways to have one propagating cluster in the vertical direction, namely it may 
or may not be allowed to wrap around the cylinder. 
Let us study these two cases independently: 

\subsubsection{Cluster allowed to wrap around the cylinder.}

Forbidding loops on the kagome lattice to wrap around the cylinder amounts 
to giving such loops a weight $0$, and therefore to fixing a twist 
$\frac{\varphi}{2} = \frac{\pi}{2}$. 

Indeed, the largest eigenvalue of the FK TM in this sector is 
found in the zero-magnetisation sector of the vertex model at twist 
$\varphi = \pi$. Alternatively, it can be obtained at zero twist from the 
following set of Bethe parameters 
\begin{equation}
(\delta m_1, \delta m_2)\;=\; (0,0)\,, \quad 
(w_1,w_2) \;=\; \left(-\frac{1}{2},\frac{1}{2}\right)\,, \quad 
\varphi\;=\;0 \,. 
\end{equation}

\subsubsection{Cluster not allowed to wrap around the cylinder.}

Forbidding the propagating cluster to wrap around the cylinder amounts to 
fixing one through-line in the O$(n)$ model. Therefore we expect to find the 
corresponding leading eigenvalues in the sector of magnetisation $m=1$. 

This is the case as, for instance, we find for $Q=3.99$ and $L=6$ the 
following set of leading eigenenergies: The first dominant states form
complex quartet
$$
\begin{array}{ccc}
f_1  &=&  0.302084 + 0.355356\, \mathrm{i}\\
f_2  &=&  0.302084 - 0.355356\, \mathrm{i}\\
f_3  &=&  0.302084 + 0.342775\, \mathrm{i}\\
f_4  &=&  0.302084 - 0.342775\, \mathrm{i}
\end{array}
$$
and the next states are given by
$$
\begin{array}{cccl}
f_5 =f_6 &=& 0.272659  & \quad (\delta m_1, \delta m_2)=(2,-1)\,, 
             (w_1,w_2)=\left(\pm\frac{1}{2},0\right)\,, \varphi=0 \\
f_7 = f_8 &=& 0.269606 & \quad (\delta m_1, \delta m_2)=(2,1) 
              \text{+ one 2-string}\,, \varphi=0 \\
f_9  &=& 0.227715 & \quad (\delta m_1, \delta m_2)=(2,-1)\,, 
             (w_1,w_2)=(0,0)\,, \varphi=0, \\
     & &           & \quad \qquad \delta_1^+ = \delta_1^-=\frac{1}{2}\\
&\ldots& & 
\end{array}  
$$

The complex quartet, as we understand it, is associated to very peculiar 
root configurations. From observations at $L=6$ and $L=9$, we see that the 
corresponding root configurations are made of:  
\begin{itemize}
 \item $\frac{2L}{3}-1$ roots with imaginary parts very close to $0$.
 \item $\frac{L}{3}-1$ roots with imaginary part very close to $\frac{\pi}{2}$.
 \item One root with imaginary part very close to $\frac{\pi}{4}$.
\end{itemize}
Such configurations are rather odd, and out of the scope of our analytical
solution. In addition, we were unfortunately unable to solve the 
corresponding BAE numerically for large system sizes, and therefore 
we could not access the corresponding conformal weights.

\subsection{Identification of the CFT}

The central charge (\ref{eq:BAE:RIV:cPotts}) is the analytic continuation to 
any real $p> 4$ of that of the $S_3$-invariant 
minimal models studied by Fateev and Zamolodchikov \cite{FZ1,FZ2}, which 
can alternatively be formulated as the conformal cosets 
\begin{equation}
\frac{SU(2)_4 \times SU(2)_{p-6}}{SU(2)_{p-2}} \,,
\end{equation}
defined for integer values of $p \geq 6$. 

The corresponding sets of conformal weights are known \cite{FZ1,FZ2} to be 
parametrised by two integers, $1\leq r \leq p-5$, 
$1\leq s \leq p-2$, and read 
\begin{equation}
h_{r,s} \;=\; \frac{\left[r p - s(p-4)\right]^2-16}{16 p (p-4)} + 
     \frac{1 - \cos^4\left( \pi\frac{r-s}{4}\right)}{12} \,.
     \label{eq:hrs}
\end{equation}
In addition there is another series of primary fields, with conformal weights 
\begin{equation}
h'_{r,s} \;=\; h_{r,s}+ 
\frac{1 + \sin^2\left( \pi\frac{r-s}{4}\right)}{3}
+ \delta_{r,1} \delta_{s,1} + \delta_{r,1} \delta_{s,2} + 
2\delta_{r,2}\delta_{s,1} \,.
 \label{eq:h'rs}
\end{equation}
Both (\ref{eq:hrs}) and (\ref{eq:h'rs}) can be written under the form 
$h_{r,s} (h'_{r,s}) = \frac{\left[r p - s(p-4)\right]^2-16}{16 p (p-4)} + 
\delta h_{r,s}$, where $\delta h_{r,s}$ encodes the sector of the associated 
$S_3$ symmetry \cite{FZ2}, and can be understood in the framework of a 
generalized CG construction as a boundary twist term 
\cite{diFrancescoSaleurZuber1988} (this is analogous to the Neveu--Schwartz 
and Ramond sectors in supersymmetric CFTs, encoding the periodic or 
antiperiodic boundary conditions for the fermionic 
field \cite{FriedanQiuShenkerSUSY}). 

Let us now compare the conformal spectrum identified in regime IV with the 
exponents (\ref{eq:hrs})/(\ref{eq:h'rs}).

\subsubsection{Thermal exponents.}

In section~\ref{sec:RIV:thermal}, we have identified a series of subleading 
exponents, $h_j + \bar{h}_j$, given by (\ref{eq:RIV:thermalj}). 
These exponents have zero momentum (in agreement with the fact that thermal 
fields should not break the translational symmetry), and therefore
\begin{equation}
h_{j} \;=\; \bar{h}_j \;=\; \frac{j}{p-4}\left(2+jp\right) \,.
\end{equation}
These are readily identified as the dimensions $h_{r,s}$ in (\ref{eq:hrs}), 
with $r=4j+1$, and $s=1$.

\subsubsection{Magnetic exponent.}

In section \ref{sec:RIV:magnetic}, we have identified the magnetic exponent as 
(\ref{eq:RIV:XH}). 
The latter cannot be written in terms of the table $h_{r,s}$ ($h'_{r,s}$) 
with fixed, $p$-independent values of $r,s$. Instead, we find that it 
coincides with $x_H = h_{r,s} + \bar{h}_{r,s} = 2 h_{r,s}$, with
\begin{equation}
r \;=\; \frac{p-4}{2} \,, \quad s \;=\; \frac{p}{2} \,. 
\end{equation} 
The fact that these indices are $p$-dependent and half-integer 
is not too surprising. Indeed, the magnetic exponent
for the usual critical $Q$-state Potts model has similar features, since it 
can be written $h_{\frac{p-1}{2},\frac{p+1}{2}}$, where $p$
here denotes the minimal model index.%
\footnote{This result can however also be written as $h_{\frac{1}{2},0}$, 
with $p$-independent indices.}

\subsubsection{Even $k$ sectors.} 

In section \ref{sec:RIV:even}, we have identified the leading conformal 
weights in the sectors with even $k\geq 2$ as given by 
eqs.~(\ref{eq:hevenk})/(\ref{eq:hbarevenk}).
Note that the expressions for $h$ and $\bar{h}$ can be recovered from each 
other by setting $(w'_1,w'_2) = (-w_1,-w_2), n'=n+2w_1+2w_2$, so we only 
focus on the former, namely (\ref{eq:hevenk}). The first term takes the form 
$\frac{\left[r p - s(p-4)\right]^2-16}{16 p (p-4)}$, with $r= 2n+2(w_1+w_2)$, 
and $s=k$. Focusing on the most relevant state, namely that with 
$n=w_1=w_2 = 0$, the second term vanishes, and we recover for 
$k\equiv0 \pmod 4$ 
an exact correspondence with the exponents (\ref{eq:hrs}), 
for which the term $\delta h_{r,s}$ is zero.  
This state has zero conformal spin and scaling dimension 
\begin{equation}
x_k \;=\; h + \bar{h} \;=\; \frac{k^2 (p-4)}{8p} - \frac{2}{p(p-4)} \,. 
\end{equation}

\subsubsection{Odd $k$ sectors.} 

In section \ref{sec:RIV:odd}, we have identified the leading conformal weights 
in the sectors with even $k\geq 2$ as given by 
eqs.~(\ref{eq:hoddk})/(\ref{eq:hbaroddk}).
As in the $k$ even case, the expressions for $h$ and $\bar{h}$ can be recovered
from each other by a relabeling of the indices, namely 
$(w'_1,w'_2) = (-w_1,-w_2), n'=n+k(w_1+w_2)$, so once again we only focus 
on the former, namely (\ref{eq:hoddk}).
The first term takes the form 
$\frac{\left[r p - s(p-4)\right]^2-16}{16 p (p-4)}$, with 
$r= k+\frac{4n}{k}+2(v_1+v_2)=\frac{4n}{k}+2(w_1+w_2)$, and $s=k$. 
In particular, we notice that fractional values of $r$ are allowed in this 
case. It is also noticeable that contrarily to the case of $k$ even, no 
well-definite set of values of the parameters $v_1,v_2,n$ seems to give rive 
to a least irrelevant state.

\subsection{Physical conclusions}

The $Q$-colouring problem of the triangular lattice $G$ is defined 
{\em sensu stricto} only for $Q \in \mathbb{N}$.
We have studied here its analytic continuation, in the form of the 
chromatic polynomial $\chi_G(Q)$,
defined for $Q \in \mathbb{R}$. Within the critical range $0 \le Q \le 4$, 
two distinct regimes (that we have called regimes~I and~IV) were identified 
\cite{Baxter86}, and we have shown that the ground state and
scaling levels of either regime are well-defined regardless of considerations 
\cite{Baxter87} about which ground state is dominant in absolute terms. 
In particular, regime~I can be defined for $0 \le Q < 4$,
and regime~IV makes sense for (at least) $2 < Q \le 4$.

We now substantiate our claims---first made in section~\ref{sec:pd_tri}---that 
regime~IV is in fact the correct analytic continuation of the colouring 
problem. To this end, we show that this regime has the correct
physical properties of a colouring problem for $Q=4$ and $Q=3$,
 and that its break-down in the limit $Q \to 2^+$ can again
be motivated by properties of the colouring problem.

The $Q=4$ vertex colourings of the triangular lattice are equivalent%
\footnote{Two models are here said to be equivalent if their set of states 
are in bijection---up to an appropriate choice of boundary conditions, such as 
choosing an appropriate (e.g.\/ even) size of the lattice in any periodic
direction and/or fixing the colour of a degree of freedom in one of the 
models---and if the corresponding Boltzmann weights are equal, up to any 
overall factor. Also, the equivalences mentioned require the lattices
to be planar, and the situation on the torus might involve further 
subtleties (see \cite[and references therein]{MoharSalas10}).}
to vertex 3-colourings of the kagome lattice and to edge 3-colourings of the 
honeycomb lattice \cite{Baxter70,MooreNewman00}. The
latter problem has been studied in its own right \cite{Baxter70} and, 
more generally, as the special case
$n=2$ of the O($n$)-type fully-packed loop (FPL) model on the hexagonal 
lattice \cite{BatchelorSuzukiYung94,KondevGierNienhuis96}. The ground state 
free energy of these problems agree 
\cite{Baxter86,Baxter70,BatchelorSuzukiYung94} and is given by 
(\ref{Baxter_g2}), and since the formulation in terms of 3-edge colourings of 
the honeycomb lattice involves only real,
non-negative Boltzmann weights \cite{Baxter70}, the Perron-Frobenius theorem 
ensures that (\ref{Baxter_g2}) correctly gives the entropy of vertex 
4-colourings of the triangular lattice.

A two-dimensional height mapping for the O($n$) FPL model was given in 
\cite{KondevGierNienhuis96}, and it
was shown that the continuum limit is described by a Liouville field theory 
(or two-component CG) involving two compactified bosons with an $n$-dependent 
radius. For $n=2$ this reduces to two free bosons,
so the central charge is $c=2$ in agreement with our result 
(\ref{eq:BAE:RIV:cPotts}) for $Q=4$ ($p \to \infty$).
A slightly different two-dimensional height mapping was given directly for 
the 4-colouring problem in \cite{MooreNewman00}. The critical exponent 
$\eta = 2 x_H = \frac{1}{3}$ was also derived and checked
numerically in \cite{MooreNewman00}, in agreement with our result 
(\ref{eq:RIV:XH}) that gives $x_H = \frac{1}{6}$ for $Q=4$. 
This provides a couple 
of convincing checks that our continuum description is indeed
that of the combinatorial colouring problem.%
\footnote{Note however the watermelons exponents of \cite{KondevGierNienhuis96}
do not compare directly to ours, since these observables are different from 
our spin-one watermelon operators.}

We next turn to the case $Q=3$. Even though this is less than $Q_0$ 
[see (\ref{eq.Q0})], we claim that
regime~IV still correctly describes the physical colouring problem. It is easy 
to see that each of the three sublattices must have a constant colour, 
so $Z = 3!$ for any size $L \in 3 \mathbb{N}$, and this is
compatible with our observation, made in section~\ref{sec:regIV.analysis.BAE}, 
that $f_L(3) = 0$.
A further confirmation comes from the magnetic exponent. Since there is a 
perfect ordering on each sublattice, the (staggered) two-point function is 
constant, so that $x_H = 0$. This indeed agrees with (\ref{eq:BAE:RIV:cPotts}) 
for $Q=3$ ($p = 6$).

Finally, consider the limit $Q \to 2^+$. It is obviously not possible 
to colour an elementary triangular face with only two colours, 
but this does not imply that the ground-state free energy must
diverge in this limit, because the cancellation mechanisms mentioned in 
section~\ref{sec:regIV.analysis.BAE} are at play at $Q=2$. 
Indeed the regime~IV expression (\ref{Baxter_g2}) stays finite in this limit.
However, the central charge (\ref{eq:BAE:RIV:cPotts}), and all critical 
exponents derived above contain a pole at $Q=2$ ($p=4$), which is concomitant 
with the break-down of the colouring interpretation.
On the other hand, it is possible to shift the ground-state energy by an 
infinite amount, so as to obtain a well-defined $Q=2$ theory in which 
each elementary face of the lattice contains
precisely one frustrated edge (i.e., that is adjacent to two equally coloured 
spins). This defines the so-called zero-temperature Ising antiferromagnet, 
which possesses a critical ground-state ensemble corresponding to a $c=1$
CFT of one free boson 
\cite{BloteHilhorst82,NienhuisHilhorstBlote84,BloteNightingale93}. It seems
an appealing idea that this CFT should be (infinitely deeply) buried inside 
regime~IV, but we have not investigated this question in details.

Further support that the critical colouring problem is described by regime~IV 
for $2 < Q \le 4$ is provided by the RSOS representation to be described in 
section~\ref{sec:anyonic}.

It is worth noticing that the results found in this paper favour Baxter's
value \cite{Baxter87} for $Q_0$ [cf. (\ref{eq.Q0})] against the alternative
value $2+\sqrt{3}$ suggested in \cite[section~6]{JacobsenSalasSokal03}. 
Furthermore, this conclusion may be a signal that Baxter's eigenvalue $g_1$ 
(\ref{Baxter_g1}) plays no role inside the critical region of the phase 
diagram in the complex $Q$-plane of the triangular-lattice chromatic polynomial
(as it was investigated in \cite{JacobsenSalasSokal03}). Notice that even 
though $g_1$ is not defined for any real $Q\in(0,4)$, it is well defined 
when $ {\rm Im}\,(Q)\neq 0$. 

Before closing this section, we notice that our results for regime~IV also 
have implications for the phase diagram in the $(Q,v)$ plane, outside the 
chromatic line $v=-1$. According to (\ref{eq:RIV:thermexps}), the thermal 
perturbation of the ground state in regime~IV is irrelevant for all $Q < 4$. 
Therefore the part of the chromatic line where regime~IV is
dominant---that is, $Q_0 < Q < 4$---is attractive under the renormalisation 
group. We conclude that the small `bubble' in figure~\ref{fig:tripd}, which 
is delimited by the bifurcation point $T$,
the two branches of $v_+$ and the vertical line $Q=4$, is in the universality 
class of regime~IV.

%
%
\section{Connection with RSOS models and anyonic chains}
\label{sec:anyonic}
\setcounter{footnote}{1}

In section~\ref{sec:pd_sq} we discussed that at the Beraha numbers, 
$Q = B_k$, all of the unphysical regime~I cancels out of the partition
function. What remains is a purely local model encompassing regime~IV. 
This local model is conveniently formulated as a
restricted solid-on-solid (RSOS) height model, which has independent interest 
because of its relation with anyonic spin chains.

\subsection{The RSOS model associated with the Izergin-Korepin vertex model}

In section~\ref{sec:mappings}, we have explained how the colouring problem on 
the triangular lattice could be mapped onto an appropriately twisted version 
of the IK model, which can be viewed as a spin-1 vertex model: the degrees 
of freedom on each lattice edge are vectors in spin-1 representations of the 
quantum group $U_\mathfrak{q}(sl_2)$, where 
$\mathfrak{q} = \mathrm{e}^{i \frac{\theta}{2}}=\mathrm{e}^{i \frac{\pi}{p}}$. 
These representations can at first stage be understood in the same way as 
classical $SU(2)$ representations, and the interaction between two edges at 
a vertex of the square lattice can therefore be decomposed into projectors 
on the spin $0$, spin $1$ and spin $2$ representations, namely 
\medskip
\begin{equation}
\begin{tikzpicture}[scale=0.6,%
     baseline={([yshift=-.5ex]current bounding box.center)}]
\draw[purple, line width=4pt] (-1,-1) -- (1,1);
\draw[purple, line width=4pt] (1,-1) -- (-1,1);
\node[below] at (-1,-1) {$a$};
\node[below] at (1,-1) {$b$};
\node[above] at (1,1) {$c$};
\node[above] at (-1,1) {$d$};
\end{tikzpicture}
\quad
\propto
\quad
\left(P_2\right)_{ab}^{cd} + \frac{\sin \left(\theta+\frac{u}{2} \right)}
       {\sin \left(  \theta-\frac{u}{2} \right)} \left(P_1\right)_{ab}^{cd}
+ \frac{\cos \left(  \frac{3\theta}{2}+\frac{u}{2} \right)}
       {\sin \left( \frac{3 \theta}{2}-\frac{u}{2} \right)} 
       \left(P_0\right)_{ab}^{cd}
 \label{eq:R:RSOS}
\end{equation}
(for convenience the lattice has been rotated by $45^\circ$ with respect to 
the figures in section \ref{sec:mappings}).

As we will now explain, this allows for a reintepretation of the model in 
terms of integer {\em heights} living on the faces of the square lattice 
(viz.\/ vertices of the dual square lattice, shown in green in the following 
figure), namely in terms of an interaction-round-a-face (IRF)
or solid-on-solid (SOS) model. The construction is adapted from the seminal 
work of Pasquier in the case of spin-$\frac{1}{2}$ models \cite{Pasquier}, 
and can be summed up as follows (for more details see \cite{VJS:a22}): 
\medskip
$$
\begin{tikzpicture}[scale=2]
  \foreach \x in {0,...,5}
{ 
\draw[green, line width=3pt](\x,0) -- (\x+0.5,-0.5) -- (\x+1,0) -- (\x+0.5,0.5) -- cycle;
\draw[purple](\x,-0.5) -- (\x+0.5,0) -- (\x+1,-0.5);
\draw[purple](\x,0.5) -- (\x+0.5,0) -- (\x+1,0.5);
\draw[purple](\x,0.5) -- (\x+0.5,1) -- (\x+1,0.5);
\node[purple] at (\x+0.3,0.8) {$1$};
\node[purple] at (\x+0.7,0.8) {$1$};
\node[purple] at (\x+0.3,0.2) {$1$};
\node[purple] at (\x+0.7,0.2) {$1$};
\node[purple] at (\x+0.3,-0.2) {$1$};
\node[purple] at (\x+0.7,-0.2) {$1$};
 }
   \foreach \x in {0,...,4}
{ 
\draw[green, line width=3pt](\x+0.5,0.5) -- (\x+1,1) -- (\x+1.5,0.5) -- (\x+1,0);
 }
 \draw[green,line width=3pt](0,1) -- (0.5,0.5);
  \draw[green, line width=3pt] (5.5,0.5) -- (6.,1);
 \draw[dashed] (0,-0.5) -- (0,1);
 \draw[dashed] (6,-0.5) -- (6,1);
 \node[below] at (0.5,-0.5) {$j_{1}$};
  \node[below] at (1,0) {$j_{2}$};
  \node[below] at (1.5,-0.5) {$j_{3}$};
    \node[below] at (2,0) {$j_{4}$};
   \node[below] at (2.5,-0.5) {$j_5$};
 \node[below] at (4,-0.6) {$\ldots$};
 \node[above] at (0.5,0.55) {$j'_{1}$};
   \node[above] at (1,1) {$j'_{2}$};
  \node[above] at (1.5,0.55) {$j'_{3}$};
     \node[above] at (2,1) {$j'_{4}$};
   \node[above] at (2.5,0.55) {$j'_5$};
 \node[above] at (4,1.1) {$\ldots$};
\end{tikzpicture}
$$
the $j_i$ and $j'_i$ living on the nodes of the square lattice are 
(half)-integers encoding the representations of $U_\mathfrak{q}(sl_2)$, 
and which must thus be compatible with the $U_\mathfrak{q}(sl_2)$ 
fusion rules: for instance, if representations $j_i$ and $j_{i+1}$ are 
separated by a lattice edge (carrying a spin-1 representation), $j_{i+1}$ 
must belong to the fusion product $j_{i} \times 1$. 

The action on the projectors $P_0,P_1,P_2$ on a given face of the dual lattice,
specified by a set of heights $j_{i-1},j_i,j_{i+1},j'_i$, is worked out by a 
change of basis which makes use of the quantum analog of Wigner's 6$j$ 
symbols and can be represented as follows :
\begin{equation}
\begin{tikzpicture}[scale=2,%
    baseline={([yshift=-.5ex]current bounding box.center)}]
\draw[purple] (0.5,0) -- (0.5,0.5) node[left] {$1$};
\draw[purple] (1.5,0) -- (1.5,0.5) node[left] {$1$};
\draw[green, line width=3pt](0,0) -- (2,0);
\node[below] at (0.,0) {$j_{i-1}$};
\node[below] at (1.,0) {$j_i$};
\node[below] at (2,0) {$j_{i+1}$};
\end{tikzpicture}
\;=\; \sum_J 
 \left\{  \begin{array}{ccc}
  j_{i-1} & 1 & j_i \\ 1 & j_{i+1} & J  \\ 
                \end{array}   \right\}_\mathfrak{q}
\begin{tikzpicture}[scale=2,baseline={([yshift=-.5ex]current bounding box.center)}]
\draw[purple] (1,0.25) -- (0.5,0.5) node[left] {$1$};
\draw[purple] (1,0.25) -- (1.5,0.5) node[right] {$1$};
\draw (1,0) -- node[right]{$J$} (1,0.25); 
\draw[green, line width=3pt](0,0) -- (2,0);
\node[below] at (0.,0) {$j_{i-1}$};
\node[below] at (1.,0) {$j_i$};
\node[below] at (2,0) {$j_{i+1}$};
\end{tikzpicture}
\,,
\label{eq:RSOS:fusion6j}
\end{equation}
and similarly for the upper part of the face. Combining it all, the action 
of a projector $P_J$ reads
\begin{equation}
\begin{tikzpicture}[scale=2,%
      baseline={([yshift=-.5ex]current bounding box.center)}]
\draw[green, line width=3pt](0,0) -- (0.5,-0.5) -- (1,0) -- (0.5,0.5) -- cycle;
\node[below] at (0.5,-0.5) {$j_i$};
\node[left] at (0.,0) {$j_{i-1}$};
\node[right] at (1.,0) {$j_{i+1}$};
\node[above] at (0.5,0.5) {$j'_i$};
\node at (0.5,0) {$P_J$};
\end{tikzpicture}
\quad 
= 
\quad
 \left\{  \begin{array}{ccc}
  j_{i-1} & 1 & j_i \\ 1 & j_{i+1} & J  \\ 
                \end{array}   \right\}_\mathfrak{q}
   \left\{  \begin{array}{ccc}
  j_{i-1} & 1 & j'_i \\ 1 & j_{i+1} & J  \\ 
                \end{array}   \right\}_\mathfrak{q}  \,.
\end{equation} 

We now specify to the case of $\mathfrak{q}$ equal to a root of unity, 
namely $p$ integer, where the fusion rules of the quantum group 
$U_\mathfrak{q}(sl_2)$ truncate, leaving us with a restricted set of 
possible representations 
\begin{equation}
j \;=\; 0,\frac{1}{2}\,1,\ldots , \frac{p}{2}-1 \,. 
\end{equation}
The resulting IRF model then becomes an RSOS model, 
with a finite-dimensional set of degrees of freedom per lattice site. The 
corresponding TM can therefore be computed, and we expect that its spectrum
is related to that of the original vertex model. 
More precisely, it was observed in \cite{VJS:a22} that the leading eigenvalues 
of the RSOS model could be obtained from those of the vertex model in the $m=0$
sector, for values of the twist $\varphi=\theta,3\theta,\ldots$. The 
ground state of the RSOS model is in particular obtained as the ground state 
in the $m=0$ sector, for $\varphi=\theta$. 
Taking for the spectral parameter the special value 
$u=\frac{\pi}{2}-\frac{\theta}{2}$ ensuring the correspondence with the 
Potts model, this state is precisely the ground state of the Potts model 
in regime~IV. 

Let us check this by comparing the spectrum of the two models (colouring model 
and RSOS) for finite-size systems. 

We start with, for instance, $p=36$, which corresponds to 
$Q\simeq 3.969616>Q_0$. 
Among the eigenvalues of the RSOS TM for $L=6$, we find that indeed: 
\begin{itemize}
\item The ground state corresponds to the ground state of the Potts model 
      in the $k=0$ sector.
\item The $11^{\text{th}}$ and $12^{\text{th}}$ eigenvalues correspond to 
      the leading doublet of the Potts model in the sector $k=1$, where 
      the cluster is allowed to wrap around the torus.
\item The $42^\text{th}$ eigenvalue corresponds to the leading eigenvalue 
      in the sector with $k=2$.
\item A very low-lying eigenvalue corresponds to the first excited state in 
      the $k=0$ sector of the Potts model. 
\item Noticeably, no eigenvalue of the RSOS model was found to correspond 
      to the leading quartet of the $k=1$ Potts sector, where the cluster 
      is not allowed to wrap around the torus.
\end{itemize}

The conclusion we can draw from these observations is that, apart from the very 
peculiar non-wrapping $k=1$ sector, the leading states of the Potts model 
in regime~IV are all recovered by the RSOS model. 
Interestingly, the conclusion still holds when we take smaller values of $p$, 
for instance $p=6$ or $p=7$, for which $Q<Q_0$: while the FK representation 
of the Potts model, and equivalently, the twisted IK vertex model are then 
dominated by regime~I, the ground state of the RSOS model is still that 
associated to regime~IV (in particular, the corresponding free energy is 
zero when $ Q = 3$). In other terms, the RSOS model, as the colouring problem, 
is described by regime~IV for any value of $3\leq Q \leq 4$. While it seems 
legitimate, regarding our previous observations for $2<Q\leq 3$, to expect 
that the correspondence still holds down to $Q=2^+$, comparison 
of the Potts and RSOS eigenspectra for $p=5$ ($Q\simeq 2.61803$) shows 
that this might actually not be the case. Indeed, what we identified as 
the continuation of the ground-state energy of regime~IV for this value of 
$Q$ was not found in the spectrum of the RSOS model, 
neither for $L=6$, nor for $L=12$.

Therefore, we can conclude that the RSOS model is, for $p \geq 6$ and for a 
certain range of values of the spectral parameter, described by the CFT 
governing the physics in regime~IV, namely the coset 
\begin{equation}
\frac{SU(2)_4 \times SU(2)_{p-6}}{SU(2)_{p-2}} \,. 
\label{eq:coset}
\end{equation}

As we shall see in the next section, these conclusions can actually be related 
to recent observations relative to the physics of spin-1 anyonic chains 
\cite{Ard}.

\subsection{Spin-1 anyonic chains, brief introduction}

The RSOS model considered in the previous section can be viewed as the 
1+1-dimensional formulation of a spin-1 anyonic chain, which is a deformation 
of the usual Heisenberg chain for $SU(2)$ spins, well suited for the description
of particles with exotic topological properties, the so-called 
non-Abelian anyons (see \cite{Ard}, or \cite{Kitaev} for a more general 
and thorough introduction).

For a given integer $\kappa$, $SU(2)_\kappa$ anyons can be thought of as 
angular momenta indexing the representations of $U_\mathfrak{q}(sl_2)$, 
with $\mathfrak{q} = \mathrm{e}^{i \frac{\pi}{p}}= 
      \mathrm{e}^{i \frac{\pi}{\kappa+2}}$, and therefore restricted to 
\begin{equation}
j \;=\; 0,\frac{1}{2},1, \ldots \frac{\kappa}{2} \,.
\end{equation}

A spin-1 chain of $SU(2)_\kappa$ anyons can be represented as the 
following `fusion tree' 
\medskip
$$
\begin{tikzpicture}
\draw[green, line width=3pt] (0,0) -- (6,0);
  \foreach \x in {0,...,5}
{ 
\draw[purple](\x+0.5,0) -- (\x+0.5,0.5);
 \node[above,purple] at (\x+0.5,0.5) {$1$};
 }
 \draw[dashed] (0,-0.25) -- (0,0.5);
 \draw[dashed] (6,-0.25) -- (6,0.5);
 \node[below] at (1,0.) {$j_{1}$};
  \node[below] at (2,0.) {$j_{2}$};
   \node[below] at (3,0.) {$j_3$};
 \node[below] at (4,-0.1) {$\ldots$};
\end{tikzpicture} 
$$
where as usual the dashed lines indicate for periodic boundary conditions, 
and as in the previous section the anyonic indices $j_i$ must be compatible 
with the $SU(2)_\kappa$ ($U_\mathfrak{q}(sl_2)$) fusion rules. 
As in spin chains with quadratic interactions, the Hamiltonian can be 
written as 
\begin{equation}
H \;=\; \sum_{i} h_{i,i+1} \,,
\label{eq:HRSOS0}
\end{equation}
where $h_{i,i+1}$ encodes the projection of the product of two $j=1$ 
representations over the decomposition 
\begin{equation}
1 \times 1 \;=\; 0 + 1 + 2 \,.
\end{equation}
In particular, we follow the notations of \cite{Ard} and write 
\begin{equation}
h_{i,i+1} \;=\; \cos\theta_{2,1} \left(P_2\right)_{i,i+1} - 
                \sin \theta_{2,1} \left(P_1\right)_{i,i+1} \,.
\label{eq:HRSOS}
\end{equation}
The connection with the fusion tree representation above is worked out 
by changes of basis which rearrange the order in which the anyons are fused, 
which we already described in (\ref{eq:RSOS:fusion6j}).

The fusion tree also provides an easy means of computing the dimension of the 
RSOS TM. The adjacency matrix (in the target space of angular momenta)
takes the form of an $M \times M$ matrix with 
rows and columns indexed by $2j+1$, shown here for $M = \kappa+1 = 6$:
\begin{equation}
A = \left(
\begin{array}{cccccc}
0&1&0&0&0&0\\
1&1&1&0&0&0\\
0&1&1&1&0&0\\
0&0&1&1&1&0\\
0&0&0&1&1&1\\
0&0&0&0&1&0\\
\end{array}
\right)
\end{equation}
Note the 0's in the upper left and lower right corner. The eigenvalues are
\begin{equation}
 \lambda_k \;=\; 4 \cos^2\left( \frac{k \pi}{2 M} \right)  - 1 \,,
\end{equation}
with $k=1,2,\ldots,M$, so the dimension of the TM for a periodic chain of 
length $L$ is ${\rm Tr}\, A^L = \sum_{k=1}^M (\lambda_k)^L$. This is dominated by the $k=1$ 
term, so the asymptotic behaviour is $(Q-1)^L$, the same as the initial 
colouring TM.

\subsection{Spin-1 anyonic chains: CFT results}

The phase diagram of spin-1 $SU(2)_\kappa$ anyonic chains was recently 
investigated in \cite{Ard},%
\footnote{ Further details have been worked out by two of us
and will be presented elsewhere \cite{VJSAnyonic}.}
giving evidence for a wealth of gapped or
critical phases as the angle $\theta_{2,1}$ is varied. An integrable TM and
a one-dimensional quantum Hamiltonian are usually connected
via the very anisotropic limit $u \to 0$ of the spectral parameter. 
This allows one to relate
the lowest Hamiltonian energy levels to the leading TM eigenvalues
for extended regions of the spectral parameter. In the present case, things are
slightly different, since the regime~IV was observed to dominate the RSOS
spectrum for a range of spectral parameters centered around the special value
$u = \frac{\pi}{2}-\frac{\theta}{2}$. This range does not contain the 
anisotropic limit $u\to0$, and in this sense, there is no straightforward 
relation to any properly defined Hamiltonian.
It is however instructive to consider the Hamiltonian $H$ derived from 
the $u\to 0$
limit of the integrable TM, which, in the vertex formulation, is dominated 
by either regimes~I or III (depending on the overall sign of $H$)
for the values of $\theta$ considered in this paper (namely 
$0 \leq \theta \leq \frac{\pi}{3}$). As shall be reported in more detail in 
\cite{VJSAnyonic}, we observe that the eigenlevels corresponding to regime~I 
are actually absent from the RSOS formulation, and that the associated RSOS 
Hamiltonian is instead dominated by those of regime~IV. 
The associated value of the parameter $\theta_{2,1}$ is 
\begin{equation}
\theta_{2,1} \;=\; \pi - 
\arctan \frac{1}{\cos\frac{2\pi}{\kappa+2}-\cos\frac{4\pi}{\kappa+2}} \,,
\end{equation}
and lies precisely inside an extended phase identified in \cite{Ard} as being 
described by the conformal coset (\ref{eq:coset}).

\section{Conclusion}
\label{sec:conclusion}
\setcounter{footnote}{1}

In this paper we have revisited the problem of $Q$-colourings of 
the triangular lattice first investigated by Baxter \cite{Baxter86,Baxter87}. 
Using Bethe Ansatz techniques, this author found three different analytical 
expressions $g_1$, $g_2$, $g_3$ for the free energy in the
thermodynamical limit. The latter two describe the critical range 
$0 \le Q \le 4$, with $g_2$ being dominant for $Q \in (Q_0,4]$ and $g_3$ 
being dominant for $Q \in [0,Q_0)$, where $Q_0 = 3.819\,671\cdots$ 
\cite{Baxter87}.

We have gone beyond these results by determining the conformal field theories 
(CFTs) describing the scaling levels above $g_3$ (regime~I) and above $g_2$ 
(regime~IV).
This was done by mapping the model to a spin-one vertex model on the square
lattice, again amenable to Bethe Ansatz analysis. In particular we have found
analytical expressions for the central charge and various critical exponents 
for both regimes.

While regime~I is well-known from the spin-one loop model on the square 
lattice \cite{BloteNienhuis89}, regime~IV is \emph{new} and emerges because 
of the particular way the vertex model must be twisted in
order to ensure the equivalence with the initial colouring problem. 
Accordingly we have dedicated the bulk of the paper to regime~IV. It 
corresponds to a system of two compact bosons and we have
identified the corresponding coset theory. We have also related its RSOS 
representation to a recently studied spin-one anyonic chain \cite{Ard}.

One of our main conclusions is that regime~IV possesses the correct 
physical properties to describe an actual colouring problem. By analytic 
continuation it can be extended to the range $Q \in (2,4]$, although the 
ground state $g_2$ is only dominant for $Q \in (Q_0,4]$.
In this sense the two CFTs coexist throughout the critical range.
We believe it to be crucial that regime~IV describes
features of the actual colouring problems at both $Q=4$ and $Q=3$,
breaks down for $Q \to 2^+$ as expected from the impossibility of two-colouring
a triangular lattice, and that its restriction in terms of a local RSOS 
height model
maintains the critical exponents of $Q$-colourings at the Beraha 
numbers (\ref{eq.Bk}) for all integer $p \ge 6$.

We have however left for future work the interesting question of what is 
the correct scaling theory above $g_2$ for $Q \in [0,2)$.

The case $Q=4$ of regime~IV is equivalent to several other models, including 
fully-packed loops (FPL) on the honeycomb lattice with fugacity $n=2$. 
The continuum limit of both these models consists simply of two free bosons.
However, while a Coulomb gas (CG) deformation of the FPL model to $n < 2$ 
is well-known \cite{KondevGierNienhuis96}, the CG description of the $Q < 4$ 
colouring problem is presently unknown. Identifying it presents
another challenge for further investigations,  and might eventually
permit one to rederive the critical exponents given here much easier 
(albeit less rigorously), without any use of the Bethe Ansatz 
\cite{KondevGierNienhuis96}.

Finally, it would be interesting to study boundary extensions of the 
$Q$-colouring problem, along the lines of \cite{JacobsenSaleur08bdrychrom},
by replacing the geometry of the cylinder (torus) with that of an annulus.
In particular, one could expect from \cite{JS08-1BTL,DJS09-2BTL} that 
restricting the number of colours allowed
on the boundary would lead to sets of continuously varying critical exponents.

\ack
We acknowledge useful discussions and correspondence with Yacine Ikhlef, 
Alan Sokal, and Paul Zinn-Justin. We also thank Alan Sokal for 
the kind permission to use computational resources (provided by Dell
Corporation) at New York University. 

J.S. is grateful for the hospitality from 2009 to 2015
of the Laboratoire de Physique Th\'eorique at the \'Ecole Normale 
Sup\'erieure where part of this work was done.  
 
The research of J.L.J. was supported in part by the Agence Nationale
de la Recherche (grant ANR-10-BLAN-0414: DIME), the Institut
Universitaire de France, and the European Research Council (through the 
advanced grant NuQFT).
The research of J.S. was supported in part by Spanish MINECO grants
FIS2012-34379 and FIS2014-57387-C3-3-P.  

\section*{References}
\bibliographystyle{iopart-num}
\bibliography{VJS}

\end{document}